%% file: dissertation.tex
\documentclass[paper=a4,pagesize=auto,DIC=off,MRes,onehalfspacing]{DissertationClass}

\usepackage{amsmath,amssymb,amsfonts,amsthm}% usual AMS math packages
\usepackage{graphicx}                       % standard graphics handler
\usepackage[table]{xcolor}                         % colours for the hyperlinks
\usepackage{framed}                         % for Mathematical Aside boxes
\usepackage{colonequals}                    % for :=
\usepackage[margin=1cm,font=small,labelfont=bf,format=plain,labelsep=endash]{caption}   % prettier captions
\usepackage[hang,flushmargin]{footmisc}     % controls footnote appearance
\usepackage{subcaption}
\usepackage{calc}                           % for calculating lengths
\usepackage{multirow}
\usepackage{hhline}
\usepackage[square,comma,numbers]{natbib}   % for proper APS style citation
\newgeometry{left=3cm,right=3cm,top=3cm,bottom=3cm}

\usepackage[
  pdfstartview={XYZ null null 1},	
  colorlinks,
  linkcolor=blue,
  urlcolor=blue,
  citecolor=blue,
  plainpages=false,
  pdfpagelabels,
  backref=page, %Pone links de retache en la bibliografía.
	pdftex,
]{hyperref}
\hypersetup{
pdftitle={Under-the-barrier electron-ion interaction during tunnel ionization},
pdfauthor={Emilio Pisanty Alatorre},
pdfkeywords={}
}
\usepackage{doi}
\RequirePackage[hyperpageref]{backref}
  \renewcommand*{\backref}[1]{}
  \renewcommand*{\backrefalt}[4]{
    \ifcase #1
       Not cited.
    \or
       Cited 1 time on p. #2.
    \else
       Cited #1 times on pp. #2.
    \fi}

\renewcommand{\Im}{\operatorname{Im}}

\renewcommand{\d}{\ensuremath{\textrm{d}}}
\newcommand{\ch}{\ensuremath{\mathcal{H}}}

\newcommand{\bra}[1]{\ensuremath{\left\langle#1\right|}}
\newcommand{\ket}[1]{\ensuremath{\left|#1\right\rangle}}
\newcommand{\bracket}[2]{\ensuremath{\left\langle#1 \vphantom{#2}\middle|  #2 \vphantom{#1}\right\rangle}}
\newcommand{\matrixel}[3]{\ensuremath{\left\langle #1 \vphantom{#2#3} \middle| #2 \middle| #3 \vphantom{#1#2} \right\rangle}}

\newcommand{\smallket}[1]{\ensuremath{|#1\rangle}}
\newcommand{\outerp}[1]{\ensuremath{\ket{#1}\!\bra{#1}}}

\newcommand{\vb}[1]{\mathbf{#1}}
  \newcommand{\vbr}{\vb{r}}
  \newcommand{\vbk}{\vb{k}}
  \newcommand{\vbp}{\vb{p}}
\newcommand{\rl}{\ensuremath{\vb{r}_\textrm{L}}}
\newcommand{\free}{\!{}_\textrm{free}}
\newcommand{\eff}{\textrm{eff}}
 
\renewcommand{\l}{\lambda} % or {{l'}} ?
\newcommand{\m}{\mu} % or {{m'}} ?
\renewcommand{\AA}{\mathcal{A}}
\newcommand{\Int}{\operatorname{Int}}
\newcommand{\ff}{{}_1 F_1}
\newcommand{\FF}{{}_2 F_1}
\newcommand{\coef}{\operatorname{Coef}}
\newcommand{\mm}{{\ensuremath{\mathsf{m}}}}
\newcommand{\nn}{{\ensuremath{\mathsf{n}}}}
\newcommand{\tauT}{{\ensuremath{\tau_{\scriptscriptstyle\textrm{T}}}}}

\DeclareSymbolFont{extraup}{U}{zavm}{m}{n}
\DeclareMathSymbol{\varheart}{\mathalpha}{extraup}{86}
\DeclareMathSymbol{\vardiamond}{\mathalpha}{extraup}{87}
\renewcommand{\labelitemi}{$\vardiamond$} \renewcommand{\labelitemii}{$\triangleright$}

\catcode`@=11
\def\nvphantom{\v@true\h@false\nph@nt}
\def\nhphantom{\v@false\h@true\nph@nt}
\def\nphantom{\v@true\h@true\nph@nt}
\def\nph@nt{\ifmmode\def\next{\mathpalette\nmathph@nt}%
  \else\let\next\nmakeph@nt\fi\next}
\def\nmakeph@nt#1{\setbox\z@\hbox{#1}\nfinph@nt}
\def\nmathph@nt#1#2{\setbox\z@\hbox{$\m@th#1{#2}$}\nfinph@nt}
\def\nfinph@nt{\setbox\tw@\null
  \ifv@ \ht\tw@\ht\z@ \dp\tw@\dp\z@\fi
  \ifh@ \wd\tw@-\wd\z@\fi \box\tw@}
\definecolor{shadecolor}{gray}{.90}
\newlength{\boxwidth}
\newsavebox{\boxcontainer}
{\endMakeFramed}
\usepackage{tocloft}
\newcommand{\listofasides}{List of Mathematical Asides}
\newlistof[chapter]{parmat}{prm}{\listofasides}
\newenvironment{parmat}[1]{%
  \refstepcounter{parmat}
  \addcontentsline{prm}{parmat}
  {\protect\numberline{\theparmat}#1}\par
  \vspace{5pt}\textbf{Mathematical Aside~\theparmat. #1}\vspace{-7pt}
  \nopagebreak[4]
    \MakeFramed {\setlength{\boxwidth}{0.95\textwidth}
    \addtolength{\boxwidth}{-2\FrameSep}
    \addtolength{\boxwidth}{-2\FrameRule}
    \setlength{\hsize}{\boxwidth} \FrameRestore}
}%
{\endMakeFramed}

\renewcommand{\theparmat}{\thechapter.\arabic{parmat}}
\title{Under-the-barrier electron-ion interaction during tunnel ionization}
\author{Emilio Pisanty Alatorre}
\date{\today}
\department{Physics}
\centre{Centre for Doctoral Training in Controlled Quantum Dynamics}
\field{Controlled Quantum Dynamics}
\supervisor{Prof. Misha Ivanov \\and Prof. Jon Marangos}
\crest{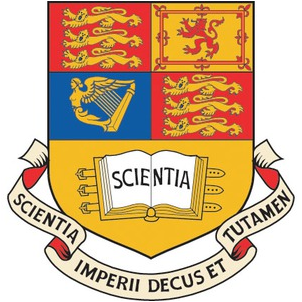}

\dedication{}

\usepackage[
disable,
author={Emilio Pisanty}
]{pdfcomment}                    % Leaving comments on pdfs

\begin{document}

\maketitle

\input{abstract}

%\makededication
\phantomsection
\tableofcontents
\addcontentsline{toc}{chapter}{Contents}
\listofparmat

%\listoftables
%\listoffigures

\input{content}

\singlespacing
%\bibliographystyle{arthur}
%\bibliography{references}{}
\input{RevisedBibliography.bbl}

\addcontentsline{toc}{chapter}{References}

\end{document}

%% file: abstract.tex
\begin{abstract}
We consider tunnel ionization of an atom or molecule in a strong field within an analytical treatment of the $R$-matrix method, in which an imaginary boundary is set up inside the classically forbidden region that acts as a source of ionized electrons. These electrons are then propagated in the semiclassical approximation, and relying on a numerical solution of the inner region, which is accessible using quantum chemical techniques, we describe the subsequent evolution of the ionized electron and the ionic core.

Importantly, we show that correlation interactions between the ionized electrons and those left behind in the core can play a role during ionization, and that interactions can occur while the electron is still inside the classically forbidden region that enable tunnelling from channels normally subject to far greater exponential suppression since they are subject to higher and wider tunnelling barriers.

This interaction can be described analytically using saddle-point methods that find the dominant contributions to the temporal, spatial and momentum integrals that make up the expressions for the ionization yield. However, these methods yield results for the angular distributions of the ionized electron that do not necessarily match experiment and which yield physically unsatisfactory solutions.

In this report we develop a formalism to evaluate these integrals exactly, which yields more precise calculations of the angular distributions while at the same time providing the language, in terms of exchange of angular momentum, with which to understand their origins and their differences from previous cases, thus giving an insight into the fundamental physics of the correlation interaction process.

\end{abstract}

%% file: content.tex
\chapter{Introduction}

\input{introduction}

\chapter{Preliminaries}
\label{chappreliminaries}
In this chapter we will lay the ground-work, as developed in references \cite{saepaper,mepaper} by Torlina, Smirnova, Walters and Ivanov, with which we will treat ionization problems as discussed in the Introduction. We will therefore start with an account of the general formalism in which this calculation is inscribed, generally known as analytical $R$-matrix theory.
%This dissertation will begin by describing the work done by Torlina, Smir\-nova and coworkers \citep{saepaper,mepaper}, and then develop newer work at suitable opportunities. We work in the framework of multi-electron dynamics but incorporate single-electron results as they become available and necessary.

\input{c1-ARMtheory}

\input{c1-hamiltonian}

\input{c1-multichannel}

\input{c1-DysonOrbital}

\input{c1-propagator-Dyson-expansion}

\chapter{Direct tunneling and the single-electron case}
\label{chapterdirect}

\input{c2-direct-tunnelling}

\input{c2-abstractRfactor}

\input{c2-hydrogenicorbitals}

\chapter{Correlation-driven ionization}
\label{chapcrossionization}

\input{c3-second-order-setup}

\input{c3-saddle-point-argument}

\section{Exact calculation of the transverse integrals}
\label{start-of-the-real-work}

\input{c3-setup-and-ansaetze}

\input{c3.s3-re-representations}

\subsection{The final transverse integral}
This solves at a stroke our previous problems with the integral over $\rho$ in equation \eqref{int1-after-k-integral}, which can now be reduced to a far simpler form in which the only change from the (manageable) $k_\perp$ integral is an increased power of the integral variable:

\begin{align}
\Int_1(\vbp,t'')  &  =
%\\  &  =
  i^{|m|+|\m|+|m+\m|} 
  \frac{i^{\m} (-1)^{\l+
	m
	}}{2^{|\m|} |\m|!} 
	\sqrt{\frac{(2\l+1)(\l+|\m|)!}{4\pi(\l-|\m|)!}}
	\frac{Q_{\l\m}}{\kappa^{|m|}} 
	e^{i(m+\m) \phi_p}
%\displaybreak[1]
\nonumber \\ & \qquad \times
  \sum_{j=0}^\infty
	\frac{1}{2\pi}
	\! \int_{-\infty}^\infty \!\!\! \d z
	\! \int_{-\infty}^\infty \!\!\! \d k_\parallel
	  \frac{K(k_\parallel)}{(-z)^{\l+|\m|+2j+1}}
	  e^{i\left(k_\parallel-p_\parallel\right)z}
	  e^{-\frac{i}{2} \int_{t_s}^{t''}\left(k_\parallel+A_\parallel(\tau)\right)^2\d\tau}
%\displaybreak[3]
\nonumber \\ & \qquad  \times
  \coef(\l,|\m|;j)
	\int_0^\infty 
	  \frac{\rho^{|m|+|\m|+2j+1}}{\xi^{|m|+1}}
	  e^{-\frac{1}{2\xi}\rho^2}
	  J_{|m+\mu|}(p_\perp \rho)
	\d \rho
	.
%\nonumber \\ & \qquad  \times
\label{int1-pre-rho-integral}
\end{align}

Our task is now, therefore, to calculate in this approximation the remaining transverse integral, which can be expressed as
\begin{align}
\Int_2(p_\perp,t'')  &  =
	\int_0^\infty 
	  \frac{\rho^{|m|+|\m|+2j+1}}{\xi^{|m|+1}}
	  e^{-\frac{1}{2\xi}\rho^2}
	  J_{|m+\mu|}(p_\perp \rho)
	\d \rho
	.
\end{align}
This integral can again be found in Gradshteyn and Ryzhik \cite[eq. 6.631.1]{gradshteyn}, but in the interest of demistifying our later results and in the spirit that whatever calculations can be done \textit{should} be done, we will calculate it directly. Additionally, it contains our earlier integral \eqref{GandRuno} as a special case.

%%%%%%%%%%%%%%%%%%%%%%%%%%%%%%%%%%%%%%%%%%%%%%%%%%%%%%%%%%%
%%%%%%%%%%%%%%%%%%%%%%%%%%%%%%%%%%%%%%%%%%%%%%%%%%%%%%%%%%%
\begin{parmat}{Explicit calculation of the $\rho$ integral}
\noindent
The calculation, given the rather esoteric-looking result, is in fact quite prosaic, and simply uses the series expansion \cite[eq. \href{http://dlmf.nist.gov/10.2.E2}{10.2.2}]{NIST-handbook} of the Bessel function:
\begin{align}
\int_0^\infty 
  x^k
 & 
	e^{-\alpha x^2}
	J_\nu(\beta x)
\d x
 =
\int_0^\infty 
  x^k
	e^{-\alpha x^2}
	\left(\frac{1}{2}\beta x\right)^\nu
	\sum_{n=0}^\infty
	  (-1)^n
		\frac{
		  \left(\frac{1}{2}\beta x\right)^{2n}
		  }{
			k!
			\Gamma(\nu+n+1)
			}
\d x
%%%%%%
\nonumber \\ & =
%%%%%%
	\sum_{n=0}^\infty
  (-1)^n
	\frac{
	  \left(\frac{1}{2}\beta \right)^{2n+\nu}
	  }{
		k!
		\Gamma(\nu+n+1)
	}
  \int_0^\infty 
    x^{k+\nu+2n}
	  e^{-\alpha x^2}
  \d x
%%%%%%
\nonumber \\ & =
%%%%%%
	\sum_{n=0}^\infty
  \frac{(-1)^n}{2}
	\frac{
	  \left(\frac{1}{2}\beta \right)^{2n+\nu}
    \alpha^{-\frac{k+\nu+2n+1}{2}}
	  }{
		k!
		\Gamma(\nu+n+1)
	}
	\Gamma\left(\frac{k+\nu+2n+1}{2}\right)
%%%%%%
\nonumber \\ & =
%%%%%%
  \frac{
	  \beta^\nu
		}{
		2^{\nu+1}
		\alpha^{\frac{k+\nu+1}{2}}
	}
	\frac{
		\Gamma\left(\frac{k+\nu+1}{2}\right)
		}{
	  \Gamma(\nu+1)
	}
	\sum_{n=0}^\infty
	\frac{
	  \Gamma(\nu+1)
	  }{
		\Gamma(\nu+n+1)
	}
	\frac{
	  \Gamma\left(\frac{k+\nu+1}{2}+n\right)
		}{
		\Gamma\left(\frac{k+\nu+1}{2}\right)
	}
	\frac{
	  1
		}{
		k!
	}
	\left(-\frac{\beta^2}{4\alpha}\right)^{n}
%%%%%%%
%\nonumber \\ & =
%%%%%%%
	%\frac{
		%\Gamma\left(\frac{k+\nu+1}{2}\right)
		%}{
	  %\Gamma(\nu+1)
	%}
  %\frac{
	  %\beta^\nu
		%}{
		%2^{\nu+1}
		%\alpha^{\frac{k+\nu+1}{2}}
	%}
%\sum_{n=0}^\infty
	%\frac{
	  %\left(\frac{k+\nu+1}{2}\right)_n
		%}{
		%(\nu+1)_n
	%}
	%\frac{1}{k!}
	%\left(-\frac{\beta^2}{4\alpha}\right)^{n}
%%%%%%
\nonumber \\ & =
%%%%%%
	\frac{
		\Gamma\left(\frac{k+\nu+1}{2}\right)
		}{
	  \Gamma(\nu+1)
	}
  \frac{
	  \beta^\nu
		}{
		2^{\nu+1}
		\alpha^{\frac{k+\nu+1}{2}}
	}
	\ff\left(\frac{k+\nu+1}{2};\nu+1;-\frac{\beta^2}{4\alpha}\right)
	.
\label{int3calculation}
\end{align}
\end{parmat}

%%%%%%%%%%%%%%%%%%%%%%%%%%%%%%%%%%%%%%%%%%%%%%%%%%%%%%%%%%%
%%%%%%%%%%%%%%%%%%%%%%%%%%%%%%%%%%%%%%%%%%%%%%%%%%%%%%%%%%%

Thus it is possible to express our integral as a hypergeometric function of type $\ff$. This is known as the \textit{confluent hypergeometric function} (also called the Kummer function and sometimes denoted $M(a,b,z)$ or $\Phi(a,b,z)$), because it can be obtained from the gaussian hypergeometric function by the limit process
$$
\ff(a;c;z)
=
\lim_{b\rightarrow\infty}
\FF\left(a,b;c;\frac{z}{b}\right),
$$ 
which merges the singularities of the latter at $z=1$ and $z=\infty$ into a single one at $\infty$.

If we now substitute the values $k=|m|+|\m|+2j+1$, $\nu=|m+\m|$, $\alpha=1/2\xi$ and $\beta=p_\perp$ into our result \eqref{int3calculation}, we obtain, then,
\begin{align}
\Int_2(p_\perp,t'')
 & 
 =
%\frac{1}{\xi^{|m|+1}}
	%\int_0^\infty 
	  %\rho^{|m|+|\m|+2j+1}
	  %e^{-\frac{1}{2\xi}\rho^2}
	  %J_{|m+\mu|}(p_\perp \rho)
	%\d \rho
%%%%%%%
%\\ & =
%%%%%%%
%\frac{1}{\xi^{|m|+1}}
	%\frac{
		%\Gamma\left(\frac{|m|+|\m|+2j+1+|m+\m|+1}{2}\right)
		%}{
	  %\Gamma(|m+\m|+1)
	%}
  %\frac{
	  %p_\perp^{|m+\m|}
		%}{
		%2^{|m+\m|+1}
		%\left(\frac{1}{2\xi}\right)^{\frac{|m|+|\m|+2j+1+|m+\m|+1}{2}}
	%}
%\nonumber \\ & \qquad  \times
	%\ff\left(\frac{|m|+|\m|+2j+1+|m+\m|+1}{2};|m+\m|+1;-\frac{1}{2}\xi p_\perp^2\right)
%%%%%%
%\\ & =
%%%%%%
		2^{\frac{|m|+|\m|-|m+\m|}{2}+j}
		\xi^{\frac{|m+\m|+|\m|-|m|}{2}+j}
	  p_\perp^{|m+\m|}
	\frac{
		\Gamma\left(\frac{|m|+|\m|+|m+\m|}{2}+j+1\right)
		}{
	  \Gamma(|m+\m|+1)
	}
\nonumber \\ & \qquad  \times
	\ff\left(\frac{|m|+|\m|+|m+\m|}{2}+j+1;|m+\m|+1;-\frac{1}{2}\xi p_\perp^2\right)
	.
\label{int2rightaftersubstitution}
\end{align}

To elucidate this relatively complex expression, we tackle first the complicated combinations of absolute values of $m$ and $\m$, one of which has already appeared in the phase of equation \eqref{int1-pre-rho-integral}. We do this on a case-by case basis, dealing first with the symmetric combinations $\frac{1}{2}\left(|m|+|\m|-|m+\m|\right)$ and $\frac{1}{2}\left(|m|+|\m|+|m+\m|\right)$:
\begin{itemize} \itemsep1pt \parskip0pt \parsep0pt \renewcommand{\labelitemi}{$\vardiamond$} \renewcommand{\labelitemii}{$\triangleright$}
	\item If $m\m\geq0$, i.e. if $m$ and $\m$ have the same sign, without loss of generality we may assume that $m=|m|\geq|\m|=\m\geq 0$, in which case
	\begin{itemize} \itemsep1pt \parskip0pt \parsep0pt
		\item $\frac{1}{2}\left(|m|+|\m|-|m+\m|\right)=\frac{1}{2}\left(m+\m-(m+\m)\right)=0$ and
		\item $\frac{1}{2}\left(|m|+|\m|+|m+\m|\right)=\frac{1}{2}\left(m+\m+(m+\m)\right)=m+\m=|m+\m|$.
	\end{itemize}
	\item If $m\m\leq 0$, then without loss of generality we assume that $m=|m|\geq|\m|=-\m\geq 0$, so that
	\begin{itemize} \itemsep1pt \parskip0pt \parsep0pt
		\item $\frac{1}{2}\left(|m|+|\m|-|m+\m|\right)=\frac{1}{2}\left(m-\m-(m+\m)\right)=-\m=|\m|=\min\{|m|,|\m|\}$ and
		\item $\frac{1}{2}\left(|m|+|\m|+|m+\m|\right)=\frac{1}{2}\left(m-\m+(m+\m)\right)=m=(m+\m)-\m=|m+\m|+\min\{|m|,|\m|\}$.
	\end{itemize}
\end{itemize}
For the asymmetric combination $\frac{1}{2}\left(|m+\m|+|\m|-|m|\right)$ we are somewhat more restricted and we therefore have
\begin{itemize} \itemsep1pt \parskip0pt \parsep0pt 
	\item If $m\m\geq0$ we can assume $m\geq0$ and $\m\geq0$ so $m+\m\geq0$ and therefore 
	\begin{itemize} \itemsep1pt \parskip0pt \parsep0pt
		\item $\frac{1}{2}\left(|m+\m|+|\m|-|m|\right)=\frac{1}{2}\left(m+\m+\m-m\right)=\m=|\m|$.
	\end{itemize}
	\item The case $m\m\leq 0$ is slightly more complicated since the cases $|m|\geq|\m|$ and $|m|\leq|\m|$ yield different results:
	\begin{itemize} \itemsep1pt \parskip0pt \parsep0pt
		\item if (without loss of generality) $m=|m|\geq|\m|=-\m\geq 0$ then 
		
		\hspace{1cm}$\frac{1}{2}\left(|m+\m|+|\m|-|m|\right)=\frac{1}{2}\left(m+\m-\m-m\right)=0,$
		
		whereas
		\item if (without loss of generality) $\m=|\m|\geq|m|=-m\geq 0$ then 
		
		\hspace{1cm}$\frac{1}{2}\left(|m+\m|+|\m|-|m|\right)=\frac{1}{2}\left(m+\m+\m+m\right)=m+\m=|m+\m|.$
	\end{itemize}
\end{itemize}
To summarize, then, we have that
\begin{subequations}
\begin{align}
\frac{|m|+|\m|-|m+\m|}{2}
&=
\begin{cases}
  0
	&\textrm{if } m\m\geq0,
	\\
	\min\{|m|,|\m|\}
	\hphantom{|m+\m|+\min\{|m|,|\m|\}}   \nhphantom{1 \min\{|m|,|\m|\}}
	&\textrm{if } m\m\leq0;
\end{cases}
\label{casesone}
\\
\frac{|m|+|\m|+|m+\m|}{2}
& =
\begin{cases}
  |m+\m|
	&\textrm{if } m\m\geq0,
	\\
	|m+\m|+\min\{|m|,|\m|\}
	\nhphantom{1}
	&\textrm{if } m\m\leq0;
\end{cases}
\label{casestwo}
\\
\frac{|m+\m|+|\m|-|m|}{2}
& =
\begin{cases}
  |\m|
	&\textrm{if } m\m\geq0,
	\\
	|m+\m|
	&\textrm{if } m\m\leq0\textrm{ and }|m|\leq|\m|,
	\\
	0 
	\hphantom{|m+\m|+\min\{|m|,|\m|\}}\nhphantom{01}
	&\textrm{if } m\m\leq0\textrm{ and }|\m|\leq|m|.
\end{cases}
\label{casesthree}
\end{align}
\end{subequations}
A table of the first of these expressions is presented later on in table \ref{table-of-power-and-degree}, along with a discussion of the physical significance of its appearance in our final results.

One particular feature of our result \eqref{int2rightaftersubstitution} that comes to light with this analysis is that the first two arguments of the hypergeometric function,
$$a=\frac{|m|+|\m|+|m+\m|}{2}+j+1\quad\textrm{and}\quad b=|m+\m|+1,$$
differ only by the integer $\nu=\frac{1}{2}\left(|m|+|\m|-|m+\m|\right)+j$. This is important, because it can easily be seen using computer algebra software that hypergeometric functions of the form $\ff(b+\nu;b;z)$ are equal to the exponential $e^z$ multiplied by a polynomial of degree $\nu$ in $z$. We will give two proofs of this fact: the first because of its intrinsic beauty and its connection to the deeper mathematics behind the hypergeometric function, and the second because it will allow us to obtain an explicit connection to the more straightforward Laguerre polynomials.

%%%%%%%%%%%%%%%%%%%%%%%%%%%%%%%%
%%%%%%%%%%%%%%%%%%%%%%%%%%%%%%%%
\begin{parmat}{Proof by differential operator }
\label{firstproofmathaside}
\noindent
The key idea to the first proof is that due to a lucky cancellation the coefficient in front of $z^n/n!$ in the series for $\ff(b+\nu;b;z)$,
$$
\frac{(b+\nu)_n}{(b)_n}
=\frac{\Gamma(b+\nu+n)}{\Gamma(b+\nu)}
 \frac{\Gamma(b)}{\Gamma(b+n)}
=\frac{(b+n)_\nu}{(b)_\nu}
,
$$
is a polynomial in $n$, and not a rational function as is normally the case. We can therefore view the $n$ inside the Pochhammer symbol $(b+n)_\nu$ as the eigenvalue of $z^n$ under the differential operator $z\frac{\d}{\d z}$, and replace $n$ by the operator:
$$
\frac{(b+\nu)_n}{(b)_n}\frac{z^n}{n!}
=\frac{\left(b+z\frac{\d}{\d z}\right)_\nu}{(b)_\nu}\frac{z^n}{n!}.
$$

We can then sum over $n$ to get $\ff(b+\nu;b;z)$ on the left-hand side, while on the right-hand side the differential operator is independent of $n$ and can be factored out to leave the series for the exponential $e^z$:
\begin{equation}
\ff(b+\nu;b;z)
=\frac{\left(b+z\frac{\d}{\d z}\right)_\nu}{(b)_\nu}e^z.
\label{explicitDOexpression}
\end{equation}
This is a slight abuse of notation, but it is clearly an exponential times a polynomial of degree $\nu$ in $z$ and it yields us a beautiful, Rodrigues-like expression for said polynomial.

This result, along with similar ones seen in the classical theory of hypergeometric functions \cite[p.~59]{Erdelyi-I}, is an expression of the deep connections between both the general and the confluent hypergeometric functions and the theory of operator calculus \cite{copeland, mathoverflow}.

\end{parmat}
%%%%%%%%%%%%%%%%%%%%%%%%%%%%%%%%
%%%%%%%%%%%%%%%%%%%%%%%%%%%%%%%%

The second proof uses more standard techniques for factoring out the exponential dependence, which we can and should handle separately, from the more interesting polynomial dynamics; in particular it relies on Kummer's transformations for hypergeometric functions.
\pagebreak

\begin{parmat}{Kummer's transformation}
\noindent
We will now prove Kummer's (first) transformation, which works essentially at the level of the hypergeometric differential equation \eqref{hypergeometric-differential-equation} and is best seen as a way to obtain new solutions to the equation starting from known ones. As such, it reads
\begin{equation}
\ff(a;b;z)=e^z\ff(b-a;b;-z).
\label{kummers-transformation}
\end{equation}
\begin{proof}[\textbf{\textup{Proof}}]$\quad$ 

\noindent
To bring the proof to the language of differential equations, we note that the function $w=\ff(a;b;z)$ is the unique solution to the hypergeometric initial value problem \cite[eq.~\href{http://dlmf.nist.gov/13.2.E1}{13.2.1}]{NIST-handbook}
\begin{equation}
  z\frac{\d^2w}{\d z^2}+(b-z)\frac{\d w}{\d z}-a\, w=0\textrm{ under } w(0)=1,\,\, \frac{\d w}{\d z}(0)=\frac{a}{b}\, .
\label{hypergeometric-IVP}
\end{equation}
To prove the transformation formula, then, all we need to do is prove that $\omega(z)=e^{z}w(-z)=e^{z}\ff(a;b;-z)$ solves the modified initial value problem corresponding to $\ff(b-a;b;z)$:
\begin{align*}
0
& \stackrel{?}{=}
z\frac{\d^2\omega}{\d z^2}+(b-z)\frac{\d \omega}{\d z}-(b-a)\, \omega
\\ & =
ze^{z}\left[w(-z)-2\frac{\d w}{\d z}(-z)+\frac{\d^2w}{\d z^2}(-z)\right]
\\ & \qquad \quad \,\,
+(b-z)e^{z}\left[w(-z)-\frac{\d w}{\d z}(-z)\right]-(b-a)e^{z}w(-z)
\\ & =
(-e^{z})\left( 
  (-z)\frac{\d^2w}{\d z^2}(-z)
	+\left(b-(-z)\right)\frac{\d w}{\d z}(-z)
	-a w(-z)
\right)
,
\end{align*}
which is zero because $w$ solves \eqref{hypergeometric-IVP}. As for the initial condition, $\omega(0)=1$ is trivial and $\frac{\d\omega}{\d z}(0)=e^{0}\left(w(0)-\frac{\d w}{\d z}(0)\right)=1-\frac{a}{b}=\frac{b-a}{a}$ only slightly less so. Since they satisfy the same initial value problem, then, we must have $\omega(z)=\ff(b-a;b;z)$, which completes the proof.
\end{proof}
\end{parmat}
%%%%%%%%%%%%%%%%%%%%%%%%%%%%%%%%%%%%%%%%%%%%%%%%%%%%%%%%%%%%%%%%%%%%%%%
%%%%%%%%%%%%%%%%%%%%%%%%%%%%%%%%%%%%%%%%%%%%%%%%%%%%%%%%%%%%%%%%%%%%%%%

Applying Kummer's transformation to our particular case, we have then that
\begin{align*}
\ff & \left(\frac{|m|+|\m|+|m+\m|}{2}+j+1  ;  |m+\m|+1  ;  -\frac{1}{2}\xi p_\perp^2\right)
\\ & \qquad\qquad\qquad\quad =
e^{-\frac{1}{2}\xi p_\perp^2}
\ff\left(-\left(\frac{|m|+|\m|-|m+\m|}{2}+j\right);|m+\m|+1;+\frac{1}{2}\xi p_\perp^2\right)
.
\label{kummer-applied}
\end{align*}
Since the first argument of the confluent hypergeometric function on the right-hand side is the nonpositive integer $-\nu$, for $\nu=\frac{1}{2}\left(|m|+|\m|-|m+\m|\right)+j\geq0$, the right hand side is of the form of a polynomial of degree $\nu$ times the desired exponential, and we get a new representation of the polynomial. This we can now simplify significantly by applying the standard connection of the Laguerre polynomials to the hypergeometric functions \cite[eq.~\href{http://dlmf.nist.gov/18.11.E2}{18.11.2}]{NIST-handbook},
\begin{subequations}
\begin{equation}
L_n^{(\alpha)}(x)=\frac{(\alpha+1)_n}{n!}\ff\left(-n;\alpha+1;x\right),
\label{Laguerre-hypergeometric}
\end{equation}
adapted to our case as
\begin{equation}
\ff\left(-\nu;|m+\m|+1;\frac{1}{2}\xi p_\perp^2\right)=\frac{\nu!}{(|m+\m|+1)_{\nu}}L_{\nu}^{(|m+\m|)}\left(\frac{1}{2}\xi p_\perp^2\right).
\label{eq:Laguerre-hypergeometric-adapted}
\end{equation}
\end{subequations}

We can then put all these results together to give a final expression for the $\rho$ integral,
\begin{align}
\Int_2(p_\perp,t'')
 & =
		%2^{\frac{|m|+|\m|-|m+\m|}{2}+j}
		%\xi^{\frac{|m+\m|+|\m|-|m|}{2}+j}
	  %p_\perp^{|m+\m|}
	%\frac{
		%\Gamma\left(\frac{|m|+|\m|+|m+\m|}{2}+j+1\right)
		%}{
	  %\Gamma(|m+\m|+1)
	%}
%\nonumber \\ & \qquad  \times
	%\ff\left(\frac{|m|+|\m|+|m+\m|}{2}+j+1;|m+\m|+1;-\frac{1}{2}\xi p_\perp^2\right)
%\\ & =
  %\frac{
	  %2^{\frac{|m|+|\m|-|m+\m|}{2}+j}
		%}{
		%(|m+\m|+1)_{\nu}
	%}
	%\frac{
		%\nu!
		%\Gamma\left(\frac{|m|+|\m|+|m+\m|}{2}+j+1\right)
		%}{
	  %\Gamma(|m+\m|+1)
	%}
%\nonumber \\ & \qquad  \times
	%\xi^{\frac{|m+\m|+|\m|-|m|}{2}+j}
	%p_\perp^{|m+\m|}
	%e^{-\frac{1}{2}\xi p_\perp^2}
	%L_{\nu}^{(|m+\m|)}\left(\frac{1}{2}\xi p_\perp^2\right)
%\\ & =
	%2^{\frac{|m|+|\m|-|m+\m|}{2}}
	%2^{j}
  %%\frac{
	  %%\Gamma(|m+\m|+1)
		%%}{
	  %%\Gamma(|m+\m|+1)
	%%}
		%%\Gamma(\nu+1)
	%%\frac{
		%%\Gamma\left(\frac{|m|+|\m|+|m+\m|}{2}+j+1\right)
		%%}{
		%%\Gamma(|m+\m|+\frac{|m|+|\m|-|m+\m|}{2}+j+1)
	%%}
		%\Gamma\left(\frac{|m|+|\m|-|m+\m|}{2}+j+1\right)
%\nonumber \\ & \qquad  \times
	%\xi^{\frac{|m+\m|+|\m|-|m|}{2}+j}
	%p_\perp^{|m+\m|}
	%e^{-\frac{1}{2}\xi p_\perp^2}
	%L_{\nu}^{(|m+\m|)}\left(\frac{1}{2}\xi p_\perp^2\right)
%\\ & =
	2^{\frac{|m|+|\m|-|m+\m|}{2}}
	\Gamma\left({\scriptstyle\frac{|m|+|\m|-|m+\m|}{2}+1}\right)
	\times
	2^{j}
  %\frac{
		%\Gamma\left(\frac{|m|+|\m|-|m+\m|}{2}+j+1\right)
		%}{
		%\Gamma\left(\frac{|m|+|\m|-|m+\m|}{2}+1\right)
	%}
	\left({\scriptstyle\frac{|m|+|\m|-|m+\m|}{2}+1}\right)_j
\nonumber \\ & \qquad  \times
	\xi^{\frac{|m+\m|+|\m|-|m|}{2}+j}
	p_\perp^{|m+\m|}
	e^{-\frac{1}{2}\xi p_\perp^2}
	L_{\nu}^{(|m+\m|)}\left(\frac{1}{2}\xi p_\perp^2\right)
	.
\label{int2finalexpression}
\end{align}
%%%%
%%%%
Here we note once again that the factor of $e^{-\frac{1}{2}\xi p_\perp^2}$ expresses the principle that, up to simple prefactors, the Fourier transform of a gaussian function is again gaussian. Substituting this result into our original integral, we have then that
\begin{align}
\Int_1(\vbp,t'')  &  =
  C_{m\m\l}
  %i^{|m|+|\m|+|m+\m|} 
  %\frac{i^{\m} (-1)^{\l+\m}}{2^{|\m|} |\m|!} 
	%2^{\frac{|m|+|\m|-|m+\m|}{2}}
	%\Gamma\left({\scriptstyle\frac{|m|+|\m|-|m+\m|}{2}+1}\right)
	%\sqrt{\frac{(2\l+1)(\l+|\m|)!}{4\pi(\l-|\m|)!}}
	\frac{Q_{\l\m}}{\kappa^{|m|}} 
	e^{i(m+\m) \phi_p}
	p_\perp^{|m+\m|}
	e^{-\frac{1}{2}\xi p_\perp^2}
\nonumber \\ &  \times
  \sum_{j=0}^\infty
	D_{j}^{m\m}
	%2^{j}
	%\left({\scriptstyle\frac{|m|+|\m|-|m+\m|}{2}+1}\right)_j
  %\coef(\l,|\m|;j)
	\xi^{\frac{|m+\m|+|\m|-|m|}{2}+j}
	L_{\scriptscriptstyle\frac{|m|+|\m|-|m+\m|}{2}+j}^{(|m+\m|)}\left(\frac{1}{2}\xi p_\perp^2\right)
\nonumber \\ &   \times
	\frac{1}{2\pi}
	\! \int_{-\infty}^\infty \!\!\! \d z
	\! \int_{-\infty}^\infty \!\!\! \d k_\parallel
	  \frac{K(k_\parallel)}{(-z)^{\l+|\m|+2j+1}}
	  e^{i\left(k_\parallel-p_\parallel\right)z}
	  e^{-\frac{i}{2} \int_{t_s}^{t''}\left(k_\parallel+A_\parallel(\tau)\right)^2\d\tau}
	,
\end{align}
where we have introduced the constants
\begin{subequations}
\begin{align}
  C_{m\m\l}
	=
  \frac{
	  i^{\m}
	  i^{|m|+|\m|+|m+\m|} 
		}{
		(-1)^{\l+
		% Corrections in revisions
		m
		}
	}
  \frac{
	  2^{\frac{|m|+|\m|-|m+\m|}{2}}
	  \Gamma\left({\scriptstyle\frac{|m|+|\m|-|m+\m|}{2}+1}\right)
	  }{
		2^{|\m|} |\m|!
	} 
	\sqrt{\frac{(2\l+1)(\l+|\m|)!}{4\pi(\l-|\m|)!}}
\end{align}
and
\begin{align}
	D_{j}^{m\m}
=
	2^{j}
	\left({\scriptstyle\frac{|m|+|\m|-|m+\m|}{2}+1}\right)_j
  \coef(\l,|\m|;j)
	,\qquad
	\textrm{so in particular }
	D_0^{m\m}=1
	.
\end{align}
\end{subequations}

\subsection{The parallel integral}
We are now quite close to a final result for the combined spatial and momentum integrals embodied in the definition, eq. \eqref{int-one-definition}, of $\Int_1(\vbp,t'')$, and we have only two loose ends to take care of. The first, and most evident, is the still unresolved parallel integral, over $z$ and $k_\parallel$. This integral, however, we can lift directly from our previous saddle-point development. The calculation proceeds exactly as in the three-dimensional case, with the exception that the prefactors are now
$$
\frac{K(k_\parallel)}{(-z)^{\l+|\m|+2j+1}}
\approx
\frac{1}{(-z)^{\l+|\m|+2j+1}}
\frac{K_0}{ \sqrt{i(k_\parallel+A_\parallel(t_s))F(t_s)}}
$$
and do not vanish in the interval of interest. We can therefore adapt the final result \eqref{final-saddle-point-result} to the one-dimensional case as
\begin{align}
	\frac{1}{2\pi}
	\! \int_{-\infty}^\infty \!\!\! \d z
 & 
	\! \int_{-\infty}^\infty \!\!\! \d k_\parallel
	  \frac{K(k_\parallel)}{(-z)^{\l+|\m|+2j+1}}
	  e^{i\left(k_\parallel-p_\parallel\right)z}
	  e^{-\frac{i}{2} \int_{t_s}^{t''}\left(k_\parallel+A_\parallel(\tau)\right)^2\d\tau}
\nonumber \\ & = 
  \frac{1}{(-z_s(p_\parallel,t''))^{\l+|\m|+2j+1}}
  \frac{K_0}{ \sqrt{i(p_\parallel+A_\parallel(t_s))F(t_s)}}
	 e^{-\frac{i}{2}\int_{t_s}^{t''} (p_\parallel+A_\parallel(\tau))^2\d\tau}
	,
\label{1D-saddle-point-integral}
\end{align}
where the position stationary point is given by 
\begin{equation}
z_s(p_\parallel,t'')=\int_{t_s}^{t''}(p_\parallel+A_\parallel(\tau))\d\tau.
\label{stationary-position-1D}
\end{equation}

The second, minor loose end we must still cover is the gaussian factor $e^{-\frac{1}{2}\xi p_\perp^2}$, which we transform into
$$
e^{-\frac{1}{2}\xi p_\perp^2}=e^{-\frac{i}{2}(t''-t_s) p_\perp^2}=e^{-\frac{i}{2}\int^{t''}_{t_s} p_\perp^2\d\tau}
$$
by remembering the shorthand $\xi=i(t''-t_s)$. This factor can then be joined with the other exponentiated momentum integral to give the unified expression
$$
e^{-\frac{1}{2}\xi p_\perp^2}
e^{-\frac{i}{2}\int_{t_s}^{t''} (p_\parallel+A_\parallel(\tau))^2\d\tau}
=
e^{-\frac{i}{2}\int_{t_s}^{t''} (\vbp+\vb{A}(\tau))^2\d\tau};
$$
it is important to keep in mind, however, that this new factor still represents the over-all gaussian dependence of the ionization yield on the transverse momentum.

\subsection{Final results and analysis}
All this gives, then, our final result for the integral we set out to calculate:
\begin{align}
\Int_1(\vbp,t'')  &  =
  %C_{m\m\l}
	%\frac{Q_{\l\m}}{\kappa^{|m|}} 
	%e^{i(m+\m) \phi_p}
	%p_\perp^{|m+\m|}
	%e^{-\frac{1}{2}\xi p_\perp^2}
%\nonumber \\ &  \times
  %\sum_{j=0}^\infty
	%D_{j}^{m\m}
	%\xi^{\frac{|m+\m|+|\m|-|m|}{2}+j}
	%L_{\scriptscriptstyle\frac{|m|+|\m|-|m+\m|}{2}+j}^{(|m+\m|)}\left(\frac{1}{2}\xi p_\perp^2\right)
%\nonumber \\ &   \times
	%\frac{1}{2\pi}
	%\! \int_{-\infty}^\infty \!\!\! \d z
	%\! \int_{-\infty}^\infty \!\!\! \d k_\parallel
	  %\frac{K(k_\parallel)}{(-z)^{\l+|\m|+2j+1}}
	  %e^{i\left(k_\parallel-p_\parallel\right)z}
	  %e^{-\frac{i}{2} \int_{t_s}^{t''}\left(k_\parallel+A_\parallel(\tau)\right)^2\d\tau}
%%%%%%
%\\  &  =
%%%%%%
  C_{m\m\l}
	\frac{Q_{\l\m}}{\kappa^{|m|}} 
	K(p_\parallel)
	e^{i(m+\m) \phi_p}
	p_\perp^{|m+\m|}
	e^{-\frac{i}{2}\int_{t_s}^{t''} (\vbp+\vb{A}(\tau))^2\d\tau}
\nonumber \\ & \qquad \times
  \sum_{j=0}^\infty
  \frac{
	  D_{j}^{m\m}
	  \xi^{\frac{|m+\m|+|\m|-|m|}{2}+j}
		}{
		(-z_s(p_\parallel,t''))^{\l+|\m|+2j+1}
	}
	L_{\scriptscriptstyle\frac{|m|+|\m|-|m+\m|}{2}+j}^{(|m+\m|)}\left(\frac{1}{2}\xi p_\perp^2\right)
	.
\label{final-int1-expression}
\end{align}

% ( \label{final-int1-expression} )
Before doing a full analysis of the physical features of this equation, we insert this result back into the expression \eqref{integral-starting-point} for the second-order ionization yield in which $\Int_1(\vbp,t'')$ first appeared, which we recall was 
\begin{align*}
a^{(2)}_\mm(\vbp,t_0)
 & =
   -i\sum_n 
	 e^{-iE_\mm t_0}
	 e^{i I_{p,\nn} t_s}
	 a_g(t_s) 
 \\ & \quad \qquad\times 
  \int \d t''
   b_\mm(t_0,t'')
   b_\nn(t'',t_s)
	 e^{i(E_\mm-E_\nn)t''}
	 e^{-\frac{i}{2} \int_{t''}^T\left(\vbp+\vb{A}(\tau)\right)^2\d\tau} 
	\Int_1(\vbp,t'')\nonumber
  .
\end{align*}
(Here we use the symbol $\mm$ to denote the channel $\ket{m}$ to avoid confusion with the magnetic number $m$ of the current section; we do the same for channel $\ket{n}$ for consistency.) Before we jump in, we notice that the exponentiated momentum integrals again join up, which in particular makes the gaussian momentum dependence be independence of the interaction time $t''$; the relevant time constant is then the tunnelling time $\tauT=\Im(t_s)$ as will be obvious shortly.

We also use this opportunity to impose a definite contour on the $t''$ time integral. Although there is still some doubt in the community as to exactly what contour is appropriate and particularly regarding how one can best justify mathematically the most physically meaningful contour choices, we will side-step those issues and use the standard choice of a contour from the starting ionization time, $t_s$, down to its real part $t_0$, exactly as in the direct amplitude's case.

Further, we now change variables to the imaginary part of the interaction time, and particularly to our old shorthand of $\xi=i(t''-t_s)$, in terms of which $t''=t_s-i\xi$. We have then that $\d t''=-i\d\xi$, and if $t''$ goes from $t_s=t_0+i\tauT$ to $t_0$, then $\xi$ should go from $0$ to $\tauT$. This gives us then
\begin{align}
a^{(2)}_\mm(\vbp,t_0)
 & =
   %-i\sum_\nn 
	 %e^{-iE_\mm t_0}
	 %e^{i I_{p,\nn} t_s}
	 %a_g(t_s) 
 %\\ & \quad \qquad\times 
  %\int \d t''
   %b_\mm(t_0,t'')
   %b_\nn(t'',t_s)
	 %e^{i(E_\mm-E_\nn)t''}
	 %e^{-\frac{i}{2} \int_{t''}^T\left(\vbp+\vb{A}(\tau)\right)^2\d\tau} 
	%\Int_1(\vbp,t'')\nonumber
%%%%%%
%\\ & =
%%%%%%
  -i
	e^{-iE_\mm t_0}
	a_g(t_s) 
	e^{-\frac{i}{2} \int_{t_s}^T\left(\vbp+\vb{A}(\tau)\right)^2\d\tau} 
	\sum_\nn 
	e^{i I_{p,\nn} t_s}
	\frac{
	  C_{m\m\l}
		}{
		\kappa^{|m|}
	} 
	K(p_\parallel)
	e^{i(m+\m) \phi_p}
	p_\perp^{|m+\m|}
\nonumber \\ &  \qquad\times 
	-i
  \sum_{j=0}^\infty
	  D_{j}^{m\m}
	\int_{0}^{\tauT} 
  \frac{
    b_\mm(t_0,t'')
    b_\nn(t'',t_s)
		Q_{\l\m}
		}{
		(-z_s(p_\parallel,t''))^{\l+|\m|+2j+1}
	}
\nonumber \\ & \qquad\qquad\qquad \times
	  L_{\scriptscriptstyle\frac{|m|+|\m|-|m+\m|}{2}+j}^{(|m+\m|)}\left(\frac{1}{2}\xi p_\perp^2\right)
	  e^{i(E_\mm-E_\nn)t''}
	  \xi^{\frac{|m+\m|+|\m|-|m|}{2}+j}
	\d\xi
  .
\label{a-2-as-a-mess}
\end{align}
% a^{(2)}_\mm(\vbp,t_0) = (... mess ...).

In this integral we consider first the role of the Stark shifts and the multipole moment $Q_{\l\m}$, which also depends on the time $t''$ since it is a representation for the interaction potential $\langle V(\vbr,t'')\rangle$. Although we had factored out the Stark shifts and the phase-accumulating energies from the definition of $\langle V(\vbr,t'')\rangle$ and $Q_{\l\m}$ for clarity, we can put them back to realize that
%%%%%%%%
\begin{align*}
b_\mm(t_0,t'')
 & 
b_\nn(t'',t_s)
e^{i(E_\mm-E_\nn)t''}
Q_{\l\m}(t'')
%%%%%
\\ & =
%%%%%
%b_\mm(t_0,t'')
%b_\nn(t'',t_s)
%e^{iE_\mm t_0}
%e^{-iE_\mm(t_0- t'')} 
%e^{-iE_\nn (t''-t_s)}
%e^{-iE_\nn t_s}
%\\ & \qquad \times
%\left[
  %\bra{m(t_0,t'')}
  %\sum_i
	  %r_i^\l
		%Y_{\l\m}(\theta_i,\phi_i)^\ast
  %\ket{n(t'',t_s)}
%\right.
%\\ & \qquad \qquad -
%%\\ & \hfill-
%\left.
%\bracket{m(t_0,t'')}{n(t'',t_s)}
  %\bra{n(t'')}
  %\sum_i
	  %r_i^\l
		%Y_{\l\m}(\theta_i,\phi_i)^\ast
	%\ket{n(t'')}
%\right]
%%%%%%
%\\ & =
%%%%%%
e^{iE_\mm t_0}
e^{-iE_\nn t_s}
\left[
  \bra{m(t_0)}
	U^{N-1}(t_0,t'')
  \left(
	\sum_i
	  r_i^\l
		Y_{\l\m}(\theta_i,\phi_i)^\ast
	\right)
  U^{N-1}(t'',t_s)
	\ket{n(t_s)}
\right.
\\ & \qquad \qquad \qquad \quad -
%\\ & \hfill-
\left.
\matrixel{m(t_0)}{U^{N-1}(t_0,t_s)}{n(t_s)}
  \bra{n(t'')}
  \sum_i
	  r_i^\l
		Y_{\l\m}(\theta_i,\phi_i)^\ast
	\ket{n(t'')}
\right]
.
\end{align*}
%%%%% Q_{\l\m} as a matrix element with propagators.
%%%%%%%%%%%%
Thus we see that given the numerical knowledge of the propagated eigenstates, it is best to simply take the multipole moment at times $t''$ halfway through the propagation between $t_s$ and $t_0$, and these can then be used to calculate the integral in \eqref{a-2-as-a-mess} directly.

However, this point of view obscures dramatically the effect of the phase $e^{i(E_\mm-E_\nn)t''}$, which is turned into an exponential damping factor by virtue of the fact that integration is along the imaginary part of $t''$:
$$
e^{i(E_\mm-E_\nn)t''}
=e^{i(E_\mm-E_\nn)(t_0+i(\tauT-\xi))}
=e^{i(E_\mm-E_\nn)t_0} e^{-\Delta I_p(\tauT-\xi)}
.
$$
Here we have introduced the change in ionization potential,
\begin{equation}
\boxed{\Delta I_p =E_\mm-E_\nn,}
\label{eq:deltaIP}
\end{equation}  
which will be positive in the cases of interest where the channel $\ket{m}$ removes an electron from deeper in the core than channel $\ket{n}$, leaving the ion in an excited state with higher energy. The presence of the exponential factor inside the integral then tells us that the correlation interaction occurs preferentially near the exit of the tunnel, near $t''=t_0$ and $\xi=\tauT$, where as little as possible of the extra exponential penalty on $E_\mm$ must be paid.

We now focus on the integral itself, and the physical significance of the different factors that appear in it. Since the integral now includes the contributions to channel $\ket{m}$ from all initial channels $\ket{n}$, this is a good point to consider the most general transition, which typically will include contributions from more than one multipolar character; this is easily done, by linearity, by including a sum over $\l$ and $\m$. We also include a sum over $m$ to allow for more complicated initial orbitals. Thus we have
\begin{align}
a^{(2)}_\mm(\vbp,t_0)
 & =
%%%%%%
%\\ & =
%%%%%%
  -i
	a_g(t_s) 
	e^{-\frac{i}{2} \int_{t_s}^T\left(\vbp+\vb{A}(\tau)\right)^2\d\tau} 
\nonumber \\ &  \quad\times 
	\sum_{\nn,m} 
	\sum_{\lambda,\mu}
	e^{-iE_\nn t_0}
	e^{i I_{p,\nn} t_s}
	\frac{
	  C_{m\m\l}
		}{
		\kappa^{|m|}
	} 
	K_\nn(p_\parallel)
	e^{i(m+\m) \phi_p}
	p_\perp^{|m+\m|}
\nonumber \\ &  \quad\times 
	-i
  \sum_{j=0}^\infty
	  D_{j}^{m\m}
	\int_{0}^{\tauT} 
    \frac{
      b_\mm(t_0,t'')
      b_\nn(t'',t_s)
	  	Q_{\l\m}(t'')
	  	}{
	  	(-z_s(p_\parallel,t''))^{\l+|\m|+2j+1}
  	}
	  \xi^{\frac{|m+\m|+|\m|-|m|}{2}+j}
\nonumber \\ & \quad \qquad\qquad\qquad\qquad \times
	  L_{\scriptscriptstyle\frac{|m|+|\m|-|m+\m|}{2}+j}^{(|m+\m|)}\left(\frac{1}{2}\xi p_\perp^2\right)
	  e^{-\Delta I_p(\tauT-\xi)}
	\d\xi
  .
	\label{semi-final-a-2-result}
\end{align}
This is then a sum over processes from all channels and with all allowed multipolarities and, particularly, all allowed magnetic quantum numbers. 

However, it is important to note that the presence of the inverse power of the trajectory $z_s$ inside the integral will strongly discourage higher multipolarities, as the electron is often far away at the times when the interaction happens predominantly. These, as discussed above, are forced to be near the exit from the classically forbidden region by the exponential factor $e^{i(E_\mm-E_\nn)t''}$. The same damping with distance occurs to contributions with $j$ bigger than 0, which were introduced as small corrections to the leading small-angle approximation at $j=0$.

If we now make a really quite crude approximation and ignore the integration step, seeing the integral as concentrated wholly near $t''=t_0$, then we can replace $\xi$ by $\tauT$ in all the slow factors. This means that the contribution to the angular dependence from each process over fixed $\nn,m,\l$ and $\m$ goes, schematically, like
\begin{align}
	e^{i(m+\m) \phi_p}
	p_\perp^{|m+\m|}
	L_{\scriptscriptstyle\frac{|m|+|\m|-|m+\m|}{2}}^{(|m+\m|)}\left(\frac{1}{2}\tauT p_\perp^2\right)
	e^{-\frac{i}{2} \int_{t_s}^T\left(\vbp+\vb{A}(\tau)\right)^2\d\tau} 
  .
\label{crude-approximation-to-outgoing-wavefunction}
\end{align}
We now see that the original gaussian dependence as given by the saddle-point method, which goes as $e^{-\frac{1}{2}\tauT p_\perp^2}$, is still present and has the same width, but it is now punctured by a zero at the centre of order $|m+\m|$. This zero is required by the phase factor $e^{i(m+\m) \phi_p}$, which indicates that this component of the wavefunction has a well-defined angular momentum of $m+\m$ around the laser polarization axis.

Additionally, the transverse wavefunction is now multiplied by a Laguerre polynomial with the nontrivial degree
$$
\frac{|m|+|\m|-|m+\m|}{2}
=
\begin{cases}
  0
	&\textrm{if } m\m\geq0,\textrm{ and}
	\\
	\min\{|m|,|\m|\}
	&\textrm{if } m\m\leq0,
\end{cases}
$$
and this will introduce this same number of zeros into the transverse wavefunction so that the final angular distribution will be a series of concentric rings. We present in table \ref{table-of-power-and-degree} the dependence of both these factors - the order of the zero in the center and the number of rings - on the magnetic quantum numbers $m$ and $\m$.

\input{power-and-degree-table2}

\pagebreak

Having said this, we must note two caveats. First, although the approximation of ignoring the integral is qualitatively quite appropriate in most usual cases, it is useful for illustrative purposes only. In general the Laguerre polynomial may be explicitly expanded and it will yield a sum of incomplete-gamma-type integrals multiplying its coefficients; these will then degrade the delicate balance required for the oscillations of the polynomial and the higher lobes will become harder to observe.

Finally, the wavefunction in equation \eqref{semi-final-a-2-result} is really a coherent sum of contributions of all possible processes: this includes all possible initial states - with their associated magnetic numbers $m$ - as well as all the multipole terms with differing $\mu$. If care is not taken to eliminate these contributions, be it by initial state preparation or by a measurement process selective to the final state's initial structure, then in general they will wash out the pattern, and their only effect will be an additional broadening of the outgoing wavefunction.

\input{example}

%\start
%\input{currentwork}

%\chapter{Conclusions}
%\input{conclusions}

%% file: introduction.tex
%\chapter{Introduction}
This report considers the ionization of an atom or a molecule subjected to a strong laser pulse in the tunnel regime, within the formalism developed by Torlina, Smirnova, Walters, Ivanov and coworkers \cite{saepaper, mepaper}. 

In particular, we do an analytical version of the $R$-matrix method wherein an imaginary boundary inside the classically forbidden region acts as a source for the ionized electrons, which are then propagated using a semiclassical trajectory-based approach that expresses the results as time integrals over the classical trajectories of the corresponding classical problem. These integrals are then evaluated using the saddle-point method by shifting the contour of integration into the complex plane, which yields a physically clear picture of ionization times.

In addition to this, Torlina, Smirnova \textit{et al.} consider the possibility of interaction between the ionized electron and those that make up the ionic core, by treating the full Coulomb interaction, as a perturbation on the self-consistent field used to propagate the outgoing electronic trajectories. This formalism then asks for matrix elements of the Coulomb interaction between the different relevant states, which are easily formulated as spatial and momentum integrals over the relevant states; these integrals can then be approximated using the saddle-point method to give definite predictions.

However, these predictions do not quite match the angular distributions observed in experiment, and they specifically predict a lack of momentum transfer during the interaction, which is physically unsatisfactory.

This work presents an alternative to the saddle-point method, which works by using a suitable decomposition of the initial state and the correlation interaction potential into functions with well-defined multipolarity, and then exactly evaluating these integrals in the transverse directions -- while keeping the saddle-point approximation in the parallel direction -- to get a full, exact calculation of the angular distribution, to which suitable approximations can then be applied.

We begin with an exposition of the method used by Torlina, Smirnova \textit{et al.} in chapter~\ref{chappreliminaries}, up to general expressions for the wavefunctions corresponding to direct tunnelling and ionization through a correlation interaction. We then solve the direct tunnelling case in chapter~\ref{chapterdirect}, obtaining the results of Torlina \textit{et al.} for hydrogenic initial orbitals while including a detour to obtain abstract expressions valid for any initial orbital.

Chapter~\ref{chapcrossionization} treats the perturbed wavefunction, with an account of the general formalism in section~\ref{cisetup} and a brief recount of the saddle-point method in section~\ref{cisetup}. Section~\ref{start-of-the-real-work} then contains the exact calculation of the transverse integrals and finishes with a simple application to the ionization of carbon dioxide.

The author thanks Misha Ivanov, Lisa Torlina, Olga Smirnova, Jon Marangos, and Silvia Alatorre, for valuable discussions and help during the preparation of this report, and acknowledges the valuable funding of the National Council of Science and Technology (CONACYT) and the Secretary of Public Education (SEP).

This version contains revisions not present in the ``official'' version submitted on 21 September 2012. These are mostly corrections in style and format, some added references suggested by the independent assessor, and corrections of some errors. In particular, a sign error of $(-1)^{m+\m}$ in the previous version has now been corrected.

%\start

%% file: c1-ARMtheory.tex
\section[Analytical R-Matrix Theory]{Analytical $R$-Matrix Theory}
The theory of the $R$ matrix is a method born in the numerical analysis of multi-electron scattering and which has recently seen wide use in the numerical treatment of strong-field ionization problems. Our use of the formalism in an analytical setting represents an attempt to make physical sense of the results and to build an intuitive picture of the ionization process while retaining an appropriate description of the main qualitative features seen in numerical and experimental results.

Applications of $R$-matrix theory to strong-field physics were developed to meet with appropriate approximations the exacting demands of strong-field phenomena on numerical analysis: the high energies involved require the use of a very fine grid, asymptotics often require a large spatial extent, full three-dimensional calculations are often required for each active electron and, most importantly, nonnegligible electron-electron interaction and exchange induces a catastrophic scaling on the dimension of the relevant Hilbert space. These characteristics, when combined, quickly make detailed calculations impossible. This is made worse in strong-field ionization by the fact that the two competing influences on the ionized electron, the ionic potential and the laser pulse, are often of the same order of magnitude, so that neither can be treated as a perturbation.

To counter that, $R$-matrix theory uses a simple idea: the relative strength of these two effects has a strong spatial dependence, and in particular the effect of the ionic potential, as well as electron-electron interaction and exchange, can be safely ignored (or treated as a perturbation) when the ionized electron is sufficiently far away from the core. To exploit this, an imaginary spherical boundary of radius $a$ is drawn well inside the classically forbidden region. Within this region the atomic and molecular hamiltonian must be solved numerically, including all relevant perturbations; outside it the laser hamiltonian can be solved exactly and the ionic potential can be treated as a perturbation.

To put this in a more precise footing, we consider the evolutions on either side of the imaginary boundary as separate problems linked by continuity conditions at the edge. This poses somewhat of a problem, because when restricted to limited regions of space the kinetic energy operator, $\hat{T}= \frac{1}{2}\nabla^2$, ceases to be hermitian. This change is due to the fact that the boundary terms in the usual integration-by-parts \textit{spiel} do not now vanish:
\begin{align*}
	0&\stackrel{\textrm{?}}{=}\int_\textrm{in}\d\vb{r} \psi_1^\ast(\vb{r})\left(\hat{T}\psi_2(\vb{r})\right)-\int_\textrm{in}\d\vb{r} \left(\hat{T}\psi_1(\vb{r})\right)^\ast\psi_2(\vb{r})\\
	&=\frac{1}{2}\int\d\Omega\int_0^a r^2 \d r\left[ \psi_1^\ast(\vb{r})\frac{1}{r}\frac{\partial ^2}{\partial r^2} r\psi_2(\vb{r})- \left(\frac{1}{r}\frac{\partial ^2}{\partial r^2} r\psi_1^\ast(\vb{r})\right)\psi_2(\vb{r})\right]\\
	&=\frac{1}{2}\int\d\Omega\left[r\psi_1^\ast(\vb{r})\frac{\partial}{\partial r} r\psi_2(\vb{r})-r\psi_2(\vb{r})\frac{\partial}{\partial r} r\psi_1^\ast(\vb{r})\right]_0^a\\
	&=\frac{a^2}{2}\int\d\Omega\left.\left(\psi_1^\ast\frac{\partial\psi_2}{\partial r} - \psi_2\frac{\partial\psi_1^\ast}{\partial r} \right)\right|_{r=a},
\end{align*}
and in general this is not zero. This can, however, be fixed by substracting the operator $\frac{1}{2}\left.\frac{\partial}{\partial r}\right|_{r=a}$ from the hamiltonian, which will eliminate each of the unwanted terms and restore hermiticity

The usual practice is to add to this the hermitian operator $\frac{1-b}{r}$, where $b$ is an unspecified constant, which of course does not affect the hermiticity of the hamiltonian. Thus we employ the so-called Bloch operator
\begin{equation}
\hat{L}^+(a)=\delta(r-a)\left(\frac{\partial}{\partial r}+\frac{1-b}{r}\right),
\label{Bloch-operator}
\end{equation}
where the delta function should be even, in the sense that $\int_0^a\delta(r-a)\d r=\int_a^\infty\delta(r-a)\d r=\frac{1}{2}$. For the outer region the operator $\hat{L}^-(a)=-\hat{L}^+(a)$ is used, so that the total hamiltonian is unaltered.

The Schr\"odinger can then be rewritten so that each region has a hermitian hamiltonian and the connection to the other region is understood as a source term:
\begin{align*}
	i\frac{\d}{\d t}\ket{\Psi}&=\hat{H}\ket{\Psi}\\
	&=\left[\hat{H}+\hat{L}^\pm(a)\right]\ket{\Psi} -\hat{L}^\pm(a)\ket{\Psi}\\
	&=\ch^\pm\ket{\Psi} -\hat{L}^\pm(a)\ket{\Psi}\!,
\end{align*}
and here the term $\hat{L}^\pm(a)\ket{\Psi}$ is strictly local at the boundary and therefore can be obtained from the other side.

In practice, in our analytic analysis, we will ignore the source terms when solving for the inner region, which amounts to neglecting the backflow of probability during ionization. This is reasonable since significant backflow only occurs during high-harmonic generation after the ionized electron's wavepacket has wholly left the ion, and can therefore be treated as a separate problem. Further, a judicious choice of $b$ makes this approximation as harmless as possible. For the outside region, the source term clearly describes the source of ionized electrons.

%% file: c1-hamiltonian.tex
\section{The hamiltonian}

The hamiltonian for the problem is given by the standard atomic and molecular one, which includes the electronic kinetic energy term, electron-electron and electron-nuclei Coulomb repulsion terms, and dipole laser coupling for each electron. Thus, for $N$ electrons,
\begin{subequations}
\label{c1.totalhamiltonian}
\begin{align}
H^N&=T_e^N+V_C^N+V_{ee}^N+V_L^N,\textrm{ where}\\
V_C^N&=-\sum_m \sum_{i=1}^N \frac{Z_m}{|\vb{R}_m-\vb{r}_i|},\\
V_{ee}^N&=\sum_{i\neq j}^N \frac{1}{|\vb{r}_i-\vb{r}_j|},\\
V_L^N&=\sum_{i=1}^N \vb{F}(t)\cdot \vb{r}_i,
\end{align}
\end{subequations}
and the nuclei are frozen at positions $\vb{R}_m$ with charges $Z_m$. Atomic units are used throughout. We consider only linearly polarized pulses, though we will not introduce special coordinate systems until relatively late.

Once the ionized electron leaves the molecule, the total hamiltonian is split into the $N-1$-electron ionic hamiltonian $H^{N-1}$, formally identical to the neutral one, and the hamiltonian for the leaving electron,
$$H_e\colonequals H^N-H^{N-1}$$
which in particular contains the entangling operator $V_{ee}$, the Coulomb repulsion between the leaving electron and the ion.

Our problem, then, is to solve the time-dependent Schr\"odinger equation for the full system, with the system initially in the molecular ground state. We restrict ourselves to solving the system in the outer region, under the assumption that ionization probabilities are low enough that the electronic wavefunction in the inner region is well approximated by the unperturbed ground state. Thus, our problem can be stated as
\begin{subequations}
\label{c1.problem}
\begin{align*}
		i\frac{\d}{\d t}\ket{\Psi(t)}&=\left[H^N+\hat{L}^-(a)\right]\ket{\Psi(t)} -\hat{L}^-(a)\ket{\Psi_g}\!,\\
		\ket{\Psi(0)}&=\ket{\Psi_g}.
\end{align*}
\end{subequations}
Here $\hat{L}^-(a)$ represents the sum of all the single-electron Bloch operators.

This problem can be solved formally in terms of the propagator $U^N$ associated with the $N$-electron effective hamiltonian for the outer region, $\ch^N=H^N+\hat{L}^-(a)$, which satisfies $i\frac{\partial}{\partial t}U^N(t,t')=H^N(t) U^N(t,t')$ and $U^N(t,t)=1$. With this we have
\begin{align}
  &\ket{\Psi(t)}=-i\int_{-\infty}^t \d t' U^N(t,t')\hat{L}^-(a)\ket{\Psi(t')}\!.
  \label{c1.formalsolution}
\end{align}
Here we take the state $\ket{\Psi(t)}$ inside the integral to be the ionic ground state, with ionization potential $I_p$ and energy $E_g=-I_p$, and an added correction $a_g(t)$ to represent ground-state depletion and Stark shifts under the laser field: $\ket{\Psi(t)}=a_g(t) e^{+iI_p t}\ket{\Psi_g}$. The problem now becomes obtaining suitable approximations for the propagator $U^N(t,t')$.

%% file: c1-multichannel.tex
\section{Multichannel formalism}
To go further, we must now develop a suitable basis in which to work, which will describe the ionic and electronic evolution in the laser field as cheaply and easily as possible and allow us to inspect in greater detail their interactions during the ionization process. For reasons explained in depth in ref. \citealp{mepaper}, the bases of choice are the quasi-static eigenstates and the eikonal-Volkov wavefunctions, which we now describe.

The quasi-static eigenstates, which we denote by $\ket{n(t)}$, are the instantaneous eigenstates of the ionic hamiltonian $H^{N-1}(t)$, which includes the laser field but treats its dependence on the time $t$ as a parameter. When the laser pulse is over they connect to the field-free ionic eigenstates $\ket{n}\free$; they are defined by the relation
\begin{equation}
H^{N-1}(t)\ket{n(t)}=E_n(t)\ket{n(t)}.
\label{c1.quasistaticstates}
\end{equation}
These states perform well for oscillating laser fields, in the sense that transitions between them are minimized during propagation, when the field frequency $\omega$ is small compared to the typical energy differences of the states involved.

More importantly, they can be found easily without solving for a TDSE propagator whilst incorporating the polarizing effect of the laser field and minimizing laser-induced transitions during the ionization step. One disadvantage is that they do require numerical diagonalization of the full laser-perturbed ionic hamiltonian, and may require multiple diagonalizations if core polarization is desired. However, numerical diagonalization will be unavoidable as soon as any atom or molecule of significant complexity is involved, since only the simplest atomic and molecular systems have worthwhile analytic solutions, either exact or approximate.

The motion of the continuum electron can be exactly described, using Volkov functions \cite{VolkovWavefunctions}, in the absence of the ionic potential. The potential can be included using semiclassical perturbation theory with respect to the classical action. The boundary radius $a$ must then be chosen such that the ionic potential is a small enough perturbation. This approach is known as the eikonal-Volkov approximation \cite{EVApaper} and produces wavefunctions with final momentum $\vbp$ of the form
%
%On another extreme, even quite complicated potentials can be treated accurately and efficiently 
%%%%
%\pdfmargincomment[color=red]{Reference?}
%%%%
%using numerical methods. For our purposes, a simple analytical treatment can be obtained by choosing a large boundary radius $a$ so that the ionic potential is small enough a perturbation that semiclassical approximations can be employed. 
%
%We therefore use final-momentum eigenfunctions in the eikonal-Volkov approximation \cite{EVApaper} (EVA), given by
\begin{align}
\bracket{\vb{r}}{\vb{k}_n(t)}&\stackrel{\textrm{EVA}}{\approx} 
\frac{1}{(2\pi)^{3/2}}e^{i\left(\vb{k}+\vb{A}(t)\right)\cdot\vb{r}} e^{-\frac{i}{2} \int_T^t\left(\vb{k}+\vb{A}(\tau)\right)^2\d\tau} e^{-i\int_T^t U_n(\rl(\tau;\vb{r},\vb{k},t),\tau)\d\tau}
\label{c1.eikonal-volkov-wavefunctions}
\end{align}
where
\begin{align}
\rl(\tau;\vb{r},\vb{k},t)&\colonequals \vb{r}+\int_t^\tau \left(\vb{k}+\vb{A}(\tau')\right)\d\tau'
\label{c1.trajectory}
\end{align}
is the classical trajectory starting at $\vb{r}$ at time $t$ which has final canonical and mechanical momentum $\vb{k}$ at a time $T$ long after the laser pulse has vanished, and $\vb{A}$ is the vector potential of the laser field, obeying $\vb{F}(t)=-\frac{\d}{\d t}\vb{A}(t)$.

These wavefunctions are channel-specific in that they respond to the mean field of the ion when it is in channel $n$, given by
$$U_n(\vb{r})\colonequals \bra{\vb{r}}\otimes\bra{n(t)}V_{ee}\ket{n(t)}\otimes\ket{\vb{r}}$$
for $V_{ee}=V_{ee}^N-V_{ee}^{N-1}$. As such, they obey the single-electron Schr\"odinger equation
$$i\frac{\d}{\d t}\ket{\vb{k}_n(t)}=H_e^n(t)\ket{\vb{k}_n(t)}\textrm{ for }H_e^n\colonequals \matrixel{n(t)}{H^N-H^{N-1}}{n(t)}.$$
This removes the main influence of the ion on the ionized electron's motion, and will allow us to focus on the transition-inducing part of the Coulomb interaction later on.
%%%
%\pdfmargincomment[color=red]{Isn't this already an entangled evolution?}	
%%%

We now implement these channels by inserting in the formal solution, expression \eqref{c1.formalsolution}, a resolution of the identity of the form
\begin{align}
	1=\int\d\vb{k}\sum_n\mathbb{A}\ket{n(t)}\otimes\ket{\vb{k}_n(t)}\bra{\vb{k}_n(t)}\otimes\bra{n(t)}\mathbb{A}
	\label{c1.resolutionofidentity}
\end{align}
where $\mathbb{A}$ is the anti-symmetrizing operator which is clearly necessary. This gives us the expression
\begin{align}
  \ket{\Psi(t)}=-i\sum_n\int\d\vb{k}\int_{-\infty}^t &\d t' U^N(t,t')\mathbb{A}\ket{n(t')}\otimes\ket{\vb{k}_n(t')}\nonumber\\
	&\qquad   \times \bra{\vb{k}_n(t')}\otimes\bra{n(t')}\mathbb{A}\hat{L}^-(a)\ket{\Psi_g}a_g(t')e^{iI_p t'},
	\label{c1.channelspecificformalsolution}
\end{align}
which is ready for further work.

%% file: c1-DysonOrbital.tex
\section{The Dyson orbital}
We begin with the matrix element involving the Bloch operator,
$$
\bra{\vb{k}_n(t')}\otimes\bra{n(t')}\mathbb{A}\hat{L}^-(a)\ket{\Psi_g}.
$$
This expression hides two summations over the different electrons: one over which electronic Bloch operator acts on $\ket{\Psi_g}$, and one, induced by the anti-symmetrizing $\mathbb{A}$, over which electron is induced into the continuum state $\ket{\vb{k}_n(t')}$. 

We can, however, neglect the contribution from the non-diagonal, exchange-like terms, in which an electron different from the one transmitted by the Bloch operator is projected into the continuum state. In terms of the characteristic momentum $\kappa_g$ of the ground state, with $\frac{1}{2}\kappa_g^2=I_p$, this can be ensured as long as $\kappa_g a\gg 1$. This signals an effective breaking of the exchange symmetry, which is due to the fact that the ionized electron is distinguishable from those left behind, and which allows us to choose which electron will tunnel out into the continuum state. This reduces the matrix element to 
$$
\bra{\vb{k}_n(t')}\otimes\bra{n(t')}\mathbb{A}\hat{L}^-(a)\ket{\Psi_g}=\frac{N}{\sqrt{N}}\bra{\vb{k}_n(t')}\hat{L}^-(a)\cdot \bracket{n(t')}{\Psi_g}
$$
where we include normalization factors of $\frac{1}{\sqrt{N}}$, due to the normalization of $\mathbb{A}$, and $N$, due to the different electrons the Bloch operator can act on. From here on we revert the Bloch symbol $\hat{L}^-(a)$ to a single-electron operator as originally introduced.

The remaining single-electron wavefunction on the right of the Bloch operator can now be recognised to be the Dyson orbital corresponding to channel $n$, which we denote by
%%%
\pdfmargincomment{Reference for Dyson orbital missing.}
%%%
\begin{equation}
\ket{n_{D}(t)}=\sqrt{N}\bracket{n(t)}{\Psi_g}.
\label{c1.dysonorbital}
\end{equation}
The matrix element in question is then left as
$$
\bra{\vb{k}_n(t')}\otimes\bra{n(t')}\mathbb{A}\hat{L}^-(a)\ket{\Psi_g}=\bra{\vb{k}_n(t')}\hat{L}^-(a)\ket{n_D(t')}.
$$

Finally, we note that the above is also valid in the single-electron case, provided that one drops the ion states and simply takes the Dyson orbital to be the ground state.

%% file: c1-propagator-Dyson-expansion.tex
\section{The propagator and its Dyson expansion}
We consider now the other half of our current solution, \eqref{c1.channelspecificformalsolution}, which involves the full propagator acting on our basis states:
$$
U^N(t,t')\mathbb{A}\ket{n(t')}\otimes\ket{\vb{k}_n(t')}.
$$
Since here the ionized electron is distinguishable as it is in a continuum wavefunction, we can ignore exchange terms and therefore drop the anti-symmetrizer $\mathbb{A}$.

Here we face the evolution under the full hamiltonian $\ch^-$ of basis states $\ket{n(t')}$ and $\ket{\vb{k}_n(t')}$ whose individual evolution, under the separated hamiltonians $H^{N-1}$ and $H_e^n$, we understand: the former follow the hamiltonian adiabatically in the small laser frequency limit of $\omega\ll\Delta E$, where $\Delta E$ is a characteristic energy spacing in the molecule, while the latter we can solve for numerically and approximate well analytically. However, these two hamiltonians fail to account for all of the system's evolution, because the ionized electron responds so far only to the ionic expectation value of the Coulomb potential.

We thus still have to account for what we will call the correlation interaction potential,
\begin{equation}
V_{ee}^n(t)\colonequals V_{ee}-\bra{n(t)}V_{ee}\ket{n(t)}.
\label{c1.correlationinteractionpotential}
\end{equation}
This we will treat as a perturbation to the previously defined hamiltonians, breaking the full evolution into
$$
\ch^-=H^{N-1}+H_e^n(t)+V_{ee}^n(t).
$$
To deal with this perturbation we will do a Dyson expansion in the correlation interaction of the full propagator.

%\newpage % Prevents a page break between ``Mathematical Parenthesis 1.1. The Dyson expansion'' and the box, if such is required.

\begin{parmat}{The Dyson expansion}
\label{pm.dyson-expansion}
\noindent
The Dyson expansion (also known as the Lippmann-Schwinger expansion) is a basic tool of time-dependent perturbation theory, which we develop here because it is not often worked out in full, despite its simplicity. The goal is the description of the propagator $U_1(t,t')$ corresponding to a perturbed hamiltonian $H_1=H_0+\Delta H$ when the evolution for the core hamiltonian $H_0$ is already known. Note that all three hamiltonians may be time-dependent.

Thus we assume knowledge of the propagator $U_0(t,t')$ which solves the problem
$$
i\frac{\partial}{\partial t} U_0(t,t')=H_0U_0(t,t')\textrm{  under  }U_0(t,t)=1,
$$
and we seek a solution $U_1(t,t')$ to the problem
$$
i\frac{\partial}{\partial t} U_1(t,t')=\left(H_0+\Delta H\right)U_1(t,t')\textrm{  under  }U_1(t,t)=1.
$$
The Dyson expansion claims that for small $\Delta H$ this solution can be found recursively as
\begin{equation}
U_1(t,t')=-i\int_{t'}^t \d t'' U_1(t,t'') \Delta H(t'') U_0(t'',t')+U_0(t,t').
\label{c1.dysonexpansion}
\end{equation}
Further, repeated application of the expansion gives a series solution for $U_1(t,t')$ in terms of known data, though the resulting series need not converge and may require renormalization procedures to work well \cite{FetterAndWalecka}.

the convergence of this series is not guaranteed and must be checked on a case-by-case basis.

\begin{proof}[\textbf{\textup{Proof}}]$\quad$ 

\noindent The proof of the expansion is by direct calculation, though one must note that it relies on both $U_0$ and $U_1$ being solutions of their given problems. Thus, we have
\begin{align*}
i\frac{\partial}{\partial t}&U_1(t,t')=i\frac{\partial}{\partial t}\left[-i\int_{t'}^t \d t'' U_1(t,t'') \Delta H(t'') U_0(t'',t')+U_0(t,t')\right]\\
&= \int_{t'}^t \d t'' i\frac{\partial U_1(t,t'')}{\partial t} \Delta H(t'') U_0(t'',t')+\left.U_1(t,t) \Delta H(t) U_0(t,t')\right.+i\frac{\partial}{\partial t}U_0(t,t')\\
&= \left(H+\Delta H\right)\int_{t'}^t \d t''  U_1(t,t'')\Delta H(t'') U_0(t'',t')+\Delta H U_0(t,t')+H_0U_0(t,t')\\
&= \left(H+\Delta H\right)\left[U_1(t,t')-U_0(t,t')\right]+\left(H_0+\Delta H \right)U_0(t,t')\\
&= \left(H_0+\Delta H )\right)U_1(t,t').
\end{align*}
The initial condition $U_1(t,t)=1$ holds trivially.
\end{proof}
\end{parmat}

We therefore apply to our channel-specific formal solution \eqref{c1.channelspecificformalsolution} a Dyson expansion with the correlation interaction as a perturbation of the separable evolution, which corresponds to the hamiltonian $H^{N-1}+H_e^n$ and whose propagator is $U^{N-1}(t,t')\otimes U_e^n(t,t')$.
%whose propagator we denote by $U_\otimes^n(t,t')$ to indicate that it is a tensor product: $U_\otimes^n(t,t')=U^{N-1}(t,t')\otimes U_e^n(t,t')$.
With this the wavefunction at a time $T$, which we take to be long after the pulse is finished, has the two components
\begin{subequations}
\label{c1.firstdecomposition}
\begin{equation}
\ket{\Psi(T)}=\smallket{\Psi^{(1)}(T)}+\smallket{\Psi^{(2)}(T)}
\label{c1.decomposition}
\end{equation}
where
\begin{align}
\smallket{\Psi^{(1)}(T)}=-i\sum_n \int\d\vb{k}\int^T\d t' & U^{N-1}(T,t')\ket{n(t')} \otimes U_e^n(T,t')\ket{\vb{k}_n(t')}\nonumber\\
&\times\matrixel{\vb{k}_n(t')}{\hat{L}^-(a)}{n_D(t')}a_g(t')e^{iI_p t'}
\label{c1.abstractdirectstate}
\end{align}
and
\begin{align}
\smallket{\Psi^{(2)}(T)}=(-i)^2\sum_n \int\d\vb{k} & \int^T\d t''\int^{t''}\d t'U^N(T,t'')V_{ee}^n(t'') U^{N-1}(t'',t')\ket{n(t')}\nonumber\\
& \otimes U_e^n(t'',t')\ket{\vb{k}_n(t')}\times\matrixel{\vb{k}_n(t')}{\hat{L}^-(a)}{n_D(t')}a_g(t')e^{iI_p t'}.
\label{c1.abstractcorrelationstate}
\end{align}
\end{subequations}

The leading-order term, $\ket{\Psi^{(1)}(T)}$, represents the direct-tunnelling contribution and will be studied in chapter~\ref{chapterdirect}. The subleading term, $\ket{\Psi^{(2)}(T)}$, is the correlation-driven signal, and will be studied in chapter~\ref{chapcrossionization}.

%% file: c2-direct-tunnelling.tex
\section{Direct ionization yield}
In this section we will focus on the direct tunnelling term, $\smallket{\Psi^{ (1)}(T)}$, which does not involve the correlation interaction. Further, in the single-active-electron case in which the core is not significantly perturbed or indeed not present, this term accounts for the full evolution as $V_{ee}^n=0$.

To bring the calculation down to more concrete quantities, we consider in particular the ionization yield with final momentum $\vb{p}$ in channel $n$: that is, we want the probability amplitude for the ion to be left in the free state $\ket{n}\free$ with the ionized electron at canonical and mechanical momentum $\vb{p}$ at some time $T$ after the laser pulse has finished. However, there is a considerable complication in that the laser pulse may cause transitions between different quasi-static eigenstates after the ionization event is over, and to evaluate this a full $N-1$-electron back-propagation of the Schr\"odinger equation is required, which is prohibitively complex.

To reach a compromise, we project on the basis of quasi-static eigenstates at a time $t_0$ shortly after the ionization step is completed. This is equivalent to projecting on the basis $U^{N-1}(T,t_0)\ket{n(t_0)}$ at time $T$ and represents a definite loss of contrast to projecting on the free states $\ket{n}\free$, but since the transitions caused by the laser are indistinguishable from those caused by the electron this loss of contrast is inevitable.

We therefore define the ionization yield, our main handle on the system's state, as
\begin{equation}
a_n(\vb{p},t_0)\colonequals\bra{\vb{p}}\otimes\bra{n(t_0)}U^{N-1}(t_0,T)\ket{\Psi(T)}\!.
\label{c1.ionizationyield}
\end{equation}
With this we attack, then, the first-order term \eqref{c1.abstractdirectstate}, which gives rise to the ionization yield
\begin{align}
	a_m^{(1)}(\vb{p},t_0)=-i\sum_n \int_C\d t' \matrixel{m(t_0)}{U^{N-1}(t_0,t')}{n(t')} \matrixel{\vb{p}_n(t')}{\hat{L}^-(a)}{n_D(t')}a_g(t')e^{iI_p t'}.
\label{setup-a1}
\end{align}
Here, as discussed in refs. \citealp{saepaper} and \citealp{mepaper}, the temporal integration has been changed to a contour starting at $-\infty$ and ending at $t_0$. This describes a single ionization burst centred at a time $t_0$.
%%%
%\pdfmargincomment{Discussion of analyticity of eigenstates and matrix elements should be done here or elsewhere.}
%%%
This freedom in the contour of integration will then allow us to pass through a saddle-point in the fast oscillating exponent arising in integral \eqref{setup-a1}, which represents the main contribution to the integral.\footnote{We note here that this freedom to change the contour of integration comes at the price of ensuring that all functions present are analytical functions of $t$. This is far from evident, particularly in the case of the quasi-static states $\ket{n(t)}$, whose time-dependence comes from the hamiltonian $H^{N-1}(t)$ of which they are eigenstates. The existence of analytic solutions can be proved \cite{KatoPerturbationTheory}, but care must still be taken during the numerical solution that all quantities in the integral are continued analytically to the complex plane.}

To bring this contribution to light, we first deal with the ionic matrix element by neglecting the possibility of laser-induced real excitations (whose timescale is significantly longer) during the ionization step:
\begin{equation}
\matrixel{m(t_0)}{U^{N-1}(t_0,t')}{n(t')}=\delta_{mn}e^{-iE_m(t_0-t')}b_m(t_0,t'),
\label{propagatedquasistaticeigenstates}
\end{equation}
where $b_m(t_0,t')$ is a slow function that accounts for Stark shifts caused by the laser field. With this we have
%\begin{align}
	%a_n^{(1)}(\vb{p},t_0)=-i e^{-iE_nt_0} \int_C\d t'b_n(t_0,t') a_g(t') \matrixel{\vb{p}_n(t')}{\hat{L}^-(a)}{n_D(t')}e^{iE_nt'}e^{iI_p t'}.
%\end{align}
\begin{align}
	a_n^{(1)}(\vb{p},t_0)=-i e^{-iE_nt_0} \int_C\d t'b_n(t_0,t') a_g(t') \matrixel{\vb{p}_n(t')}{\hat{L}^-(a)}{n_D(t')}e^{i I_{p,n} t'},
	\label{saeyieldbeginning}
\end{align}
where 
\begin{equation}
\boxed{I_{p,n}=I_p+E_n=E_n-E_g,}
\label{eq:IPdefinition}
\end{equation}
expression which we are now ready to attack as a single-electron problem. (In the absence or neglectability of the core excitations, we have $E_n=0$ since the core does not accumulate meaningful phase, and this reduces indeed to the single-electron result from ref. \citealp{saepaper}.)

%% file: c2-abstractRfactor.tex
\section{The Volkov action}
In this section we deviate from the course taken by Torlina and Smirnova \citep{saepaper} in that we proceed to do the temporal integration before we explicitly evaluate the Bloch matrix element, which requires an explicit expression for the Dyson orbital $\ket{n_D}$. Thus we will obtain a general, symbolic expression for the orbital-dependent factor by doing the time integral using the same saddle-point method \cite[pp.~489-497]{Arfken} as Torlina and Smirnova, and only then will we substitute an explicit hydrogenic expression for the orbital to check that it integrates correctly to the previously obtained results.

To obtain the Volkov action in the exponent, we must nevertheless formulate the matrix element as a position-space integral by inserting the resolution of identity $\int\d\vb{r}\outerp{\vbr}=1$. This leaves, using the EVA approximation as in eq. \eqref{c1.eikonal-volkov-wavefunctions}, the expression
\begin{align}
	a_n^{(1)}(\vb{p},t_0)= & -i e^{-iE_nt_0} \int_C\d t'b_n(t_0,t') a_g(t') 
	%\bracket{\vb{p}_n(t')}{\vbr}
	e^{\frac{i}{2} \int_T^{t'}\left(\vbp+\vb{A}(\tau)\right)^2\d\tau}e^{i I_{p,n} t'}     \nonumber \\
	& \times
	\int\frac{\d\vbr}{(2\pi)^{3/2}}
	e^{-i\left(\vbp+\vb{A}(t')\right)\cdot\vb{r}} 
	e^{i\int_T^{t'} U_n(\rl(\tau;\vb{r},\vbp,t'),\tau)\d\tau}
	\matrixel{\vbr}{\hat{L}^-(a)}{n_D(t')}.
\end{align}
For simplicity, from here onwards we drop the arguments of the functions $b_n$, $a_g$ and $U_n$ unless they play an active role. This leaves us with the expression

\begin{align}
	a_n^{(1)}(\vb{p},t_0)= & -i e^{-iE_nt_0} \int_C\d t' a_g b_n
	%\bracket{\vb{p}_n(t')}{\vbr}
	e^{\frac{i}{2} \int_T^{t'}\left(\vbp+\vb{A}(\tau)\right)^2\d\tau}e^{i I_{p,n} t'}     \nonumber \\
	& \times
	\int\frac{\d\vbr}{(2\pi)^{3/2}}
	e^{-i\left(\vbp+\vb{A}(t')\right)\cdot\vb{r}} 
	e^{i\int_T^{t'} U_n\d\tau}
	\matrixel{\vbr}{\hat{L}^-(a)}{n_D(t')}.
\end{align}

To perform the temporal integral without evaluating the Bloch matrix element we face the major hurdle that the Dyson orbital depends on our time variable $t'$. This we resolve by noting that this temporal dependence comes from that of the quasi-static state $\ket{n(t)}$ and this represents perturbations on the ionic states on the timescales of the laser period. Thus for our purposes it can be replaced by its value at the saddle point, which we will denote by $t_a$. Analogously, we replace $t'$ by $t_a$ in the slow functions $b_n$, $a_g$ and $U_n$.
%%%
%\pdfmargincomment{Should there be a mathematical parenthesis here explaining the saddle-point method? Just a reference?}
%%%

Under these approximations, then, we have
\begin{align}
	a_n^{(1)}(\vb{p},t_0)= & -i e^{-iE_nt_0}a_g(t_a) b_n(t_0,t_a)
	\int\frac{\d\vbr}{(2\pi)^{3/2}} 
	\matrixel{\vbr}{\hat{L}^-(a)}{n_D(t_a)}
	e^{i\int_T^{t_a} U_n(\rl(\tau;\vb{r},\vbp,t_a),\tau)\d\tau}     \nonumber \\
	& \times
	\int_C\d t' 
	e^{\frac{i}{2} \int_T^{t'}\left(\vbp+\vb{A}(\tau)\right)^2\d\tau}e^{i I_{p,n} t'}
	e^{-i\left(\vbp+\vb{A}(t')\right)\cdot\vb{r}} .
\end{align}
To keep the notation simple, we will retain formally the full $\d\vbr$ integral as well as the Bloch operator $\hat{L}^{-}(a)=\delta(r-a)\left(\frac{\partial}{ \partial r}+\frac{1-b}{r}\right)$, but it is important to note that the presence of the delta function restricts the integral to the boundary surface, and we can therefore replace $r$ by $a$ save when derivatives are involved: throughout the following discussion, the length of $\vbr$ will be $a$.

The method of stationary phase demands now that the exponent inside the temporal integral be zero, which gives us the equation
\begin{align*}
	0 & = -i\frac{\d S_V}{\d t'}(t_a)
		  = \frac{\d}{\d t'}\left[\frac{i}{2} \int_T^{t'}\left(\vbp+\vb{A}(\tau)\right)^2\d\tau+i I_{p,n} t'-i\left(\vbp+\vb{A}(t')\right)\cdot\vb{r}\right]_{t_a}\\
	  & = i\left(\frac{1}{2} \left(\vbp+\vb{A}(t_a)\right)^2+ I_{p,n} + \vb{F}(t_a)\cdot\vbr\right).
\end{align*}
The last term in this equation, equal to $F(t_a)a\cos(\theta)$ where $\theta$ is the angle between the position vector $\vbr$ and the laser polarization, is inconvenient as it makes the saddle point depend on the position integration variable $\vbr$. To overcome this, we ask that the boundary radius $a$ be small enough that the saddle point is not significantly shifted; that is, we require that
\begin{equation}
|Fa|\ll I_{p,n},
\label{c1.Fa.Ip.restriction}
\end{equation}
for all channels of interest. This is equivalent to putting the boundary sphere well within the tunnelling barrier and is therefore an acceptable approximation.

Having done this, we can therefore take our main saddle point $t_s$ as that which satisfies the equation
\begin{equation}
\frac{1}{2} \left(\vbp+\vb{A}(t_s)\right)^2+ I_{p,n}=0
\label{c1.ts-equation}
\end{equation}
and then taking $t_a=t_s+\Delta t_a$ slightly displaced from it. The above definition of $t_s$ cannot be simplified significantly and must be left as it is, though it must be supplemented by the condition $\Im(t_s)>0$, which is imposed due to global considerations on the integration contour.

The saddle-point displacement $\Delta t_a$, on the other hand, can be expressed at least approximately in terms of the small parameter ${Fa}/{I_{p,n}}$. To do this, we express the vector potential at $t_a$ as
$$
\vb{A}(t_a)=\vb{A}(t_s+\Delta t_a)\approx\vb{A}(t_s)+\Delta t_a \frac{\d\vb{A}}{\d t}(t_s)=\vb{A}(t_s)-\Delta t_a\vb{F}(t_s).
$$
Under this expansion and a similar one for $\vb{F}(t_a)$, our equation for $t_a$ then reads
\begin{align*}
	0 & = \frac{1}{2} \left(\vbp+\vb{A}(t_s)-\Delta t_a\vb{F}(t_s)\right)^2+ I_{p,n} + \vb{F}(t_s)\cdot\vbr +\Delta t_a\frac{\d\vb{F}}{\d t}(t_s)\cdot\vbr\nonumber \\
	  & = \frac{1}{2} \left(\vbp+\vb{A}(t_s)\right)^2+ I_{p,n}  -\Delta t_a\left(\vbp+\vb{A}(t_s)\right)\cdot\vb{F}(t_s) + \vb{F}(t_s)\cdot\vbr \nonumber\\
		  & \qquad+\frac{1}{2} \left(\Delta t_a\vb{F}(t_s)\right)^2  +\Delta t_a\frac{\d\vb{F}}{\d t}(t_s)\cdot\vbr.
\end{align*}
Here the first two terms vanish and the last two terms, of order $\left(\frac{Fa}{I_{p,n}}\right)^2$, can be neglected, which leaves us with 
\begin{equation*}
\Delta t_a = \frac{\vb{F}(t_s)\cdot\vbr}{\left(\vbp+\vb{A}(t_s)\right)\cdot\vb{F}(t_s)}=\frac{a\cos(\theta)}{v_\parallel(t_s)},
\end{equation*}
in terms of the parallel component of the velocity vector $\vb{v}(t)=\vbp+\vb{A}(t)$.%, which obeys the relation $v_\parallel^2(t_s)+v_\perp^2=-2I_{p,n}$.

As a brief aside, we note that this velocity vector obeys the equation $\vb{v}^2(t_s)=-2I_{p,n}\equalscolon-\kappa^2$, so that its transverse component is given by $v_\parallel(t_s)=\pm i \kappa\sqrt{1+p_\perp^2/\kappa^2}=\pm i \kappa_\eff$. Here it can be shown that for sinusoidal pulses the condition that $\Im(t_s)>0$ translates into the sign choice of $v_\parallel(t_s)=-i\kappa_\eff$.

We can now complete the stationary-phase approximation for the temporal integral, which in this case indicates that we replace all times by the stationary point, $t_a$, and add a factor of $\sqrt{2\pi/i S_V''(t_a)}$. With this, then, we have
\begin{align*}
	a_n^{(1)}(\vb{p},t_0)= & -i e^{-iE_nt_0}a_g(t_a) b_n(t_0,t_a)
	\int\frac{\d\vbr}{2\pi} 
	\matrixel{\vbr}{\hat{L}^-(a)}{n_D(t_a)}
	e^{i\int_T^{t_a} U_n\d\tau} %U_n(\rl(\tau;\vb{r},\vbp,t_a),\tau)
	\frac{e^{-iS_V(t_a)}}{\sqrt{iS_V''(t_a)}}   \nonumber\\
	%%%
	\approx & 
	-i \frac{e^{-iE_nt_0}}{2\pi}
	\int\frac{\d\vbr}{\sqrt{iS_V''(t_a)}} 
	a_g(t_a) b_n(t_0,t_a)
	\matrixel{\vbr}{\hat{L}^-(a)}{n_D(t_a)}   \nonumber\\
	&  \qquad \times
	e^{i\int_T^{t_a} U_n\d\tau} %U_n(\rl(\tau;\vb{r},\vbp,t_a),\tau)  
	e^{-iS_V(t_s)-i\Delta t_aS_V'(t_s)}
\end{align*}
and here we have
$$
-iS_V'(t_s)  = \frac{i}{2} \left(\vbp+\vb{A}(t_s)\right)^2  +i I_{p,n} t_s +i\vb{F}(t_s)\cdot\vb{r}=iF(t_s)a\cos(\theta)
$$
and we can approximate $a_g(t_a)\approx a_g(t_s)$, $b_n(t_0,t_a)\approx b_n(t_0,t_s)$ and $S_V''(t_a)\approx S_V''(t_s)$ so that
%{\allowdisplaybreaks
\begin{align}
	a_n^{(1)}(\vb{p},t_0)= & 
	-i \frac{e^{-iE_nt_0}}{2\pi}
	\int\frac{\d\vbr}{\sqrt{iS_V''(t_s)}} 
	a_g(t_s) b_n(t_0,t_s)
	\matrixel{\vbr}{\hat{L}^-(a)}{n_D(t_a)}
	e^{i\int_T^{t_a} U_n\d\tau} %U_n(\rl(\tau;\vb{r},\vbp,t_a),\tau)     
	\nonumber\\ 	&  \qquad \times
	e^{\frac{i}{2} \int_T^{t_s}\left(\vbp+\vb{A}(\tau)\right)^2\d\tau+i I_{p,n} t_s-i\left(\vbp+\vb{A}(t_s)\right)\cdot\vb{r}}
	e^{i\frac{a\cos(\theta)}{v_\parallel(t_s)} F(t_s)a\cos(\theta)} 
	\displaybreak[1]
	%%%
	\nonumber \\ = & 
	%%%
	-i\frac{a_g(t_s) b_n(t_0,t_s)e^{-iE_nt_0}}{2\pi\sqrt{iS_V''(t_s)}} 
	e^{-\frac{i}{2} \int_{t_s}^T\left(\vbp+\vb{A}(\tau)\right)^2\d\tau+i I_{p,n} t_s}
	\nonumber\\ 	&  \qquad \times
	\int\d\vbr
	\matrixel{\vbr}{\hat{L}^-(a)}{n_D(t_a)}
	e^{i\int_T^{t_a}U_n(\rl(\tau;\vb{r},\vbp,t_a),\tau) \d\tau}     % U_n\d\tau}
	e^{-i\left(\vbp+\vb{A}(t_s)\right)\cdot\vb{r}}
	%e^{i\frac{F(t_s)}{v_\parallel(t_s)} a^2\cos^2(\theta)} 
	.
	\label{c1.almostfinalRfactor}
\end{align}
%}
Here we have neglected the final term in $a^2$, which is of order $F a^2/\kappa$ for $\kappa=\sqrt{2I_{p,n}}$; this represents the strongest assumption on our boundary and together with the asymptotic-regime requirement that $\kappa a \gg 1$ implies the weaker condition $Fa\ll\frac{\kappa}{a}\sim\frac{I_{p,n}}{\kappa a} \ll I_p$.

This last equation \eqref{c1.almostfinalRfactor} needs only one final manipulation to get it into as neat and usable a form as it can, and this is the elimination of the explicit dependence on the boundary radius $a$. Since we are calculating a physically measurable quantity, the ionization yield, and the boundary radius is external to the physical problem we are attacking, the result does not depend on $a$ and its explicit appearance should be avoided as far as possible.

The main dependence of this result on the boundary radius is through the wavefunction of the Dyson orbital, which must be evaluated at the boundary:
\begin{align*}
\int\d\vbr	\matrixel{\vbr}{\hat{L}^-(a)}{n_D(t_a)}&=\int\d\vbr\,\delta(r-a)\left(\frac{\partial}{\partial r}+\frac{1-b}{r}\right) \bracket{\vbr}{n_D(t_a)}\\
 &= a^2 \int\d\Omega\left.\left(\frac{\partial}{\partial r}+\frac{1-b}{r}\right)\right|_{r=a} \bracket{\vbr}{n_D(t_a)}.
\end{align*}
(The Dyson orbital's dependence on $t_a$, on the other hand, is not strong, and we can harmlessly replace $t_a$ by $t_s$ in it.) This is dependence generally cancelled by the exponentiated integral of the ionic potential, 
\begin{equation}
e^{i\int_T^{t_a}U_n(\rl(\tau;\vb{r},\vbp,t_a),\tau) \d\tau}=\exp\left[{-i\int_{t_a}^TU_n\left(\vb{r}+\int_{ t_a}^\tau \vb{v}(\tau')\d\tau',\tau\right) \d\tau}\right],
%t_s+\Delta t_a
\label{c1.U_n-integral}
\end{equation}
which generally represents the fact that in the asymptotic region the probability of more than one electron being present, and the total wavefunction is an (antisymmetrized) tensor product of an ion-core state and a single-electron ground orbital, which will behave in the WKB regime as
\begin{equation}
\psi_g(a) \propto e^{-i\int_{t_\kappa}^{t_a}U\d\tau}.
\label{WKB-formula}
\end{equation}

General results regarding this connection are hard to establish analytically -- although numerical methods based on ensembles of wavepackets propagated classically are quite successful \cite{Tannor} --, which is due partly to the difficulty of constructing WKB methods on more than one spatial dimension and also to the fact that the Dyson orbital $\ket{n_D(t)}=\sqrt{N}\bracket{n(t)}{\Psi_g}$ depends in far more detail on the neutral and ionic hamiltonians than through the mean value $U_n=\matrixel{n(t)}{V_{ee}}{n(t)}$.

Thus, we will not attempt to prove any formal independence; instead, we will transform both integrals in eq. \eqref{c1.U_n-integral} to simpler forms which, under suitable approximations, will enable us to extricate the $a$-dependence from the dynamics of interest at this point. The key realization for this is that the extra time interval $\Delta t_a=a\cos(\theta)/v_\parallel$ is essentially the time it takes an electron going at parallel velocity $v_\parallel$ to advance a length $z=a\cos(\theta)$ along the axis of the laser's polarization. Therefore, we can essentially back-propagate the inner integral to the time $t_s$:
\begin{align*}
\vb{r}+\int_{t_s+\Delta t_a}^\tau& \vb{v}(\tau')\d\tau'  =   \vb{r}   -   \int_{t_s}^{t_s+\Delta t_a} \vb{v}(\tau')\d\tau'   +   \int_{t_s}^\tau \vb{v}(\tau')\d\tau'\approx   \vb{r}   -   \Delta t_a \vb{v}(t_s)   +   \int_{t_s}^\tau \vb{v}(\tau')\d\tau'  .
 %\\& =   \vb{r}   -   \frac{\vb{F}(t_s)\cdot\vbr}{\vb{F}(t_s)\cdot\vb{v}(t_s)} \vb{v}(t_s)   +   \int_{t_s}^\tau \vb{v}(\tau')\d\tau'
\end{align*}
%\Delta t_a = \frac{\vb{F}(t_s)\cdot\vbr}{\left(\vbp+\vb{A}(t_s)\right)\cdot\vb{F}(t_s)}=\frac{a\cos(\theta)}{v_\parallel(t_s)},

%%% The next paragraph developed not on the notebook and directly for this on 12.09.2012.
The first two terms are best considered in a reference frame in which the laser field points along the positive $z$ axis, which we will adopt whenever we require coordinate calculations. In this frame we have
\begin{align*}
\vb{r} - \Delta t_a \vb{v}(t_s)=\vb{r} - \frac{\vb{F}(t_s)\cdot\vbr}{\vb{F}(t_s)\cdot\vb{v}(t_s)} \vb{v}(t_s) = \begin{pmatrix}x\\y\\z\end{pmatrix} - \frac{z}{v_z}\begin{pmatrix}v_x\\v_y\\v_z \end{pmatrix}=\frac{1}{v_z}\begin{pmatrix}v_z x-v_x z\\v_z y-v_y z\\0\end{pmatrix},
\end{align*}
where the components are strongly reminiscent of a vector cross product. Thus, as long as the main contributions to the spherical integral come from points that satisfy
$$\frac{x}{z}\approx \frac{v_x}{v_z}\textrm{ and }\frac{y}{z}\approx \frac{v_y}{v_z}$$
we can ignore the remnant $\vb{r} - \Delta t_a \vb{v}(t_s)$, which makes the inner integral $\vbr$-independent. It is important to note here, though, that $v_z(t_s) = p_z+A_z(t_s)$ is a complex quantity whereas the transverse components are simply the transverse momentum and are therefore real, as are in this setting all three position coordinates. Nevertheless, this approximation is good enough that we will ignore the remnant.

With this, then, we can approximate the inner integral in expression \eqref{c1.U_n-integral} as
\begin{align*}
e^{i\int_T^{t_a}U_n\d\tau}%&=\exp\left[{-i\int_{t_a}^TU_n\left(\vb{r}+\int_{ t_a}^\tau \vb{v}(\tau')\d\tau',\tau\right) \d\tau}\right]\\
& \approx \exp\left[{-i\int_{t_a}^TU_n\left(\int_{ t_s}^\tau \vb{v}(\tau')\d\tau',\tau\right) \d\tau}\right].
\end{align*}
The final step is then to replace the $a$-dependent limit $t_a$ by some limit $t_\kappa$ that can depend only on the initial state so that we can factor out the WKB dependence of \eqref{WKB-formula} from the dynamics of interest; informally, $t_\kappa$ is the time of entrance to the tunnel. This leaves then, finally,
\begin{align*}
e^{i\int_T^{t_a}U_n(\rl(\tau;\vb{r},\vbp,t_a),\tau)\d\tau}& \approx e^{-i\int_{t_a}^{t_\kappa}U_n\left(\int_{ t_s}^\tau \vb{v}(\tau')\d\tau',\tau\right) \d\tau}e^{-i\int_{t_\kappa}^TU_n\left(\int_{ t_s}^\tau \vb{v}(\tau')\d\tau',\tau\right) \d\tau},
\end{align*}
where the first factor is retained to cancel the $a$-dependence of the Dyson orbital, and the second one represents an overall Coulomb correction, in the semiclassical approximation, to our wavefunction.

This finalizes, then, our analysis of the first-order ionization yield for a general Dyson orbital. Our work can then be summarized as
\begin{equation}
	a_n^{(1)}(\vb{p},t_0)
	 = 
	a_g(t_s) b_n(t_0,t_s)e^{-iE_nt_0}
	e^{-\frac{i}{2} \int_{t_s}^T\left(\vbp+\vb{A}(\tau)\right)^2\d\tau+i I_{p,n} t_s}
	e^{-i\int_{t_\kappa}^TU_n\left(\int_{ t_s}^\tau \vb{v}(\tau')\d\tau',\tau\right) \d\tau}
	R(\vbp),
	\label{first-order-yield-abstract}
\end{equation}
where
\begin{equation}
	R(\vbp)  =
	\frac{-ia^2 }{\sqrt{iS_V''(t_s)}} 
	\int\frac{\d\Omega}{2\pi}
	e^{-i\int_{t_a}^{t_\kappa}U_n\left(\int_{ t_s}^\tau \vb{v}(\tau')\d\tau',\tau\right) \d\tau}
	e^{-i\left(\vbp+\vb{A}(t_s)\right)\cdot\vb{r}}
	\left.\left(\frac{\partial}{\partial r}+\frac{1-b}{r}\right)\right|_{a} \bracket{\vbr}{n_D(t_s)}
	.
\end{equation}
is a factor encoding all the dependence on the structure of the initial state, as well as most of the dependence on $\vbp$.
%all the dependence on $\vbp$ except for the gaussian kernel caused by the imaginary part of $t_s$ in the factor $e^{-\frac{i}{2} \int_{t_s}^T\left(\vbp+\vb{A}(\tau)\right)^2\d\tau}$. 
This factor is effectively a Fourier transform over the spherical boundary of the tunnelled state, which is transmitted across the boundary by the Bloch operator as $\matrixel{\vbr}{\hat{L}^-(a)}{n_D(t_s)}$.

For completeness, we include here a brief analysis of the Volkov action's second derivative, $S_V''$. We know already the first derivative, 
\begin{align*}
	-i\frac{\d S_V}{\d t'}
	  & = i\left(\frac{1}{2} \left(\vbp+\vb{A}(t')\right)^2+ I_{p,n} + \vb{F}(t')\cdot\vbr\right),
\end{align*}
from which we can calculate
\begin{align*}
	\frac{\d^2 S_V}{\d t'^2}
	  & = -\left(\left(\vbp+\vb{A}(t')\right)\cdot\frac{\d\vb{A}}{\d t'} + \frac{\d\vb{F}(t')}{\d t'}\cdot\vbr\right)
		= \left(\vb{v}(t')\cdot\vb{F}(t') - \frac{\d\vb{F}(t')}{\d t'}\cdot\vbr\right).
\end{align*}
The second term does depend on $\vbr$ but it is precisely this dependence that makes it neglectable with respect to the first, as long as the product $a^2\omega$ is of order unity, for $\omega$ the pulse's carrier frequency. Under this approximation, then, $S_V''(t_s)$ depends only on the parallel momentum $p_\parallel$, and is equal in a semiclassical interpretation to the power delivered by the laser pulse to the ionized electron.

%% file: c2-hydrogenicorbitals.tex
\section{Hydrogenic Dyson orbitals}
We will now consider the case that the Dyson orbital in question has the structure of a hydrogenic state with well-defined angular momentum. Thus we consider the specific case in which 
\begin{align}
\bracket{\vbr}{n_D(t)} & = \varphi(r)Y_{lm}(\theta,\phi),\textrm{ for}\\
      \varphi(r) & = C_{\kappa l} \kappa^{3/2} \frac{e^{-\kappa r}}{\kappa r}(\kappa r)^{Q/\kappa}\textrm{, where of course}\\
   Y_{lm}(\theta,\phi) & = N_{lm} P_l^m(\cos(\theta))\frac{e^{im\phi}}{\sqrt{2\pi}} \textrm{ with } N_{lm}=\sqrt{\frac{(2l+1)(l-m)!}{2(l+m)!}};
\end{align}
%%%
%\pdfmargincomment{Citation missing for the radial asymptotic formula.}
%%%
see \cite{PPT} and \cite[eq. \href{http://dlmf.nist.gov/14.30.E1}{14.30.1}]{NIST-handbook}. Here $Q$ is the net positive charge on the ion, and $C_{\kappa l}$ is a constant that depends, among other things, on how concentrated the Dyson orbital is in regions close to the ion, so that it must be evaluated numerically. We will perform the angular integration as well as the explicit elimination of the boundary radius $a$, and confirm that the results coincide with those of reference \citealp{saepaper}.

The first step is to deal with the action of the Bloch operator, where we can ignore the angular part of the wavefunction and are therefore only interested in the combination
\begin{align*}
	\left.\left(\frac{\partial}{\partial r}+\frac{1-b}{r}\right)\right|_{a} \varphi(r)
	& = C_{\kappa l} \kappa^{3/2}\left.\left(\frac{\partial}{\partial r}+\frac{1-b}{r}\right)\right|_{a} \left[ \frac{e^{-\kappa r}}{\kappa r}(\kappa r)^{Q/\kappa}\right]\\
	& = C_{\kappa l} \kappa^{3/2}\left.\left(-\kappa-\frac{1}{a}+\frac{Q/\kappa}{a}+\frac{1-b}{a}\right)\right. \left[ \frac{e^{-\kappa a}}{\kappa a}(\kappa a)^{Q/\kappa}\right]\\
	& = -\kappa \varphi(a),
\end{align*}
where we have fixed $b=Q/\kappa$.

We now consider the exponentiated ionic-potential integral, where in the asymptotic regime we can replace the potential by its monopole component:
$$U_n(r,\theta,\phi;\tau)\approx -\frac{Q}{r}.$$
As such, its integral is given by 
\begin{align*}
-i\int_{t_a}^{t_\kappa}U_n\left(\int_{ t_s}^\tau \vb{v}(\tau')\d\tau',\tau\right) \d\tau = \int_{t_a}^{t_\kappa} \frac{-iQ\d\tau}{\sqrt{r(\tau)^2}}
\end{align*}
where we have defined $r(\tau)\colonequals \int_{ t_s}^\tau \hat{\vbr}\cdot\vb{v}(\tau')\d\tau'$, for which $\frac{\d r}{\d\tau}=\hat{\vbr}\cdot \vb{v}(\tau).$ To proceed, we now perform two approximations: we first assume that the velocity will not change significantly during the time interval being integrated, so we can replace $\vb{v}(\tau)$ by $\vb{v}(t_s)$, and we further suppose that the tunnelling velocity will be primarily directed towards negative $z$ (since the field points towards positive $z$). These two together imply that $\frac{\d r}{\d\tau}=-v_\parallel(t_s)\approx+i\kappa$ and therefore that
\begin{align*}
  %-i\int_{t_a}^{t_\kappa}U_n\left(\int_{ t_s}^\tau \vb{v}(\tau')\d\tau',\tau\right) \d\tau 
  %= 
  \int_{t_a}^{t_\kappa} \frac{-iQ\d\tau}{\sqrt{r(\tau)^2}}
  =
	-\frac{Q}{\kappa}\int_{a}^{r_\kappa} \frac{\d r}{\sqrt{r^2}}
  =
	+\frac{Q}{\kappa}\int_{a}^{r_\kappa} \frac{\d r}{r}
	=
	\frac{Q}{\kappa}\ln\left(\frac{r_\kappa}{a}\right)
	=
	-\frac{Q}{\kappa}\ln\left(\kappa a\right),
\end{align*}
by choosing $r_\kappa=1/\kappa$.

Our $R$ factor is then given by
\begin{align*}
	R(\vbp) & =
	\frac{-ia^2 }{\sqrt{iS_V''(t_s)}} 
	\int\frac{\d\Omega}{2\pi}
	\left(\kappa a\right)^{-{Q}/{\kappa}}
	\left.e^{-i\left(\vbp+\vb{A}(t_s)\right)\cdot\vb{r}}\right|_a
	(-\kappa) \varphi(a)Y_{lm}(\theta,\phi)
	%%%
	\displaybreak[1] \\  & =
	%%%
	 C_{\kappa l} \kappa^{3/2}
	\frac{-ia^2 (-\kappa)}{\sqrt{iS_V''(t_s)}} 
	\int_0^\pi\sin(\theta)\d\theta\int_0^{2\pi}\frac{\d\phi}{2\pi}
	%\left(\kappa a\right)^{-{Q}/{\kappa}}
	e^{-i\left[\left(p_\parallel+A(t_s)\right)a\cos(\theta)+p_\perp a \sin(\theta)\cos(\phi-\phi_p)\right]}
	%\left(\vbp+\vb{A}(t_s)\right)\cdot\vb{r}}
	\\ & \qquad \qquad \qquad\qquad\qquad\qquad\qquad\qquad\times
	\frac{e^{-\kappa a}}{\kappa a}Y_{lm}(\theta,\phi)
	%%%
	\displaybreak[1] \\  & =
	%%%
	\frac{C_{\kappa l} \kappa^{3/2}}{(2\pi)^{3/2}}
	\frac{ia }{\sqrt{iS_V''(t_s)}} 
	\int_0^\pi\sin(\theta)\d\theta
	N_{lm}P_l^m(\cos(\theta))
	e^{-\kappa a}
	e^{-iv_\parallel(t_s)a\cos(\theta)}
	\\ & \qquad\qquad\qquad\qquad\qquad\qquad\times
	\int_0^{2\pi}
	e^{-ip_\perp a \sin(\theta)\cos(\phi-\phi_p)}
	e^{im(\phi-\phi_p)}
	\d\phi
	e^{im\phi_p}
	%%%
	\displaybreak[1] \\  & =
	%%%
	\frac{C_{\kappa l} \kappa^{3/2}}{\sqrt{2\pi}}
	\frac{ia }{\sqrt{iS_V''(t_s)}} 
	\int_0^\pi\sin(\theta)\d\theta
	N_{lm}P_l^m(\cos(\theta))
	e^{-\kappa a}
	e^{-\kappa a\cos(\theta)}
	\\ & \qquad\qquad\qquad\qquad\qquad\qquad\times
	e^{im\phi_p} (-i)^m
	J_m(p_\perp a \sin(\theta))
	.
\end{align*}
Here we apply again the approximation of small tunnelling angles, which means that the integral over $\theta$ is concentrated near $\theta\approx\pi$, which is the case in the limit $\kappa a\gg1$. In this limit, then, we can take leading-order expansions in $\pi-\theta$ on all the functions involved: assuming that $m\geq0$, we have
\begin{subequations}
\begin{align}
	\sin(\theta) & \approx \pi-\theta,\\
	e^{-\kappa a}	e^{-\kappa a\cos(\theta)}& \approx e^{-\frac{1}{2}\kappa a (\pi-\theta)^2},\\
	J_m(p_\perp a \sin(\theta)) & \approx  \left(\frac{p_\perp a}{2}\right)^{m} \frac{(\pi-\theta)^{m}}{\Gamma(m+1)}, \\
  N_{lm}P_l^m(\cos(\theta)) & \approx (-1)^l \sqrt{\frac{(2l+1)(l+m)!}{2(l-m)!}}\frac{(\pi-\theta)^{m}}{2^m m!};
\end{align}
\end{subequations}
%\pdfmargincomment{This needs references.}
see \cite[eqs.~\href{http://dlmf.nist.gov/10.2.E2}{10.2.2} and \href{http://dlmf.nist.gov/14.3.E4}{14.3.4}]{NIST-handbook}. With this, then, the angular integral reduces to 
\begin{align*}
	R(\vbp) & \approx
	\frac{C_{\kappa l} \kappa^{3/2}}{\sqrt{2\pi}}
	\frac{ia e^{im\phi_p} (-i)^m}{\sqrt{iS_V''(t_s)}} 
	 \frac{(-1)^l}{\Gamma(m+1)}
	\sqrt{\frac{(2l+1)(l+m)!}{2(l-m)!}}
	\left(\frac{p_\perp a}{2}\right)^{m}
	\\ & \qquad \times
	\int_0^\pi\d\theta
	%(\pi-\theta)
	\frac{(\pi-\theta)^{2m+1}}{2^m m!}
	e^{-\frac{1}{2}\kappa a (\pi-\theta)^2}
	%(\pi-\theta)^{m}
	%%%
	\displaybreak[1] \\  & \approx
	%%%
	\frac{C_{\kappa l} \kappa^{3/2}}{\sqrt{2\pi}}
	\frac{ia e^{im\phi_p} (-i)^m}{\sqrt{iS_V''(t_s)}} 
	 \frac{(-1)^l}{\Gamma(m+1)} 
	\frac{1}{2^m m!}
	\sqrt{\frac{(2l+1)(l+m)!}{2(l-m)!}}
	\left(\frac{p_\perp a}{2}\right)^{m}
	\\ & \qquad \times
	\int_0^\infty
	\frac{1}{2}\left(\frac{2}{\kappa a}\right)^{m+1}
	\eta^m
	e^{-\eta}
	\d\eta
	\qquad\qquad\textrm{ for } \eta=\frac{1}{2}\kappa a (\pi-\theta)^2
	%%%%
	%\\  & =
	%%%%
	%\frac{C_{\kappa l} \kappa^{3/2}}{\sqrt{2\pi}}
	%\frac{ia e^{im\phi_p} (-i)^m}{\sqrt{iS_V''(t_s)}} 
	 %\frac{(-1)^l}{\Gamma(m+1)}
	%\sqrt{\frac{(2l+1)(l+m)!}{2(l-m)!}}
	%\left(\frac{p_\perp a}{2}\right)^{m}
	%\\ & \qquad \times
	%\frac{1}{2}\left(\frac{2}{\kappa a}\right)^m
	%\Gamma(m+1)
	%%%
	\displaybreak[1] \\  & =
	%%%
	\frac{ (-1)^{l+m} i^{m+1}}{\sqrt{iS_V''(t_s)}} 
	\frac{1}{2^m m!}
	\sqrt{\frac{(2l+1)(l+m)!}{4\pi(l-m)!}}
	e^{im\phi_p}  \left(\frac{p_\perp }{\kappa}\right)^{m}
	C_{\kappa l} \kappa^{1/2}
	.
\end{align*}
This coincides with the results of Torlina and Smirnova, and concludes the calculation.

For ease of handling, we encapsulate all the constants into a single factor,
\begin{equation}
K_0\colonequals  C_{\kappa l}\sqrt{\kappa}\frac{(-1)^{l+m} i^{m+1}}{2^m m!}	\sqrt{\frac{(2l+1)(l+m)!}{4\pi(l-m)!}}.
\label{K0definition}
\end{equation}
We also separate the dependence on the parallel momentum into a slow factor defined as ${K(p_\parallel)=\frac{K_0}{ \sqrt{iS_V''(t_s)}} } $, for which we then have
\begin{equation}
R(\vbp)=K(p_\parallel)e^{im\phi_p}  \left(\frac{p_\perp }{\kappa}\right)^{m}.
\label{Kdefinition}
\end{equation}

Putting it all together, the ionization yield for hydrogenic orbitals is given by
%\begin{align}
	%a_n^{(1)}(\vb{p},t_0)
	%& = 
	%a_g(t_s) b_n(t_0,t_s)
	%R(\vbp)
	%e^{-iE_nt_0}
	%e^{i I_{p,n} t_s}
	%e^{-\frac{i}{2} \int_{t_s}^T\left(\vbp+\vb{A}(\tau)\right)^2\d\tau}
	%e^{-i\int_{t_\kappa}^TU_n\left(\int_{ t_s}^\tau \vb{v}(\tau')\d\tau',\tau\right) \d\tau}
	%.
  %\label{hydrogenic-direct-ionization-yield}
%\end{align}
%
\begin{align}
	a_n^{(1)}(\vb{p},t_0)
	& = 
	a_g(t_s) b_n(t_0,t_s)
	\frac{K_0}{ \sqrt{iS_V''(t_s)}}
	e^{im\phi_p}  \left(\frac{p_\perp }{\kappa}\right)^{m}
	e^{-iE_nt_0}
	\nonumber \\ & \qquad \times
	e^{-\frac{i}{2} \int_{t_s}^T\left(\vbp+\vb{A}(\tau)\right)^2\d\tau}
	e^{i I_{p,n} t_s}
	e^{-i\int_{t_\kappa}^TU_n\left(\int_{ t_s}^\tau \vb{v}(\tau')\d\tau',\tau\right) \d\tau},
  \label{hydrogenic-direct-ionization-yield}
\end{align}
though we will in general prefer the more abstract expression \eqref{first-order-yield-abstract}. Here we notice in particular that the modulus of this ionization yield is reduced by the term $\left|e^{i I_{p,n} t_s}\right|=e^{-I_{p,n}\tauT}$ for $\tauT=\Im(t_s)$. This damps exponentially the tunnelling amplitudes from orbitals further down the potential well, which have higher ionization potential, and as we will see this will enable us to see much more clearly the second-order contribution, to which we now turn.

%% file: c3-second-order-setup.tex
\section{The second-order ionization yield}
\label{cisetup}

We now turn to the second term in the Dyson expansion, given in eq. \eqref{c1.abstractcorrelationstate} and equal to
\begin{align*}
\smallket{\Psi^{(2)}(T)}=(-i)^2\sum_n \int\d\vb{k} & \int^T\d t''\int^{t''}\d t'U^N(T,t'')V_{ee}^n(t'') U^{N-1}(t'',t')\ket{n(t')}\nonumber\\
& \otimes U_e^n(t'',t')\ket{\vb{k}_n(t')}\times\matrixel{\vb{k}_n(t')}{\hat{L}^-(a)}{n_D(t')}a_g(t')e^{iI_p t'}.
\end{align*}
Here we truncate the Dyson series at this point by replacing the full propagator by its tensor version: we approximate $U^N(T,t'')\approx U^{N-1}(T,t'')\otimes U^m_e(T,t'')$. This represents the definitive assumption that the correlation interaction is a weak perturbation on each channel's propagation, and neglects quadratic terms in this perturbation. Such an assumption, of course, needs to be verified \textit{a posteriori} by checking that the perturbed amplitudes are indeed much smaller than the ones for direct tunnelling, and can always be corrected by including further terms in the series, although the calculation quickly becomes impractical

We want to calculate, then, the corresponding second-order ionization yield, which can be expressed as
\begin{align*}
a^{(2)}_m&(\vbp,t_0)
=\bra{\vb{p}}\otimes\bra{m(t_0)}U^{N-1}(t_0,T)\smallket{\Psi^{(2)}(T)}\\
&=-i\sum_n \int\!\d\vb{k}  
\int^T\!\!\!\d t''  
\int^{t''}\!\!\!\d t' 
\bra{m(t_0)}U^{N-1}(t_0,t'')
 \bra{\vbp_m(t'')}V_{ee}^n(t'')\ket{\vb{k}_n(t'')} \nonumber\\
& \qquad\qquad\qquad \times 
(-i)U^{N-1}(t'',t_s)
U^{N-1}(t_s,t')\ket{n(t')}
\matrixel{\vb{k}_n(t')}{\hat{L}^-(a)}{n_D(t')}
a_g(t')e^{iI_p t'}.
\end{align*}
Here we note that the inner integral, over $t'$, is essentially identical to the one from the direct-ionization case discussed above. To bring it into closer agreement, we rephrase eq. \eqref{propagatedquasistaticeigenstates} as
\begin{equation}
{U^{N-1}(t_s,t')}\ket{n(t')}=e^{-iE_n(t_s-t')}b_n(t_s,t')\ket{n(t_s)}.
\label{fullypropagatedquasistaticeigenstates}
\end{equation}
Pulling the $t'$ integral inwards, our first task is to calculate
\begin{align*}
	-i\int^{t''}\!\!\!\d t' &  U^{N-1}(t_s,t')  \ket{n(t')}\matrixel{\vb{k}_n(t')}{\hat{L}^-(a)}{n_D(t')}a_g(t')e^{iI_p t'}  
	\\ 	& = 
	-i\int^{t''}\!\!\!\d t' e^{-iE_n(t_s-t')}b_n(t_s,t')\ket{n(t_s)}\matrixel{\vb{k}_n(t')}{\hat{L}^-(a)}{n_D(t')}a_g(t')e^{iI_p t'}
	\\ 	& = 
	\ket{n(t_s)}\times(-i)e^{-iE_nt_s}\int^{t''}\!\!\!\d t'b_n(t_s,t')a_g(t')\matrixel{\vb{k}_n(t')}{\hat{L}^-(a)}{n_D(t')}e^{i I_{p,n} t'}
	%\\  & =	\ket{n(t_s)}a^{(1)}(\vbp,t_s)
	,
\end{align*}
and this is precisely the direct-tunnelling amplitude as given by equation \eqref{saeyieldbeginning}, as long as the real part of the upper limit, $t''$ exceeds that of the saddle point; otherwise, the integral will be zero, which will allow us to restrict the interval of the $t''$ integral to physically meaningful times.

We can therefore simply use the result from the direct case, eq. \eqref{hydrogenic-direct-ionization-yield}, and insert it into our expression for the second-order term. Thus we seek
\begin{align*}
a^{(2)}_m(\vbp,t_0) 
 & =
   -i\sum_n 
   \int\!\d\vb{k}  
   \int^T\!\!\!\d t''  
   \bra{m(t_0)}U^{N-1}(t_0,t'') 
   \bra{\vbp_m(t'')}V_{ee}^n(t'')\ket{\vb{k}_n(t'')} 
\nonumber \\ & \,\,\,\,\, \times 
   U^{N-1}(t'',t_s)
   (-i)\int^{t''}\!\!\!\d t' 
   U^{N-1}(t_s,t')\ket{n(t')}
   \matrixel{\vb{k}_n(t')}{\hat{L}^-(a)}{n_D(t')}a_g(t')e^{iI_p t'}
%%%%
%\\ & =
%%%%
   %-i\sum_n 
   %\int\!\d\vb{k}  
   %\int^T\!\!\!\d t''  
   %\bra{n(t_0)}U^{N-1}(t_0,t'') 
   %\bra{\vbp_m(t'')}V_{ee}^n(t'')\ket{\vb{k}_n(t'')} 
%\nonumber \\ & \qquad\quad \times 
   %U^{N-1}(t'',t_s)
   %\ket{n(t_s)}e^{-iE_nt_s}
	%\times(-i)
	%\int^{t''}\!\!\!\d t'
	%b_n(t_s,t')%=b_n(t_s,t_0)b_n(t_0,t')
	%a_g(t')
	%\matrixel{\vb{k}_n(t')}{\hat{L}^-(a)}{n_D(t')}
	%e^{i I_{p,n} t'}
%%%
\\ & =
%%%
   -i\sum_n 
   \int\!\d\vb{k}  
   \int^T\!\!\!\d t''  
   \bra{m(t_0)}U^{N-1}(t_0,t'') 
   \bra{\vbp_m(t'')}V_{ee}^n(t'')\ket{\vb{k}_n(t'')} 
\nonumber \\ & \qquad\quad \times 
   U^{N-1}(t'',t_s)
   \ket{n(t_s)}
	 e^{-iE_nt_s}e^{iE_nt_0} b_n(t_s,t_0)
	 a^{(1)}(\vbk,t_s)
  .
\end{align*}
To proceed further, we now insert once again a resolution of the identity of the form $\int\d\vbr\outerp{\vbr}=1$ into the correlation interaction's matrix element, which turns it into a position-dependent ionic operator $V_{ee}^n(t'',\vbr)$ and brings to light the phases associated with the two eikonal Volkov wavefunctions $\ket{\vbp_m(t'')}$ and $\ket{\vbk_n(t'')}$.

This transforms the ionization yield into 
\begin{align*}
a^{(2)}_m(\vbp,t_0)
 & =
   -i\sum_n 
   \int \d t''\!
   \int\!\d\vb{k} 
	 \int\!\d\vbr  
   \bra{m(t_0)}U^{N-1}(t_0,t'') 
	 V_{ee}^n(t'',\vbr)
	 U^{N-1}(t'',t_s)\ket{n(t_s)}
\nonumber \\ & \qquad\quad \times 
   \bracket{\vbp_m(t'')}{\vbr}
	 \bracket{\vbr}{\vb{k}_n(t'')} 
	 a^{(1)}(\vbk,t_s)
	e^{iE_n(t_0-t_s)} b_n(t_s,t_0)
%%%%
%\\ &  =
%%%%
  .
\end{align*}
This we now simplify further by explicitly factoring out the energy and Stark shift phases associated with the quasistatic eigenstates propagated during the ionization time in the correlation interaction's matrix product, 
\begin{equation}
U^{N-1}(t'',t_s)\ket{n(t_s)}
\equalscolon 
e^{-iE_n(t''-t_s)}
b_n(t'',t_s)
\ket{n(t'',t_s)}
,
\label{definition-propagated-eigenstates}
\end{equation}
and analogously for $U^{N-1}(t'',t_0)\ket{m(t_0)}$. Of course, the new states $\ket{n(t'',t_s)}$ will most likely have components on more than one quasi-static eigenstates and they must be found by numerically propagating the states during the ionization step in imaginary time. This is nevertheless acceptable and is part of the numerical effort that must go into calculating the correlation interaction's matrix element, which we will denote by the shorthand
\begin{equation}
\langle V(\vbr)\rangle
\colonequals 
   \bra{m(t_0,t'')}
	 V_{ee}^n(t'',\vbr)
   \ket{n(t'',t_s)}
	.
\label{shorthand-for-correlation-interaction-matrix-element}
\end{equation}
This hides the important dependence of $\langle V(\vbr)\rangle$ on $m$, $n$, $t_0$, $t_s$ and $t''$, but it will not be used for long.

Mixing all these ingredients back in, along with the explicit expression \eqref{c1.eikonal-volkov-wavefunctions} for the Volkov wavefunctions, we obtain the formidable integral
%\begin{align*}
%a^{(2)}_m(\vbp,t_0)
 %& =
   %-i\sum_n 
   %\int \d t''\!
   %\int\!\d\vb{k} 
	 %\int\!\d\vbr  
	 %%U^{N-1}(t'',t_0)\ket{m(t_0)}
   %%=
	 %%e^{-iE_m(t''-t_0)}
   %%b_n(t'',t_0)
   %%\ket{n(t'',t_0)}
	 %%\bra{m(t_0)}U^{N-1}(t_0,t'')
	 %e^{iE_m(t''-t_0)}
	 %e^{-iE_n(t''-t_s)}
   %b_m(t_0,t'')
   %b_n(t'',t_s)
   %\langle V(\vbr) \rangle
%\nonumber \\ & \qquad\quad \times 
   %\frac{1}{(2\pi)^{3/2}}
	 %e^{-i\left(\vbp+\vb{A}(t'')\right)\cdot\vb{r}}
	 %e^{+\frac{i}{2} \int_T^{t''}\left(\vbp+\vb{A}(\tau)\right)^2\d\tau} 
	 %e^{+i\int_T^{t''} U_m(\rl(\tau;\vb{r},\vbp,t''),\tau)\d\tau}
%\nonumber \\ & \qquad\quad \times 
   %\frac{1}{(2\pi)^{3/2}}
	 %e^{i\left(\vb{k}+\vb{A}(t'')\right)\cdot\vb{r}} 
	 %e^{-\frac{i}{2} \int_T^{t''}\left(\vb{k}+\vb{A}(\tau)\right)^2\d\tau} 
	 %e^{-i\int_T^{t''} U_n(\rl(\tau;\vb{r},\vb{k},t''),\tau)\d\tau}
%\nonumber \\ & \qquad\quad \times 
	 %%a^{(1)}(\vbk,t_s)
	 %a_g(t_s) b_n(t_0,t_s)e^{-iE_nt_0}
	 %e^{-\frac{i}{2} \int_{t_s}^T\left(\vbk+\vb{A}(\tau)\right)^2\d\tau+i I_{p,n} t_s}
	 %e^{-i\int_{t_\kappa}^TU_n\left(\int_{ t_s}^\tau (\vbk+\vb{A}(\tau'))\d\tau',\tau\right) \d\tau}
	 %R(\vbk)
	 %e^{-iE_nt_s}e^{iE_nt_0} b_n(t_s,t_0)
%\end{align*}
%
%
\begin{align}
a^{(2)}_m(\vbp,t_0)
 & =
   -i\sum_n 
	 e^{-iE_mt_0}
	 e^{i I_{p,n} t_s}
  \int \d t''
   b_m(t_0,t'')
   b_n(t'',t_s)
	 e^{i(E_m-E_n)t''}
	 e^{-\frac{i}{2} \int_{t''}^T\left(\vbp+\vb{A}(\tau)\right)^2\d\tau} 
\nonumber \\ & \quad \times 
	 a_g(t_s) 
   \frac{1}{(2\pi)^{3}}
	 \int\!\d\vbr  \!
   \int\!\d\vb{k} \,
	 e^{i\left(\vbk-\vbp\right)\cdot\vb{r}} 
	 R_n(\vbk)
   \langle V(\vbr) \rangle
	 e^{-\frac{i}{2} \int_{t_s}^{t''}\left(\vbk+\vb{A}(\tau)\right)^2\d\tau}
	 e^{-iW_{mn}^C(\vbr,\vbk,t'')}
  .
\label{integral-starting-point-old}
\end{align}
Here we have encapsulated the Coulomb correction as
\begin{align}
W_{mn}^C(t'',\vbr,\vbk)
& =
 \int_{t''}^T U_m(\rl(\tau;\vb{r},\vbp,t''),\tau)\d\tau
\nonumber \\ & \qquad 
-\int_{t''}^T U_n(\rl(\tau;\vb{r},\vb{k},t''),\tau)\d\tau
+\int_{t_\kappa}^TU_n\left( \rl(\tau;\vb{0},\vbk,t_s) ,\tau\right) \d\tau
\label{coulombcorrectionphase}
\end{align}
to simplify the notation as far as possible; wherever necessary, we will treat it as a slow prefactor.

%% file: c3-saddle-point-argument.tex
\section{The saddle-point argument}
\label{saddlepointargument}
We will now discuss the method of Torlina, Ivanov \textit{et al}. \cite{mepaper} of dealing, using a saddle-point argument, with the spatial and momentum integrals in expression \eqref{integral-starting-point-old} for the second-order wavefunction. We will rephrase some parts of the argument - in particular, performing all the necessary algebraic manipulations \textit{before} applying the saddle-point approximation - but the mathematical essentials of the method will not be changed. We will then discuss some physical drawbacks of the method, and suggest some ways that these come about. Afterwards, in section \ref{start-of-the-real-work}, we will develop a new take on the transverse integrals of equation \eqref{integral-starting-point-old} and evaluate them exactly, which brings out new, distinct physical features and predictions.

We begin by considering the momentum integral in equation \eqref{integral-starting-point-old}. Despite its complicated looks, if one ignores the prefactors it is structurally very simple, and although $\vbk$ appears inside an inner integral the actual dependence is simply quadratic:
\begin{align*}
-\frac{i}{2} \int_{t_s}^{t''}\left(\vbk+\vb{A}(\tau)\right)^2\d\tau
& =
  -\frac{i}{2} \int_{t_s}^{t''}\left(\vbk^2+2\vbk\cdot\vb{A}(\tau)+\vb{A}(\tau)^2\right)\d\tau
\\ & =
  -\frac{i}{2}(t''-t_s)\vbk^2 -i\vbk\cdot\int_{t_s}^{t''}\vb{A}(\tau)\d\tau-\frac{i}{2} \int_{t_s}^{t''}\vb{A}(\tau)^2\d\tau
	.
\end{align*}
%%%
%\pdfmargincomment{Cite PPT? Justify this here? elsewhere?}
%%%
Further, the $t''$ integral now goes from $t_s$, which has a positive imaginary part, to the real time $t_0$, as is argued in detail in ref. \cite{mepaper}; this implies that $t''-t_s$ has negative imaginary part, and for purposes of analytic continuation it has positive real part. In this case one usually takes $t_0$ as the real part of $t_s$ and therefore the coefficient $-i(t''-t_s)$ is a negative real number, so that the $\vbk$ integral has a gaussian kernel.

We deal with this gaussian kernel using the standard routine of completing squares, which must be done to accomodate the Fourier-transform-like phases arising from the $\vbr$-dependent terms. Thus we transform the exponent in equation \eqref{integral-starting-point-old}, using the shorthands $\AA(t'')=\int_{t_s}^{t''}\vb{A}(\tau)\d\tau$ and $\AA^2(t'')=\int_{t_s}^{t''}\vb{A}(\tau)^2\d\tau$, into
\begin{align*}
i(\vbk- & \vbp)\cdot\vbr-\frac{i}{2} \int_{t_s}^{t''}\left(\vbk+\vb{A}(\tau)\right)^2\d\tau
=
  i(\vbk-\vbp)\cdot\vbr
	-\frac{i}{2}(t''-t_s)\vbk^2 
	-i\vbk\cdot\AA(t'') 
	-\frac{i}{2}\AA^2(t'')
\\  & =
	-\frac{i}{2}(t''-t_s)\vbk^2 
	+i\vbk\cdot\left(\vbr-\AA(t'') \right)
	-\frac{i}{2}\AA^2(t'')
  -i\vbp\cdot\vbr
\\  & =
	-\frac{i}{2}(t''-t_s)
	  \left(
	    \vbk 
			-\frac{\vbr-\AA(t'')}{t''-t_s}
		\right)^2 
	+\frac{i}{2}(t''-t_s)
	  \left(
		\frac{\vbr-\AA(t'')}{t''-t_s}
		\right)^2
	-\frac{i}{2}\AA^2(t'')
  -i\vbp\cdot\vbr
\\  & =
	-\frac{i}{2}(t''-t_s)
	  \left(
	    \vbk 
			-\vbk_s
			%-\frac{\vbr-\AA(t'')}{t''-t_s}
		\right)^2 
	+\frac{i/2}{t''-t_s}
	  \left(
		\vbr^2-2\vbr\cdot\AA(t'')+\AA(t'')^2
		\right)
	-\frac{i}{2}\AA^2(t'')
  -i\vbp\cdot\vbr
\\  & =
	-\frac{i}{2}(t''-t_s)
	  \left(
	    \vbk 
			-\vbk_s
			%-\frac{\vbr-\AA(t'')}{t''-t_s}
		\right)^2 
	+\frac{i/2}{t''-t_s}
	  \left[
		\vbr^2-2\vbr\cdot\int_{t_s}^{t''}(\vbp+\vb{A}(\tau))\d\tau+\AA(t'')^2
		\right]
	-\frac{i}{2}\AA^2(t'')
\\  & =
	+\frac{i/2}{t''-t_s}
	  \left[
		  \left(
			\vbr
			-\int_{t_s}^{t''}(\vbp+\vb{A}(\tau))\d\tau
			\right)^2
			-\left(
			  \int_{t_s}^{t''}(\vbp+\vb{A}(\tau))\d\tau
			\right)^2
			+\AA(t'')^2
		\right]
\\ &\qquad\qquad
	-\frac{i}{2}\AA^2(t'')
	-\frac{i}{2}(t''-t_s)
	  \left(
	    \vbk 
			-\vbk_s
		\right)^2 
\\  & =
	\frac{i/2}{t''-t_s}
	  \left[
		  \left(
			\vbr
			-\vbr_s
			\right)^2
			  -\left(
			    \int_{t_s}^{t''}\vbp\d\tau
			  \right)^2
			  -2\int_{t_s}^{t''}\vbp\d\tau\cdot\AA(t'')
		\right]
\\ &\qquad\qquad
	-\frac{i}{2}\AA^2(t'')
	-\frac{i}{2}(t''-t_s)
	  \left(
	    \vbk 
			-\vbk_s
		\right)^2 
\\  & =
	-\frac{i}{2}(t''-t_s)
	  \left(
	    \vbk 
			-\vbk_s
		\right)^2
	+\frac{i/2}{t''-t_s}
		  \left(
			\vbr
			-\vbr_s
			\right)^2
	-\frac{i}{2}(t''-t_s)\vbp^2
	-i\vbp\cdot\AA(t'')
	-\frac{i}{2}\AA^2(t'').
\end{align*}

Here we have introduced the extra shorthands
\begin{equation}
\vbk_s(\vbr)=\frac{\vbr-\int_{t_s}^{t''}\vb{A}(\tau)\d\tau}{t''-t_s}
\textrm{ and }
\vbr_s(\vbp)=\int_{t_s}^{t''}(\vbp+\vb{A}(\tau))\d\tau,
\label{stationary-momentum-and-position}
\end{equation}
%%%
%\pdfmargincomment{Rephrase?}
%%%
which will be seen to have a deeper significance as the saddle points of the $\vbk$ and $\vbr$ integrals; 
they are also the final momentum of the classical trajectory that starts at the origin at time $t_s$
and the trajectory that starts at the origin at time $t_s$ and has final momentum $\vbp$, respectively.
In terms of these shorthands, the (admittedly long and tedious) calculation above gives the exact identity
\begin{align}
i(\vbk-  \vbp)\cdot\vbr-\frac{i}{2} \int_{t_s}^{t''}\left(\vbk+\vb{A}(\tau)\right)^2\d\tau
& =
	-\frac{i}{2}(t''-t_s)
	  \left(
	    \vbk 
			-\vbk_s(\vbr)
		\right)^2
	+\frac{i/2}{t''-t_s}
		  \left(
			\vbr
			-\vbr_s(\vbp)
			\right)^2
\nonumber\\ & \qquad\quad
	-\frac{i}{2}\int_{t_s}^{t''} (\vbp+\vb{A}(\tau))^2\d\tau.
\label{exact-saddle-point-identity}
\end{align}

Thus the integrals over $\vbk$ and then over $\vbr$ both have gaussian kernels and, what's more, the two kernels have reciprocal widths, so it could be understood as a single kernel with constant area in phase space. One must be careful, however, not to push this idea too far: the saddle point for $\vbk$ depends on $\vbr$, and if one tries to unravel this dependence the seemingly elliptic quadratic form \eqref{exact-saddle-point-identity} defaults to its original parabolic character, which of course it must.

It is now easy to compute the saddle-point argument. We consider the spatial and momentum integrals,
\begin{align*}
   \frac{1}{(2\pi)^{3}}
	 \int\!\d\vbr  \!
   \int\!\d\vb{k} \,
&   
	 e^{i\left(\vbk-\vbp\right)\cdot\vb{r}} 
	 R_n(\vbk)
   \langle V(\vbr) \rangle
	 e^{-\frac{i}{2} \int_{t_s}^{t''}\left(\vbk+\vb{A}(\tau)\right)^2\d\tau}
	 e^{-iW_{mn}^C(\vbr,\vbk,t'')}
\\ & =
   \frac{1}{(2\pi)^{3}}
	 e^{-\frac{i}{2}\int_{t_s}^{t''} (\vbp+\vb{A}(\tau))^2\d\tau}
 \int\!\d\vbr  
   \langle V(\vbr) \rangle
	 e^{\frac{i/2}{t''-t_s}\left(\vbr-\vbr_s(\vbp)\right)^2}
\\ & \qquad \qquad\qquad   \times
 \int\!\d\vb{k} \,
	 R_n(\vbk)
	 e^{-iW_{mn}^C(\vbr,\vbk,t'')}
	 %e^{i\left(\vbk-\vbp\right)\cdot\vb{r}} 
	 %e^{-\frac{i}{2} \int_{t_s}^{t''}\left(\vbk+\vb{A}(\tau)\right)^2\d\tau}
	 %
	 e^{-\frac{i}{2}(t''-t_s)\left(\vbk -\vbk_s(\vbr)\right)^2}
  .
\end{align*}
If we now assume that the factors $R_n(\vbk)$ and $\langle V(\vbr)\rangle$ are slow functions with respect to the momentum and length scales dictated by the time $t''-t_s$, we can substitute their arguments for the saddle points and simply integrate the gaussian kernels on their own. This is even simpler than usual since the kernel widths cancel out, and the factor of $\sqrt{2\pi}$ per spatial and momentum dimension exactly cancels out the factor of $1/(2\pi)^3$ already present; even better, the final stationary momentum simplifies to
$$
\vbk_s(\vbr_s(\vbp))
=\frac{\vbr_s(\vbp)-\int_{t_s}^{t''}\vb{A}(\tau)\d\tau}{t''-t_s}
=\frac{\int_{t_s}^{t''}(\vbp+\vb{A}(\tau))\d\tau-\int_{t_s}^{t''}\vb{A}(\tau)\d\tau}{t''-t_s}
=\vbp.
$$
Thus, the final value of the integral under consideration is simply
\begin{equation}
   \langle V(\vbr_s(\vbp)) \rangle
	 R_n(\vbp)
	 e^{-\frac{i}{2}\int_{t_s}^{t''} (\vbp+\vb{A}(\tau))^2\d\tau}
	 e^{-iW_{mn}^C(\vbr_s,\vbp,t'')}
	.
\label{final-saddle-point-result}
\end{equation}

Here, however, the method's deficiencies have now been brought to the surface. On one side, the equality of the pre-interaction momentum $\vbk$ and the final momentum $\vbp$ means that no momentum is exchanged during the correlation interaction, even though the ion is lifted to a state with a higher energy that can only come from the electron's motion. Additionally, the ionization yield's functional dependence on $\vbp$ is only very slightly affected by the interaction and therefore the wavefunction for the outgoing electron bears only slight traces of the details of the transition moment and the structure of the channels $\ket{n}$ and $\ket{m}$.
%%%
%\pdfmargincomment{This should damn probably have to be expanded.}
%%%

This behaviour indicates a breakdown of the assumption that the prefactors $R_n(\vbk)$ and $\langle V(\vbr)\rangle$ are slow functions, particularly since they may have zeros on or near the stationary point, and a more careful treatment of the saddle point method in these circumstances is required.

%% file: c3-setup-and-ansaetze.tex
In view of these shortcomings of the saddle-point method, in this section we will develop an alternative approach to the spatial and momentum integrals that appear in the second-order ionization yield. The key realization at the core of this development is that given reasonable expressions for the prefactors $R_n(\vbk)$ and $\langle V(\vbr)\rangle$, all four of the transverse integrals are analytically tractable. (The parallel integrals, on the other hand, involve the vector potential $\vb{A}$ which we will maintain unspecified, so that our method will hold for pulses of arbitrary shape, with the corresponding setback that both parallel integrals must be evaluated in the saddle-point approximation, which in this case is fully justified.)

Thus, we consider the integral
\begin{align}
\Int_1(\vbp,t'')=
   \frac{1}{(2\pi)^{3}}
	 \int\!\d\vbr  \!
   \int\!\d\vb{k} \,
	 e^{i\left(\vbk-\vbp\right)\cdot\vb{r}} 
	 R_n(\vbk)
   \langle V(\vbr) \rangle
	 e^{-\frac{i}{2} \int_{t_s}^{t''}\left(\vbk+\vb{A}(\tau)\right)^2\d\tau}
	 e^{-iW_{mn}^C(\vbr,\vbk,t'')},
\label{int-one-definition}
\end{align}
in terms of which the second-order ionization yield from equation \eqref{integral-starting-point-old} can be expressed as
\begin{align}
a^{(2)}_m(\vbp,t_0)
 & =
   -i\sum_n 
	 e^{-iE_mt_0}
	 e^{i I_{p,n} t_s}
	 a_g(t_s) 
\label{integral-starting-point}
\\ & \quad \qquad\times 
  \int \d t''
   b_m(t_0,t'')
   b_n(t'',t_s)
	 e^{i(E_m-E_n)t''}
	 e^{-\frac{i}{2} \int_{t''}^T\left(\vbp+\vb{A}(\tau)\right)^2\d\tau} 
	\Int_1(\vbp,t'')\nonumber
  .
\end{align}

\subsection{Models for the initial orbital and the interaction potential}
In order to evaluate this analytically, however, we need reasonable models for both prefactors. These models should be as easy to handle as possible to ensure that the transverse integrals remain analytically accessible, while at the same time capturing the essence of the probable dependence of both factors and remaining general enough that they are applicable to as wide a range of cases as possible.

For the $R$ factor we have already such a satisfactory model, in the form of the explicit calculation we made in the case of a hydrogenic orbital. Since the $R$ factor depends only on the values of the Dyson orbital's wavefunction at the spherical boundary of radius $a$, far away from the molecule, it can always be expressed as a sum of such hydrogenic states, which have well-defined angular momentum properties. Further, unless the Dyson orbital has a fairly complicated angular dependence, there will typically be rather few such hydrogenic states with significant contributions. However, in what follows we will not perform this expansion and consider instead a single hydrogenic orbital; the general result will then follow automatically by linearity. Our ansatz for the $R$ factor is then given by equation\eqref{Kdefinition},
\begin{equation}
R_n(\vbk)=K(k_\parallel)e^{im\phi_k}  \left(\frac{k_\perp }{\kappa}\right)^{|m|},
\label{reKdefinition}
\end{equation}
for some integer $m$.

For the correlation potential we will adopt a similar strategy. Being after all an electrostatic potential, $\langle V(\vbr)\rangle$ as a function of position can quite meaningfully be expanded in a multipole expansion \cite[pp. 145-150]{Jackson}. We will thus treat a pure multipolar field,
\begin{equation}
\langle V(\vbr)\rangle =\frac{Q_{\l\m}}{r^{\l+1}} Y_{\l\m}(\theta,\phi_r), 
\label{multipole-ansatz}
\end{equation}
where $\l$ and $\m$ are integers with $-\l\leq\m\leq\l$; the general case then also follows by linearity and one generally hopes that only a few multipole terms will have significant contributions.\footnote{It is important to note, however, that the relative importance of the different terms in a multipolar series of the form $\Phi (\vbr)=\sum_{l,m}Q_{lm} r^{-(l+1)}Y_{lm}(\theta,\phi)$ depends on the distance $r$ from the system, so that our assertion that only few terms will contribute will need to be reevaluated once we restore the multipolar series and evaluate the radius $r$ at the classical trajectory's parallel component, to account for how large the radius actually is.} If one already has the functional dependence of the potential $\langle V(\vbr)\rangle$ then the multipoles can be recovered from it by integration; otherwise, the moments are best obtained by applying the expansion in spherical harmonics \cite[p. 111]{Jackson} before taking the matrix elements:
\begin{align*}
\left\langle \sum_i \frac{1}{\left|\vbr-\vbr_i\right|} \right\rangle
& = 
\left\langle
  \sum_i
	\sum_{\l=0}^\infty
	\sum_{\m=-\l}^\l
	  \frac{4\pi}{2\l+1}
		\frac{r_i^\l}{r^{\l+1}}
		Y_{\l\m}(\theta,\phi)
		Y_{\l\m}(\theta_i,\phi_i)^\ast
\right\rangle
\\ & = 
	\sum_{\l=0}^\infty
	\sum_{\m=-\l}^\l
	  \frac{4\pi}{2\l+1}
		\frac{Y_{\l\m}(\theta,\phi)}{r^{\l+1}}
\left\langle
  \sum_i
	  r_i^\l
		Y_{\l\m}(\theta_i,\phi_i)^\ast
\right\rangle
,
\end{align*}
and therefore
\begin{align*}
\langle V(\vbr)\rangle
 & =
\bra{m(t_0,t'')}
\left(
\sum_i \frac{1}{\left|\vbr-\vbr_i\right|}
    -\bra{n(t'')}
		\sum_i \frac{1}{\left|\vbr-\vbr_i\right|}
		\ket{n(t'')}
\right)
\ket{n(t'',t_s)}
%\\ & =
%\bra{m(t_0,t'')}
%\sum_i \frac{1}{\left|\vbr-\vbr_i\right|}
%\ket{n(t'',t_s)}
%\\ & \qquad\qquad-
%\bracket{m(t_0,t'')}{n(t'',t_s)}
  %\bra{n(t'')}
	%\sum_i \frac{1}{\left|\vbr-\vbr_i\right|}
	%\ket{n(t'')}
\\ & =
\sum_{\l=0}^\infty
\sum_{\m=-\l}^\l
  \frac{4\pi}{2\l+1}
	\frac{Y_{\l\m}(\theta,\phi)}{r^{\l+1}}
\left[
  \bra{m(t_0,t'')}
  \sum_i
	  r_i^\l
		Y_{\l\m}(\theta_i,\phi_i)^\ast
  \ket{n(t'',t_s)}
\right.
\\ & \qquad\qquad\qquad\qquad\qquad\quad -
%\\ & \hfill-
\left.
\bracket{m(t_0,t'')}{n(t'',t_s)}
  \bra{n(t'')}
  \sum_i
	  r_i^\l
		Y_{\l\m}(\theta_i,\phi_i)^\ast
	\ket{n(t'')}
\right].
\end{align*}
The multipole moments $Q_{\l\m}$ can thus be read off as the factors in square brackets, and they constitute a very specific question to ask of the numerical effort to calculate the quasi-static eigenstates and their sub-cycle propagation, particularly since they can be reduced to linear combinations of the multipole moments of the basis functions of the numerical method.

Before we dive into the calculation, a word about the physical content of the models we have postulated, \eqref{reKdefinition} and \eqref{multipole-ansatz}, is in order. In particular, we have chosen both an initial state and a transition with well-defined angular momentum properties, which shifts the paradigm from the (linear) momentum transfers mentioned at the end of section \ref{saddlepointargument} to the transfer of angular momentum about the axis of laser polarization. Both these views are of course equivalent - they simply reflect different bases to use for the transverse coordinates - but we will find the phase-matching associated with the transfers far more accessible and physically clear. Finally, we note that by taking an initial state with angular momentum $m$ and a transition of angular momentum $\mu$ out of the ionic state, we impose an angular momentum of $m+\m$ on the outgoing wavefunction, which has a definite and visible effect (see for instance equation \eqref{crude-approximation-to-outgoing-wavefunction}). 
%%%%
%\pdfmargincomment{This could possibly be clearer.}
%%%%

%% file: c3.s3-re-representations
\subsection{Coordinate representation for the integral}
Now that we have both our Ans\"atze, then, we can formulate a concrete goal for our calculation: using cylindrical coordinates, $(\rho,\phi_r,z)$ and $(k_\perp,\phi_k,k_\parallel)$, for both integrals (though borrowing the spherical coordinates $\theta$ and $r$) with the laser polarization towards positive $z$ we have
\begin{align}
\Int_1(\vbp,t'')  &  =
  \frac{1}{(2\pi)^{3}}
	%\int\!\d\vbr  \!
	\! \int_{-\infty}^\infty \!\!\! \d z
	\! \int_0^\infty \!\!\! \rho\,\d \rho
	\! \int_0^{2\pi} \!\!\! \d\phi_r
  %\int\!\d\vb{k} \,
	\! \int_{-\infty}^\infty \!\!\! \d k_\parallel
	\! \int_0^\infty \!\!\! k_\perp\d k_\perp
	\! \int_0^{2\pi} \!\!\! \d\phi_k
	e^{i\left(\vbk-\vbp\right)\cdot\vb{r}} 
	%R_n(\vbk)
\\ & \qquad  \times
	K(k_\parallel)e^{im\phi_k}  \left(\frac{k_\perp }{\kappa}\right)^{|m|}
  %\langle V(\vbr) \rangle
	\frac{Q_{\l\m}}{r^{\l+1}} Y_{\l\m}(\theta,\phi_r)
	e^{-\frac{i}{2} \int_{t_s}^{t''}\left(\vbk+\vb{A}(\tau)\right)^2\d\tau}
	e^{-iW_{mn}^C(\vbr,\vbk,t'')}
.\nonumber 
\end{align}
We will now drop the Coulomb correction $e^{-iW_{mn}^C(\vbr,\vbk,t'')}$, since it depends nonlinearly on the ionic potential and thus completely impedes further progress. In a variation of the saddle-point method, though, we will analytically evaluate the remaining integrals and after an analysis of where the main contributions to the transverse integrals come from will form suitable saddle-point approximations for this slow factor. However, such an analysis will be left out of this work for reasons of time.
%%%
%\pdfmargincomment{Is this actually in the thesis?}
%%%

Expanding now all the factors, including the identity $Y_{\l\m}(\theta,\phi_r)=\frac{N_{\l\m}}{\sqrt{2\pi}}P_\l^\m(\cos(\theta))e^{i\m \phi_r}$ for the spherical harmonic, we find that
\begin{align}
\Int_1(\vbp,t'')  &  =
  \frac{1}{(2\pi)^{3}} 
	\frac{1}{\sqrt{2\pi}}
	\frac{Q_{\l\m}}{\kappa^{|m|}} 
	\! \int_{-\infty}^\infty \!\!\! \d z
	\! \int_{-\infty}^\infty \!\!\! \d k_\parallel
	K(k_\parallel)
	e^{i\left(k_\parallel-p_\parallel\right)z}
	e^{-\frac{i}{2} \int_{t_s}^{t''}\left(k_\parallel+A_\parallel(\tau)\right)^2\d\tau}
\nonumber \\ & \qquad  \times
	\! \int_0^\infty \!\!\! \,\d \rho
	\frac{\rho}{r^{\l+1}}
	N_{\l\m}P_\l^\m(\cos(\theta))
	\! \int_0^\infty \!\!\! \d k_\perp^{\vphantom{|m|}}
	k_\perp^{{|m|}+1}
	e^{-\frac{i}{2}(t''-t_s)k_\perp^2}
\nonumber \\ & \qquad  \times
	\! \int_0^{2\pi} \!\!\! 
	\d\phi_r
	  e^{-ip_\perp \rho\cos(\phi_r-\phi_p)}
	  e^{i\m \phi_r}
	\! \int_0^{2\pi} \!\!\! 
	\d\phi_k
	  e^{ik_\perp \rho\cos(\phi_r-\phi_k)}
	  e^{im\phi_k}
.
\label{int1-with-phi-integrals}
\end{align}
We face here four hurdles in the form of the four transverse integrals, in $\phi_k$, $\phi_r$, $k_\perp$ and $\rho$, of which the first three are relatively simple whilst the fourth will be more involved and will require further approximations and decompositions. 

The first two, however, are both versions of Bessel's first integral \cite[p. 140]{Jackson}, which serves as a definition for the Bessel functions of integer order as good as any other:
\begin{equation}
J_\nu(z)=\frac{1}{2\pi i^\nu}\int_0^{2\pi} e^{iz\cos(\phi)}e^{i\nu\phi}\d\phi.
\label{Bessels-first-integral}
\end{equation}
In order to get a uniform representation with respect to reflections in the $x$-$z$ plane (i.e. change of sign in $m$ and $\m$), we will use this in the form $\frac{1}{2\pi }\int_0^{2\pi} e^{iz\cos(\phi)}e^{i\nu\phi}\d\phi=i^{|\nu|}J_{|\nu|}(z)$. The first integral must be offset by $\phi_r$, which brings an additional phase into the $\phi_r$ integral, after which the last line of equation \eqref{int1-with-phi-integrals} reads
\begin{align*}
	2\pi i^{|m|}
	&
	J_{|m|}(k_\perp \rho)
	e^{+i(m+\m) (\phi_p+\pi)}
	\! \int_0^{2\pi} \!\!\! 
	  e^{ip_\perp \rho\cos(\phi_r-\phi_p-\pi)}
	  e^{i(m+\m) (\phi_r-\phi_p-\pi)}
	\d\phi_r
\\	& =
	(2\pi)^2 i^{|m|+|m+\mu|}
	e^{i(m+\m) (\phi_p+\pi)}
	J_{|m|}(k_\perp \rho)
	J_{|m+\mu|}(p_\perp \rho).
\end{align*}

We will now tackle the integral over $k_\perp$. To set this up, we first simplify our current results using the identity $(-1)^{m+\m}i^{|m+\m|}=(-i)^{|m+\m|}$, and we introduce the shorthand 
\begin{equation}
\boxed{\xi=i(t''-t_s),}
\label{xi-definition}
\end{equation}
which as we saw in the last section will be a positive real number in the integration path.\footnote{It is probably advisable for the reader to keep a note of this shorthand on a readily accessible place. We note also that this differs from the $xi$ used in Ref.~\cite{mepaper}, which is equal to $\tauT-\xi$ in this setting.} With this, then, we have
\begin{align}
\Int_1(\vbp,t'')  &  =
  \frac{i^{|m|}(-i)^{|m+\mu|}}{(2\pi)^{3/2}} 
	\frac{Q_{\l\m}}{\kappa^{|m|}} 
	\! \int_{-\infty}^\infty \!\!\! \d z
	\! \int_{-\infty}^\infty \!\!\! \d k_\parallel
	K(k_\parallel)
	e^{i\left(k_\parallel-p_\parallel\right)z}
	e^{-\frac{i}{2} \int_{t_s}^{t''}\left(k_\parallel+A_\parallel(\tau)\right)^2\d\tau}
\nonumber \\ & \qquad  \times
	e^{i(m+\m) \phi_p}
	\! \int_0^\infty \!\!\! \,\d \rho
	\frac{\rho}{r^{\l+1}}
	N_{\l\m}P_\l^\m(\cos(\theta))
	J_{|m+\mu|}(p_\perp \rho)
\nonumber \\ & \qquad  \times
	\! \int_0^\infty \!\!\! 
	  k_\perp^{{|m|}+1}
	  e^{-\frac{1}{2}\xi k_\perp^2}
	  J_{|m|}(k_\perp \rho)
	\d k_\perp%^{\vphantom{|m|}}
.
\label{int1-after-phi-integrals}
\end{align}
Although this integral looks complicated, it can be done analytically and can be found, e.g. in Gradshteyn and Ryzhik \cite[p. 706, eq. 6.631.4]{gradshteyn}:
\begin{equation}
	\! \int_0^\infty \!\!\! 
	  x^{{\nu}+1}
	  e^{-\alpha x^2}
	  J_{\nu}(\beta x)
	\d x
=
  \frac{\beta^\nu}{(2\alpha)^{\nu+1}}e^{-\frac{\beta^2}{4\alpha}}.
\label{GandRuno}
\end{equation}
We will accept this as true for now, though later on we will prove a stronger version of this result.
%%%
%\pdfmargincomment{Make sure you do. Or change this.}
%%%
The result of the $k_\perp$ integral is then $\frac{\rho^{|m|}}{\xi^{|m|+1}}e^{-\frac{1}{2\xi}\rho^2}$, which expresses the general principle that Fourier transforms of gaussian functions are again gaussian, with reciprocal widths. (On the other hand, equation~\eqref{GandRuno} can be seen as simply a Laplace transform of a Bessel function.) Substituting this result in, we are now left with the expression

\begin{align}
\Int_1(\vbp,t'')  &  =
  \frac{i^{|m|}(-i)^{|m+\mu|}}{(2\pi)^{3/2}} 
	\frac{Q_{\l\m}}{\kappa^{|m|}} 
	\! \int_{-\infty}^\infty \!\!\! \d z
	\! \int_{-\infty}^\infty \!\!\! \d k_\parallel
	K(k_\parallel)
	e^{i\left(k_\parallel-p_\parallel\right)z}
	e^{-\frac{i}{2} \int_{t_s}^{t''}\left(k_\parallel+A_\parallel(\tau)\right)^2\d\tau}
\times\displaybreak[1]\nonumber \\ & \qquad  \times
	e^{i(m+\m) \phi_p}
	\int_0^\infty 
	  \frac{N_{\l\m}P_\l^\m(\cos(\theta))}{r^{\l+1}}
	  \frac{\rho^{|m|+1}}{\xi^{|m|+1}}
	  e^{-\frac{1}{2\xi}\rho^2}
	  J_{|m+\mu|}(p_\perp \rho)
	\d \rho
%\nonumber \\ & \qquad  \times
.
\label{int1-after-k-integral}
\end{align}

\subsection{Series representation for the interaction potential}

Unfortunately, as it stands the integral over $\rho$ is not analytically tractable. It is structurally quite similar to the $k_\perp$ integration except for the rather more complicated prefactor of 
{\setlength{\abovedisplayskip}{1.5mm}\setlength{\belowdisplayskip}{1.5mm}
$${N_{\l\m}P_\l^\m(\cos(\theta))}\big/{r^{\l+1}},$$
}
which makes it too complicated to handle. To deal with this, we will use the approximation of small tunnelling angles to reduce the integral to something more manageable.

The key insight of this approximation is that near the negative-$z$ pole of the unit sphere, the associated Legendre function, originally brought in by the electrostatic interaction, behaves like
$$
P_\l^\m(\cos(\theta))\propto \sin^{|\m|}(\theta),
$$
and multiplying both numerator and denominator by $r^{|\m|}$ then changes the Legendre function to a factor of $\rho^{|\m|}$, with which integration can proceed.

To bring this on a more formal footing, we note that the associated Legendre function, given for positive $\m$ by \cite[eq. \href{http://dlmf.nist.gov/14.7.E8}{14.7.8}]{NIST-handbook}
$$
P_\l^\m(\cos(\theta))\colonequals\frac{(-1)^\m}{2^\l\l!} \sin^{\m}(\theta) \frac{\d^{\l+\m}}{\d (\cos(\theta))^{\l+\m}}\left(\cos^2(\theta)-1\right)^{\l}
$$
%%%
%\pdfmargincomment{Done!}
%%%
is given by the power of $\sin(\theta)$ mentioned above multiplied by a polynomial in $\cos(\theta)$ which is nonzero for $\theta=\pi$. Explicit representations for this polynomial are relatively hard to come by, as are series expansions around the poles of the unit sphere. Nevertheless, a useful expansion can be found in refs. \cite[p. 54]{MagnusAndOberhettinger} and \cite[eq.~\href{http://dlmf.nist.gov/14.3.E4}{14.3.4}]{NIST-handbook}, given by 
\begin{equation}
P_\l^\m(x)=(-1)^\m \left(1-x^2\right)^{\m/2}
\frac{(\l+\m)!}{2^\m \m! (\l-\m)!} 
\FF\left(\m-\l,\m+\l+1;\m+1;\frac{1-x}{2}\right),
\label{Legendre-as-hypergeometric}
\end{equation}
where $\m\geq0$ and $\FF$ is Gauss's hypergeometric function.

%%%%%%%%%%%%%%%%%%%%%%%%%%%%%%%%%%%%%%%%%%%%%%%%%%%%
%%%%%%%%%%%%%%%%%%%%%%%%%%%%%%%%%%%%%%%%%%%%%%%%%%%%
\begin{parmat}{The (Gauss) hypergeometric function}
\label{aside-HGfunction}
\noindent
In essence, hypergeometric functions are those power series
$$f(z)=\sum_{n=0}^\infty c_n z^n$$
for which the ratio of successive coefficients is a rational function of $n$: that is, power series that satisfy a relation of the form
$$\frac{c_{n+1}}{c_n}=\frac{A(n)}{B(n)}$$
for polynomials $A$ and $B$. Since this includes many common coefficients such as factorials, powers and rational functions, hypergeometric functions include most of the special functions of mathematical physics.

This property is in fact a strong restriction on $f$; we will now give a short account of how this comes about as well as some basic properties, to help readers not familiar with these functions get a working grasp on them. (For more details we refer the reader to the standard texts by Askey \cite{Askey} and Erdelyi \cite{Erdelyi-I}, or for a summary of results to the Digital Library of Mathematical Functions \cite[chapter~\href{http://dlmf.nist.gov/15}{15}]{NIST-handbook}.)

After appropriate rescalings, one can suppose that both polynomials $A$ and $B$ have unit leading coefficients and that the initial term in the series is also 1. Putting in the factors of the polynomials this means a relation of the form
\begin{equation}
	c_{n+1}=\frac{(a_1+n)\cdots(a_p+n)}{(b_1+n)\cdots(b_q+n)}\frac{c_n}{n+1}
	\label{HG-coefficients-recurrence}
\end{equation}
holds. (It is standard practice to include $n+1$ as a factor of $B$; this can be undone by including it as a factor of $A$.) This is enough to completely specify the series: we know the first term and can get each coefficient from the last. Thus,
$$
f(z)=1+\frac{a_1\cdots a_p}{b_1\cdots b_q}\frac{z}{1!}+\frac{a_1(a_1+1)\cdots a_p(a_p+1)}{b_1(b_1+1)\cdots b_q(b_q+1)}\frac{z^2}{2!}+\cdots.
$$

We see then that $f$ is essentially determined by the $p+q$ complex numbers $a_1,\ldots,a_p$ and $b_1,\ldots,b_q$, which are the zeros of $A$ and $B$ up to a sign. The series above is called the generalized hypergeometric function and is usually denoted ${}_p F_q(a_1,\ldots,a_p;b_1,\ldots,b_q;z)$; note the semicolons separating arguments of different types. This can be brought into more convenient terminology by introducing the Pochhammer symbol
\begin{equation}
(a)_n=\frac{\Gamma(a+n)}{\Gamma(a)}=a(a+1)\cdots(a+n-1)\textrm{ with }(a)_0=1,
\label{Pochhammer-symbol-def}
\end{equation}
in terms of which one can write
\begin{equation}
{}_p F_q(a_1,\ldots,a_p;b_1,\ldots,b_q;z)
=
\sum_{n=0}^\infty \frac{(a_1)_n\cdots(a_p)_n}{(b_1)_n\cdots(b_q)_n}\frac{z^n}{n!}.
\label{generalized-hypergeometric-def}
\end{equation}
This series converges everywhere when $p\leq q$ and nowhere when $p>q+1$. When $p=q+1$ it converges in general only for $|z|<1$ and it can be analytically continued to the whole complex plane; in that case it has a branch point at $z=1$ whose branch cut is usually taken on the interval $[1,\infty)$.

One must also note that if $b$ is a negative integer then $(b)_n$ is zero for all sufficiently large $n$ and therefore ${}_p F_q$ is not defined if any of the $b_i$ are zero or a negative integer. Similarly, if any of the $a_i$ is a negative integer then the series terminates and the hypergeometric function is a polynomial, one instance of which is our particular example, eq. \eqref{Legendre-as-hypergeometric}. This is also the reason for the restriction to $\m\geq 0$ in our example.

The hypergeometric function most commonly encountered in mathematical physics is the gaussian hypergeometric function $\FF(a,b;c;z)$, so named since Gauss pioneered the study of the corresponding series. This has a profound reason: using the identity $z\frac{\d}{\d z}z^n=nz^n$, the recurrence relation \eqref{HG-coefficients-recurrence} can be rephrased as
$$
\left(c-1+z\frac{\d}{\d z}\right)z\frac{\d}{\d z}
c_{n+1} z^{n+1}
%=
%(c+n)(1+n)
%c_{n+1} z^{n+1}
%=
%(a+n)(b+n)
%c_nz^n
=
\left(a+z\frac{\d}{\d z}\right)\left(b+z\frac{\d}{\d z}\right)
c_nz^n,
$$
which when summed over $n$ means that $w=\FF$ solves the differential equation
$$
\left(c-1+z\frac{\d}{\d z}\right)z\frac{\d}{\d z}
(w-1)
=
\left(a+z\frac{\d}{\d z}\right)\left(b+z\frac{\d}{\d z}\right)
w,
$$
or equivalently
\begin{equation}
z(1-z)\frac{\d^2w}{\d z^2}+[c-(a+b+1)z]\frac{\d w}{\d z}-ab\, w=0.
\label{hypergeometric-differential-equation}
\end{equation}

This differential equation  (called, not surprisingly, the hypergeometric equation) is important for mathematical physics in its generality: it is in essence the only second-order differential equation with up to three regular singular points, a characteristic which describes most equations described in practice. Thus most special functions of physics are particular cases of this hypergeometric function. This broad generality, on the other hand, also precludes the possibility of simple insights into its properties, since they cover much of the range of what is physically possible.

\end{parmat}

%%%%%%%%%%%%%%%%%%%%%%%%%%%%%%%%%%%%%%%%%%%%%%%%%%%%
%%%%%%%%%%%%%%%%%%%%%%%%%%%%%%%%%%%%%%%%%%%%%%%%%%%%

Returning now to our problem, the expansion \eqref{Legendre-as-hypergeometric} now allows us to expand the Legendre function in the potential term ${N_{\l\m}P_\l^\m(\cos(\theta))}/{r^{\l+1}}$ as a series of terms of which only the first few will be important in the approximation of small tunnelling angles. 

Here we note briefly that first a small change is needed, since the expansion above describes small deviations from the \textit{north} pole of the sphere, at which $\cos(\theta)=1$. We use, therefore, the reflection formula \cite[eq. \href{http://dlmf.nist.gov/14.7.E17}{14.7.17}]{NIST-handbook}
$$
P_\l^\m(x)=(-1)^{\l-\m} P_\l^\m(-x).
$$
We also use the fact that in the branch with $z<0$ we can write $\cos(\theta)=z/r=z/\sqrt{z^2+\rho^2}=-1/\sqrt{1+\rho^2/z^2}$.

With both these ingredients, then, we can reformulate the potential term as
\begin{align*}
	\frac{N_{\l\m}P_\l^\m(\cos(\theta))}{r^{\l+1}}
 &	=	
	\frac{\rho^\m}{r^{\l+\m+1}}
  \frac{(-1)^{\l}}{2^\m \m!} 
	\sqrt{\frac{(2\l+1)(\l+\m)!}{2(\l-\m)!}}
\\ & \qquad \times
  \FF\left(\m-\l,\m+\l+1;\m+1;\frac{1}{2}-\frac{1}{2}\frac{1}{\sqrt{1+\rho^2/z^2}}\right)
\end{align*}
for $\m\geq0$. This now needs only two adjustments, the first of which is the lifting of this last restriction by applying the reflection formula
$$
P_\l^{-\m}(x)
=
(-1)^\m
\frac{(\l-\m)!}{(\l+\m)!}
P_\l^\m(x)
$$
to get for negative $\m$
$$
i^\m N_{\l\m}P_\l^\m(\cos(\theta))
=
%N=\sqrt{\frac{(2\l+1)(\l-\m)!}{2(\l+\m)!}}
(-i)^\m
\sqrt{\frac{(2\l+1)(\l+\m)!}{2(\l-\m)!}}
P_\l^{-\m}(\cos(\theta))
=i^{|\m|} N_{\l|\m|}P_\l^{|\m|}(\cos(\theta)),
$$
and thus a direct way to apply our previous results, which now read
\begin{align}
	i^{\m}\frac{N_{\l\m}P_\l^\m(\cos(\theta))}{r^{\l+1}}
 &	=	
	\frac{\rho^{|\m|}}{r^{\l+|\m|+1}}
  \frac{(-1)^{\l}i^{|\m|}}{2^{|\m|} |\m|!} 
	\sqrt{\frac{(2\l+1)(\l+|\m|)!}{2(\l-|\m|)!}}
\nonumber \\ & \qquad \times
  \FF\left(|\m|-\l,|\m|+\l+1;|\m|+1;\frac{1}{2}-\frac{1}{2}\frac{1}{\sqrt{1+\rho^2/z^2}}\right).
\end{align}
for $-\l\leq\m\leq\l$.

The second adjustment is somewhat more serious, and it reflects the fact that since we are applying the small-angles approximation in cylindrical coordinates we would ideally like to get a series in $\rho/z$ and not in $1+\cos(\theta)$. Making things worse, the relatively complicated dependence of the argument of the hypergeometric function is compounded with the presence of the term $r^{-(\l+|\m|+1)}=\left(z^2+\rho^2\right)^{-\frac{\l+|\m|+1}{2}}$. Both these shortcomings we will solve in an obvious fashion: after factoring out the main dependence, as $(-z)^{-(\l+|\m|+1)}$, of the radial term (where the minus sign comes from the fact that for an electric field along the $+z$ axis, the electron tunnels towards \textit{negative} $z$), we will ask for a series representation of the combined factor:
\begin{align}
\frac{\FF\left(|\m|-\l,|\m|+\l+1;|\m|+1;\frac{1}{2}-\frac{1}{2}\frac{1}{\sqrt{1+\rho^2/z^2}}\right)}{\left(1+\frac{\rho^2}{z^2}\right)^{\l+|\m|+1}}
=
\sum_{j=0}^\infty
\coef(\l,|\m|;j)
\frac{\rho^{2j}}{z^{2j}}
,
\label{coef-def}
\end{align}
where in the small-tunnelling-angles approximation we expect, in fact, only the first term to contribute appreciably. The subleading term can be used to correct the results if needed and to corroborate the approximation is correct.

Although this implicit definition of the series coefficients $\coef(\l,|\m|;j)$ looks quite convoluted, they can be easily implemented using Wolfram Mathematica's \footnote{All the results in this work were computed with Wolfram Mathematica 8.0.4.0 running on Microsoft Windows (64-bit).} \texttt{Series} function to calculate a list of the coefficients in the series up to some specified cutoff:
\vspace{-3.5mm}
\begin{shaded}
\includegraphics[scale=0.9]{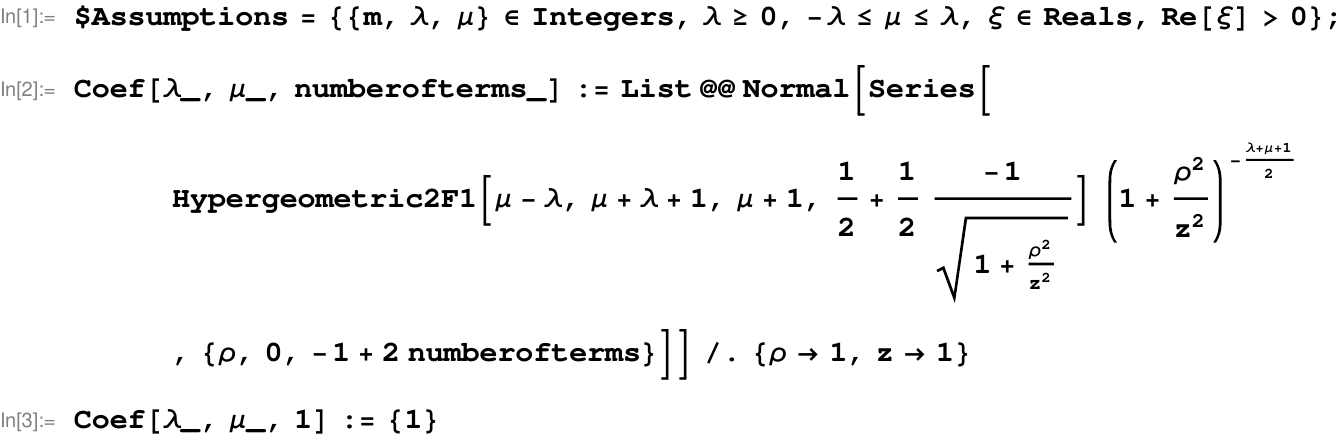}
\end{shaded}
\vspace{-3.5mm}
\noindent the first few outputs of which are
\vspace{-3.5mm}
\begin{shaded}
\includegraphics[scale=0.9]{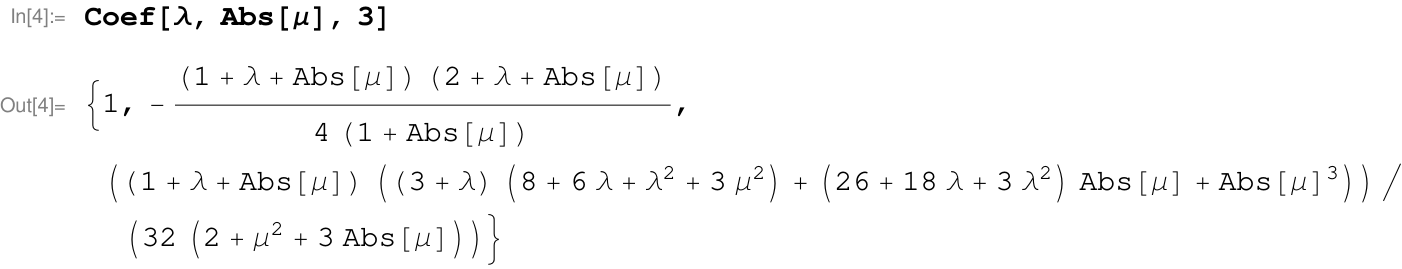}
\end{shaded}
\vspace{-3.5mm}
\noindent and can be checked by hand to be the correct coefficients.

%% file: power-and-degree-table2.tex
\newlength{\celldim}
\setlength{\celldim}{1.3em}
\newlength{\fontheight}
\settoheight{\fontheight}{A}
\newlength{\extraheight}
\setlength{\extraheight}{\celldim - \fontheight}
\newcolumntype{S}{
  @{}
  >{\centering\arraybackslash\columncolor{white}[0pt][0pt]}
  p{\celldim}
  %<{}
  <{\rule[-0.5\extraheight]{0pt}%
  {\fontheight + \extraheight-30pt}}
  @{} 
}
\setlength{\arrayrulewidth}{1pt}

\definecolor{color0}{rgb}{	0.83	,	0.83	,	0.91	}
\definecolor{color1}{rgb}{	0.66	,	0.66	,	0.83	}
\definecolor{color2}{rgb}{	0.50	,	0.50	,	0.75	}
\definecolor{color3}{rgb}{	0.33	,	0.33	,	0.66	}
\definecolor{color4}{rgb}{	0.17	,	0.17	,	0.58	}
\definecolor{color5}{rgb}{	0.00	,	0.00	,	0.50	}
\definecolor{color6}{rgb}{	0.00	,	0.00	,	0.50	}
\definecolor{color7}{rgb}{	0.00	,	0.00	,	0.50	}
\definecolor{color8}{rgb}{	0.00	,	0.00	,	0.50	}
\definecolor{color9}{rgb}{	0.00	,	0.00	,	0.50	}
\definecolor{color10}{rgb}{	0.00	,	0.00	,	0.50	}
%%%%%%%%%%%%%%%%%%%%%%%%%%%%%%%%%%%%%%%%%%%%%%%%%
\definecolor{text0}{rgb}{	0.00	,	0.00	,	0.00	}
\definecolor{text1}{rgb}{	0.00	,	0.00	,	0.00	}
\definecolor{text2}{rgb}{	0.00	,	0.00	,	0.00	}
\definecolor{text3}{rgb}{	0.70	,	0.70	,	0.70	}
\definecolor{text4}{rgb}{	0.70	,	0.70	,	0.70	}
\definecolor{text5}{rgb}{	0.70	,	0.70	,	0.70	}
\definecolor{text6}{rgb}{	0.70	,	0.70	,	0.70	}
\definecolor{text7}{rgb}{	0.70	,	0.70	,	0.70	}
\definecolor{text8}{rgb}{	0.70	,	0.70	,	0.70	}
\definecolor{text9}{rgb}{	0.70	,	0.70	,	0.70	}
\definecolor{text10}{rgb}{	0.70	,	0.70	,	0.70	}
%%%%%%%%%%%%%%%%%%%%%%%%%%%%%%%%%%%%%%%%%%%%%%%%%
\definecolor{othercolor0}{rgb}{	0.83	,	0.83	,	0.91	}
\definecolor{othercolor1}{rgb}{	0.75	,	0.75	,	0.87	}
\definecolor{othercolor2}{rgb}{	0.66	,	0.66	,	0.83	}
\definecolor{othercolor3}{rgb}{	0.58	,	0.58	,	0.79	}
\definecolor{othercolor4}{rgb}{	0.50	,	0.50	,	0.75	}
\definecolor{othercolor5}{rgb}{	0.42		0.42		0.71	}
\definecolor{othercolor6}{rgb}{	0.33		0.33		0.66	}
\definecolor{othercolor7}{rgb}{	0.25		0.25		0.62	}
\definecolor{othercolor8}{rgb}{	0.17		0.17		0.58	}
\definecolor{othercolor9}{rgb}{	0.08		0.08		0.54	}
\definecolor{othercolor10}{rgb}{	0.00	,	0.00	,	0.50	}
%%%%%%%%%%%%%%%%%%%%%%%%%%%%%%%%%%%%%%%%%%%%%%%%%
\definecolor{othertext0}{rgb}{			0.00	,	0.00	,	0.00	}
\definecolor{othertext1}{rgb}{			0.00	,	0.00	,	0.00	}
\definecolor{othertext2}{rgb}{			0.00	,	0.00	,	0.00	}
\definecolor{othertext3}{rgb}{			0.00	,	0.00	,	0.00	}
\definecolor{othertext4}{rgb}{			0.00	,	0.00	,	0.00	}
\definecolor{othertext5}{rgb}{			0.70	,	0.70	,	0.70	}
\definecolor{othertext6}{rgb}{			0.70	,	0.70	,	0.70	}
\definecolor{othertext7}{rgb}{			0.70	,	0.70	,	0.70	}
\definecolor{othertext8}{rgb}{			0.70	,	0.70	,	0.70	}
\definecolor{othertext9}{rgb}{			0.70	,	0.70	,	0.70	}
\definecolor{othertext10}{rgb}{			0.70	,	0.70	,	0.70	}

\begin{table}[ht]
	\centering

%{\vrule width 1pt}

\begin{tabular}{c|S*{11}{S} p{3mm} *{11}{S}} %\cline{2-13} \cline{15-25}
\multicolumn{1}{c}{} & \multicolumn{1}{c}{} &\multicolumn{11}{c}{$\m$} & &\multicolumn{11}{c}{$\m$} \\ \hhline{~~*{11}{-}~*{11}{-}}
\multicolumn{1}{c}{} & \multicolumn{1}{c}{} & -5	&	-4	&	-3	&	-2	&	-1	&	0	&	1	&	2	&	3	&	4	&	5	& &	-5	&	-4	&	-3	&	-2	&	-1	&	0	&	1	&	2	&	3	&	4	&	5	\\ 
%\hhline{~~*{11}{-}~*{11}{-}}
\multirow{11}{*}{$m$}
&	-5		& \cellcolor{color0} \textcolor{text0}{	0	} & \cellcolor{color0} \textcolor{text0}{	0	} & \cellcolor{color0} \textcolor{text0}{	0	} & \cellcolor{color0} \textcolor{text0}{	0	} & \cellcolor{color0} \textcolor{text0}{	0	} & \cellcolor{color0} \textcolor{text0}{	0	} & \cellcolor{color1} \textcolor{text1}{	1	} & \cellcolor{color2} \textcolor{text2}{	2	} & \cellcolor{color3} \textcolor{text3}{	3	} & \cellcolor{color4} \textcolor{text4}{	4	} & \cellcolor{color5} \textcolor{text5}{	5	} & & \cellcolor{othercolor10} \textcolor{othertext10}{	10	} & \cellcolor{othercolor9} \textcolor{othertext9}{	9	} & \cellcolor{othercolor8} \textcolor{othertext8}{	8	} & \cellcolor{othercolor7} \textcolor{othertext7}{	7	} & \cellcolor{othercolor6} \textcolor{othertext6}{	6	} & \cellcolor{othercolor5} \textcolor{othertext5}{	5	} & \cellcolor{othercolor4} \textcolor{othertext4}{	4	} & \cellcolor{othercolor3} \textcolor{othertext3}{	3	} & \cellcolor{othercolor2} \textcolor{othertext2}{	2	} & \cellcolor{othercolor1} \textcolor{othertext1}{	1	} & \cellcolor{othercolor0} \textcolor{othertext0}{	0	} \\
&	-4		& \cellcolor{color0} \textcolor{text0}{	0	} & \cellcolor{color0} \textcolor{text0}{	0	} & \cellcolor{color0} \textcolor{text0}{	0	} & \cellcolor{color0} \textcolor{text0}{	0	} & \cellcolor{color0} \textcolor{text0}{	0	} & \cellcolor{color0} \textcolor{text0}{	0	} & \cellcolor{color1} \textcolor{text1}{	1	} & \cellcolor{color2} \textcolor{text2}{	2	} & \cellcolor{color3} \textcolor{text3}{	3	} & \cellcolor{color4} \textcolor{text4}{	4	} & \cellcolor{color4} \textcolor{text4}{	4	} & & \cellcolor{othercolor9} \textcolor{othertext9}{	9	} & \cellcolor{othercolor8} \textcolor{othertext8}{	8	} & \cellcolor{othercolor7} \textcolor{othertext7}{	7	} & \cellcolor{othercolor6} \textcolor{othertext6}{	6	} & \cellcolor{othercolor5} \textcolor{othertext5}{	5	} & \cellcolor{othercolor4} \textcolor{othertext4}{	4	} & \cellcolor{othercolor3} \textcolor{othertext3}{	3	} & \cellcolor{othercolor2} \textcolor{othertext2}{	2	} & \cellcolor{othercolor1} \textcolor{othertext1}{	1	} & \cellcolor{othercolor0} \textcolor{othertext0}{	0	} & \cellcolor{othercolor1} \textcolor{othertext1}{	1	} \\
&	-3		& \cellcolor{color0} \textcolor{text0}{	0	} & \cellcolor{color0} \textcolor{text0}{	0	} & \cellcolor{color0} \textcolor{text0}{	0	} & \cellcolor{color0} \textcolor{text0}{	0	} & \cellcolor{color0} \textcolor{text0}{	0	} & \cellcolor{color0} \textcolor{text0}{	0	} & \cellcolor{color1} \textcolor{text1}{	1	} & \cellcolor{color2} \textcolor{text2}{	2	} & \cellcolor{color3} \textcolor{text3}{	3	} & \cellcolor{color3} \textcolor{text3}{	3	} & \cellcolor{color3} \textcolor{text3}{	3	} & & \cellcolor{othercolor8} \textcolor{othertext8}{	8	} & \cellcolor{othercolor7} \textcolor{othertext7}{	7	} & \cellcolor{othercolor6} \textcolor{othertext6}{	6	} & \cellcolor{othercolor5} \textcolor{othertext5}{	5	} & \cellcolor{othercolor4} \textcolor{othertext4}{	4	} & \cellcolor{othercolor3} \textcolor{othertext3}{	3	} & \cellcolor{othercolor2} \textcolor{othertext2}{	2	} & \cellcolor{othercolor1} \textcolor{othertext1}{	1	} & \cellcolor{othercolor0} \textcolor{othertext0}{	0	} & \cellcolor{othercolor1} \textcolor{othertext1}{	1	} & \cellcolor{othercolor2} \textcolor{othertext2}{	2	} \\
&	-2		& \cellcolor{color0} \textcolor{text0}{	0	} & \cellcolor{color0} \textcolor{text0}{	0	} & \cellcolor{color0} \textcolor{text0}{	0	} & \cellcolor{color0} \textcolor{text0}{	0	} & \cellcolor{color0} \textcolor{text0}{	0	} & \cellcolor{color0} \textcolor{text0}{	0	} & \cellcolor{color1} \textcolor{text1}{	1	} & \cellcolor{color2} \textcolor{text2}{	2	} & \cellcolor{color2} \textcolor{text2}{	2	} & \cellcolor{color2} \textcolor{text2}{	2	} & \cellcolor{color2} \textcolor{text2}{	2	} & & \cellcolor{othercolor7} \textcolor{othertext7}{	7	} & \cellcolor{othercolor6} \textcolor{othertext6}{	6	} & \cellcolor{othercolor5} \textcolor{othertext5}{	5	} & \cellcolor{othercolor4} \textcolor{othertext4}{	4	} & \cellcolor{othercolor3} \textcolor{othertext3}{	3	} & \cellcolor{othercolor2} \textcolor{othertext2}{	2	} & \cellcolor{othercolor1} \textcolor{othertext1}{	1	} & \cellcolor{othercolor0} \textcolor{othertext0}{	0	} & \cellcolor{othercolor1} \textcolor{othertext1}{	1	} & \cellcolor{othercolor2} \textcolor{othertext2}{	2	} & \cellcolor{othercolor3} \textcolor{othertext3}{	3	} \\
&	-1		& \cellcolor{color0} \textcolor{text0}{	0	} & \cellcolor{color0} \textcolor{text0}{	0	} & \cellcolor{color0} \textcolor{text0}{	0	} & \cellcolor{color0} \textcolor{text0}{	0	} & \cellcolor{color0} \textcolor{text0}{	0	} & \cellcolor{color0} \textcolor{text0}{	0	} & \cellcolor{color1} \textcolor{text1}{	1	} & \cellcolor{color1} \textcolor{text1}{	1	} & \cellcolor{color1} \textcolor{text1}{	1	} & \cellcolor{color1} \textcolor{text1}{	1	} & \cellcolor{color1} \textcolor{text1}{	1	} & & \cellcolor{othercolor6} \textcolor{othertext6}{	6	} & \cellcolor{othercolor5} \textcolor{othertext5}{	5	} & \cellcolor{othercolor4} \textcolor{othertext4}{	4	} & \cellcolor{othercolor3} \textcolor{othertext3}{	3	} & \cellcolor{othercolor2} \textcolor{othertext2}{	2	} & \cellcolor{othercolor1} \textcolor{othertext1}{	1	} & \cellcolor{othercolor0} \textcolor{othertext0}{	0	} & \cellcolor{othercolor1} \textcolor{othertext1}{	1	} & \cellcolor{othercolor2} \textcolor{othertext2}{	2	} & \cellcolor{othercolor3} \textcolor{othertext3}{	3	} & \cellcolor{othercolor4} \textcolor{othertext4}{	4	} \\
&	0		& \cellcolor{color0} \textcolor{text0}{	0	} & \cellcolor{color0} \textcolor{text0}{	0	} & \cellcolor{color0} \textcolor{text0}{	0	} & \cellcolor{color0} \textcolor{text0}{	0	} & \cellcolor{color0} \textcolor{text0}{	0	} & \cellcolor{color0} \textcolor{text0}{	0	} & \cellcolor{color0} \textcolor{text0}{	0	} & \cellcolor{color0} \textcolor{text0}{	0	} & \cellcolor{color0} \textcolor{text0}{	0	} & \cellcolor{color0} \textcolor{text0}{	0	} & \cellcolor{color0} \textcolor{text0}{	0	} & & \cellcolor{othercolor5} \textcolor{othertext5}{	5	} & \cellcolor{othercolor4} \textcolor{othertext4}{	4	} & \cellcolor{othercolor3} \textcolor{othertext3}{	3	} & \cellcolor{othercolor2} \textcolor{othertext2}{	2	} & \cellcolor{othercolor1} \textcolor{othertext1}{	1	} & \cellcolor{othercolor0} \textcolor{othertext0}{	0	} & \cellcolor{othercolor1} \textcolor{othertext1}{	1	} & \cellcolor{othercolor2} \textcolor{othertext2}{	2	} & \cellcolor{othercolor3} \textcolor{othertext3}{	3	} & \cellcolor{othercolor4} \textcolor{othertext4}{	4	} & \cellcolor{othercolor5} \textcolor{othertext5}{	5	} \\
&	1		& \cellcolor{color1} \textcolor{text1}{	1	} & \cellcolor{color1} \textcolor{text1}{	1	} & \cellcolor{color1} \textcolor{text1}{	1	} & \cellcolor{color1} \textcolor{text1}{	1	} & \cellcolor{color1} \textcolor{text1}{	1	} & \cellcolor{color0} \textcolor{text0}{	0	} & \cellcolor{color0} \textcolor{text0}{	0	} & \cellcolor{color0} \textcolor{text0}{	0	} & \cellcolor{color0} \textcolor{text0}{	0	} & \cellcolor{color0} \textcolor{text0}{	0	} & \cellcolor{color0} \textcolor{text0}{	0	} & & \cellcolor{othercolor4} \textcolor{othertext4}{	4	} & \cellcolor{othercolor3} \textcolor{othertext3}{	3	} & \cellcolor{othercolor2} \textcolor{othertext2}{	2	} & \cellcolor{othercolor1} \textcolor{othertext1}{	1	} & \cellcolor{othercolor0} \textcolor{othertext0}{	0	} & \cellcolor{othercolor1} \textcolor{othertext1}{	1	} & \cellcolor{othercolor2} \textcolor{othertext2}{	2	} & \cellcolor{othercolor3} \textcolor{othertext3}{	3	} & \cellcolor{othercolor4} \textcolor{othertext4}{	4	} & \cellcolor{othercolor5} \textcolor{othertext5}{	5	} & \cellcolor{othercolor6} \textcolor{othertext6}{	6	} \\
&	2		& \cellcolor{color2} \textcolor{text2}{	2	} & \cellcolor{color2} \textcolor{text2}{	2	} & \cellcolor{color2} \textcolor{text2}{	2	} & \cellcolor{color2} \textcolor{text2}{	2	} & \cellcolor{color1} \textcolor{text1}{	1	} & \cellcolor{color0} \textcolor{text0}{	0	} & \cellcolor{color0} \textcolor{text0}{	0	} & \cellcolor{color0} \textcolor{text0}{	0	} & \cellcolor{color0} \textcolor{text0}{	0	} & \cellcolor{color0} \textcolor{text0}{	0	} & \cellcolor{color0} \textcolor{text0}{	0	} & & \cellcolor{othercolor3} \textcolor{othertext3}{	3	} & \cellcolor{othercolor2} \textcolor{othertext2}{	2	} & \cellcolor{othercolor1} \textcolor{othertext1}{	1	} & \cellcolor{othercolor0} \textcolor{othertext0}{	0	} & \cellcolor{othercolor1} \textcolor{othertext1}{	1	} & \cellcolor{othercolor2} \textcolor{othertext2}{	2	} & \cellcolor{othercolor3} \textcolor{othertext3}{	3	} & \cellcolor{othercolor4} \textcolor{othertext4}{	4	} & \cellcolor{othercolor5} \textcolor{othertext5}{	5	} & \cellcolor{othercolor6} \textcolor{othertext6}{	6	} & \cellcolor{othercolor7} \textcolor{othertext7}{	7	} \\
&	3		& \cellcolor{color3} \textcolor{text3}{	3	} & \cellcolor{color3} \textcolor{text3}{	3	} & \cellcolor{color3} \textcolor{text3}{	3	} & \cellcolor{color2} \textcolor{text2}{	2	} & \cellcolor{color1} \textcolor{text1}{	1	} & \cellcolor{color0} \textcolor{text0}{	0	} & \cellcolor{color0} \textcolor{text0}{	0	} & \cellcolor{color0} \textcolor{text0}{	0	} & \cellcolor{color0} \textcolor{text0}{	0	} & \cellcolor{color0} \textcolor{text0}{	0	} & \cellcolor{color0} \textcolor{text0}{	0	} & & \cellcolor{othercolor2} \textcolor{othertext2}{	2	} & \cellcolor{othercolor1} \textcolor{othertext1}{	1	} & \cellcolor{othercolor0} \textcolor{othertext0}{	0	} & \cellcolor{othercolor1} \textcolor{othertext1}{	1	} & \cellcolor{othercolor2} \textcolor{othertext2}{	2	} & \cellcolor{othercolor3} \textcolor{othertext3}{	3	} & \cellcolor{othercolor4} \textcolor{othertext4}{	4	} & \cellcolor{othercolor5} \textcolor{othertext5}{	5	} & \cellcolor{othercolor6} \textcolor{othertext6}{	6	} & \cellcolor{othercolor7} \textcolor{othertext7}{	7	} & \cellcolor{othercolor8} \textcolor{othertext8}{	8	} \\
&	4		& \cellcolor{color4} \textcolor{text4}{	4	} & \cellcolor{color4} \textcolor{text4}{	4	} & \cellcolor{color3} \textcolor{text3}{	3	} & \cellcolor{color2} \textcolor{text2}{	2	} & \cellcolor{color1} \textcolor{text1}{	1	} & \cellcolor{color0} \textcolor{text0}{	0	} & \cellcolor{color0} \textcolor{text0}{	0	} & \cellcolor{color0} \textcolor{text0}{	0	} & \cellcolor{color0} \textcolor{text0}{	0	} & \cellcolor{color0} \textcolor{text0}{	0	} & \cellcolor{color0} \textcolor{text0}{	0	} & & \cellcolor{othercolor1} \textcolor{othertext1}{	1	} & \cellcolor{othercolor0} \textcolor{othertext0}{	0	} & \cellcolor{othercolor1} \textcolor{othertext1}{	1	} & \cellcolor{othercolor2} \textcolor{othertext2}{	2	} & \cellcolor{othercolor3} \textcolor{othertext3}{	3	} & \cellcolor{othercolor4} \textcolor{othertext4}{	4	} & \cellcolor{othercolor5} \textcolor{othertext5}{	5	} & \cellcolor{othercolor6} \textcolor{othertext6}{	6	} & \cellcolor{othercolor7} \textcolor{othertext7}{	7	} & \cellcolor{othercolor8} \textcolor{othertext8}{	8	} & \cellcolor{othercolor9} \textcolor{othertext9}{	9	} \\
&	5		& \cellcolor{color5} \textcolor{text5}{	5	} & \cellcolor{color4} \textcolor{text4}{	4	} & \cellcolor{color3} \textcolor{text3}{	3	} & \cellcolor{color2} \textcolor{text2}{	2	} & \cellcolor{color1} \textcolor{text1}{	1	} & \cellcolor{color0} \textcolor{text0}{	0	} & \cellcolor{color0} \textcolor{text0}{	0	} & \cellcolor{color0} \textcolor{text0}{	0	} & \cellcolor{color0} \textcolor{text0}{	0	} & \cellcolor{color0} \textcolor{text0}{	0	} & \cellcolor{color0} \textcolor{text0}{	0	} & & \cellcolor{othercolor0} \textcolor{othertext0}{	0	} & \cellcolor{othercolor1} \textcolor{othertext1}{	1	} & \cellcolor{othercolor2} \textcolor{othertext2}{	2	} & \cellcolor{othercolor3} \textcolor{othertext3}{	3	} & \cellcolor{othercolor4} \textcolor{othertext4}{	4	} & \cellcolor{othercolor5} \textcolor{othertext5}{	5	} & \cellcolor{othercolor6} \textcolor{othertext6}{	6	} & \cellcolor{othercolor7} \textcolor{othertext7}{	7	} & \cellcolor{othercolor8} \textcolor{othertext8}{	8	} & \cellcolor{othercolor9} \textcolor{othertext9}{	9	} & \cellcolor{othercolor10} \textcolor{othertext10}{	10	} \\

 \multicolumn{25}{c}{}\\[-0.8em]
 \multicolumn{2}{c}{} &\multicolumn{11}{c}{$\frac{|m|+|\m|-|m+\m|}{2}$} & &\multicolumn{11}{c}{$|m+\m|$} 
\end{tabular}
	\caption{
	Degree of the Laguerre polynomial in equation \eqref{semi-final-a-2-result} for $j=0$, which gives the number of zeros in the transverse wavefunction, and the power of $p_\perp$, which gives the depth of the central hole. Importantly, there is some overlap between their nontrivial regions.}
	\label{table-of-power-and-degree}
\end{table}

%% file: example.tex
\subsection{An example: ionization of carbon dioxide}
\begin{figure}[hb]
\newlength{\subfiglength}
\setlength{\subfiglength}{4.5cm}
        \centering
        \begin{subfigure}{\subfiglength}
                \centering
                \includegraphics[width=\textwidth]{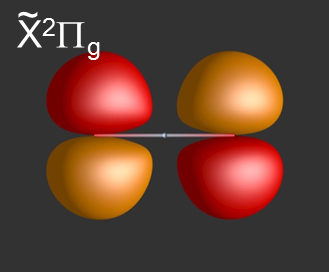}
                \caption{}
                \label{CO2X}
        \end{subfigure}%
        ~ 
        \begin{subfigure}{\subfiglength}
                \centering
                \includegraphics[width=\textwidth]{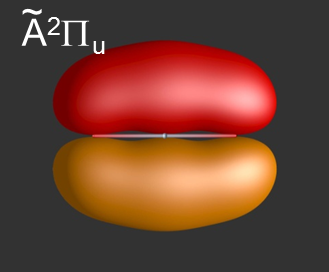}
                \caption{}
                \label{CO2A}
        \end{subfigure}
        ~ 
        \begin{subfigure}{\subfiglength}
                \centering
                \includegraphics[width=\textwidth]{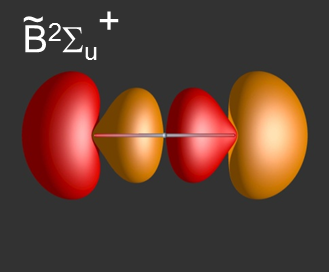}
                \caption{}
                \label{CO2B}
        \end{subfigure}
\caption{Schematic representation of the wavefunctions of the Dyson orbitals for the three main channels in the ionization of a CO$_2$ molecule, denoted in order of increasing energy by X, A and B. Here red and orange denote differing signs of the wavefunction, and the molecular notation uses the greek letters $\Sigma$ and $\Pi$ to denote well-defined magnetic angular momentum number of 0 and $\pm 1$, resp., about the internuclear axis.}
				\label{CO2orbitals}
\end{figure}

We will close this study with a brief illustration of how our results function and what new physical features they show in a typical strong-field ionization problem by considering the ionization of a CO$_2$ molecule. In this case the main channels for ionization are depicted graphically in figure~\ref{CO2orbitals}.

We consider in particular ionization parallel to the internuclear axis in channel B, both directly and in correlation-driven ionization, from which the dominant contribution will come from the lowest-energy state, channel X. In this situation it is quite clear that the direct tunnelling has a magnetic quantum number of $m=0$, and that the initial tunnelling state from channel X has $m=\pm 1$. The correlation interaction potential will also, in this case, have a definite dipole multipolarity with $\m=\pm 1$.

This means that we will have two meaningful contributions available for the cross-ionization case, each with a distinct wavefunction. If $m=\m=\pm1$, then we will have $|m+\m|=2$ but $\frac{1}{2}\left(|m|+|\m|-|m+\m|\right)=0$, so that the wavefunction will have a central zero of order two and no secondary rings; if, on the other hand, $m=-\m$, then we will have $|m+\m|=0$ but $\frac{1}{2}\left(|m|+|\m|-|m+\m|\right)=1$ and the wavefunction will have one secondary ring and no central zero. Thus, the two contributions are quite different from each other and from that of the direct-tunnelling case, which has no rings and no central zero.

\begin{figure}[ht]
%\newlength{\subfiglength}
\setlength{\subfiglength}{4.5cm}
        \centering
        \begin{subfigure}{\subfiglength}
                \centering
                \includegraphics[width=\textwidth]{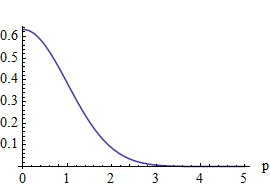}
                \caption{$m=0$, direct ionization in channel B.}
                \label{CO2d}
        \end{subfigure}%
        ~ 
        \begin{subfigure}{\subfiglength}
                \centering
                \includegraphics[width=\textwidth]{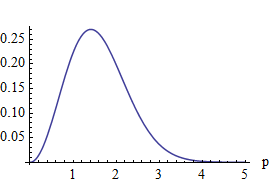}
                \caption{$m=\m=1$, channel X to channel B.}
                \label{CO2cp}
        \end{subfigure}
        ~ 
        \begin{subfigure}{\subfiglength}
                \centering
                \includegraphics[width=\textwidth]{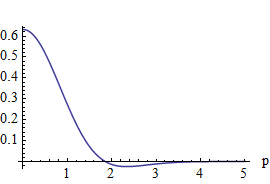}
                \caption{$m=1$, $\m=-1$, channel~X to channel B.}
                \label{CO2ca}
        \end{subfigure}
\caption{Angular distributions, in arbitrary units, for the tunnelling in channel~B of carbon dioxide, parallel to the internuclear axis. We disregard constants and ignore the slow factors in the integral in eq.~\eqref{semi-final-a-2-result}, so we graph only the integral $p_\perp^{|m+\m|}e^{-\frac{1}{2}\tauT p^2}\int_{0}^{\tauT}\xi^{\frac{|m+\m|+|\m|-|m|}{2}}L_{\scriptscriptstyle\frac{|m|+|\m|-|m+\m|}{2}}^{(|m+\m|)}\left(\frac{1}{2}\xi p_\perp^2\right)	  e^{-\Delta I_p(\tauT-\xi)}	\d\xi$, for $\Delta I_p=\tauT=1$.}
				\label{CO2angulardistributions}
\end{figure}

We show in figure \ref{CO2angulardistributions} schematic graphs of these wavefunctions, mainly to show their qualitative differences, which are quite marked. Although the ingredients omitted from the calculation of these angular distributions will degrade the contrast of the different features, they will not change them qualitatively. 

However, the wavefunction for the correlation-driven case will be a coherent sum of both the latter wavefunctions unless a specific effort is made to enforce the initial magnetic number and to measure the angular momentum of the remaining ion. Schemes for which can be readily devised using magnetic fields, but they are nevertheless not necessarily within reach of current experiments. 

If no such distinction is made then the wavefunction for the cross-ionization will qualitatively look quite similar to the direct one -- no central zero and no secondary ring -- but it will be noticeably broader, on a scale which is probably within reach of current detector resolution, and, most importantly, it will be exponentially enhanced from the direct ionization amplitude since it must not pay the full exponential penalty $e^{-I_{p,\textrm{B}} \tauT_\textrm{,B}}$ which damps the direct amplitude.

Thus we see that even with simplistic calculations and with quite standard examples from the field of strong-field ionization, the formalism developed in this work is able to show qualitative differences -- which in all likelihood will become precise, quantitative differences once detailed calculations are done for specific examples -- in the angular distributions of the ionized electron. Further, we have now an intuitive understanding of the underlying processes and mechanisms that are responsible for these differences, understanding which is in general absent from the more precise numerical calculations.

Of course, current work is focused on finding appropriate examples in which these features will be the most evident, and on performing the full calculations to obtain definite physical predictions comparable with current experiments, as well as the investigation of schemes for the more detailed measurement of the initial and final state of the ion which would permit resolving the different multipolarities and therefore observing the more interesting angular distributions.

%% file: RevisedBibliography.bbl
%%Changes to DLMF, cosmetic change to (eds.) in Erdelyi.
%% Removed all \newblock's and \bibAnnoteFile's.

%% file: dissertation.bbl
\begin{thebibliography}{10}
\providecommand{\url}[1]{\texttt{#1}}
\providecommand{\urlprefix}{URL }
\expandafter\ifx\csname urlstyle\endcsname\relax
  \providecommand{\doi}[1]{doi:\discretionary{}{}{}#1}\else
  \providecommand{\doi}{doi:\discretionary{}{}{}\begingroup
  \urlstyle{rm}\Url}\fi
\providecommand{\selectlanguage}[1]{\relax}
\providecommand{\bibAnnoteFile}[1]{%
  \IfFileExists{#1}{\begin{quotation}\noindent\textsc{Key:} #1\\
  \textsc{Annotation:}\ \input{#1}\end{quotation}}{}}
\providecommand{\bibAnnote}[2]{%
  \begin{quotation}\noindent\textsc{Key:} #1\\
  \textsc{Annotation:}\ #2\end{quotation}}
\providecommand{\eprint}[2][]{\url{#2}}

\bibitem{saepaper}
\textsc{L.~Torlina and O.~Smirnova}.
 Time-dependent analytical {$R$}-matrix approach for strong field dynamics. {I}. {O}ne-electron systems. \href{http://dx.doi.org/10.1103/PhysRevA.86.043408}{\emph{Phys. Rev. A} \textbf{86} no.~4 (2012), p. 043\,408}.



\bibitem{mepaper}
\textsc{L.~Torlina, M.~Ivanov, Z.~B. Walters and O.~Smir{\-}nova}.
 Time-dependent analytical {$R$}-matrix approach for strong field dynamics. {II}. {M}any-electron systems. \href{http://dx.doi.org/10.1103/PhysRevA.86.043409}{\emph{Phys. Rev. A} \textbf{86} no.~4 (2012), p. 043\,409}.


\bibitem{VolkovWavefunctions}
\textsc{J.~Bergou}.
 Wavefunctions of a free electron in an external field and their application in intense field interactions. i. non-relativistic treatment\href{http://dx.doi.org/10.1088/0305-4470/13/8/029}{.\emph{Journal of Physics A: Mathematical and General} \textbf{13} no.~8 (1980), p. 2817}.


\bibitem{EVApaper}
\textsc{O.~Smirnova, M.~Spanner and M.~Ivanov}. Analytical solutions for strong field-driven atomic and molecular
  one- and two-electron continua and applications to strong-field problems\href{http://dx.doi.org/10.1103/PhysRevA.77.033407}{. \emph{Phys. Rev. A} \textbf{77} no.~3 (2008), p. 033\,407}.


\bibitem{FetterAndWalecka}
\textsc{A.~L. Fetter and J.~D. Walecka}. \emph{Quantum Theory of Many-Particle Systems} (McGraw-Hill, USA,
  1971).


\bibitem{KatoPerturbationTheory}
\textsc{T.~Kat{\=o}}.
 \emph{Perturbation theory for linear operators}, vol. 132 of \emph{Grundlehren der mathematischen Wissenschaften} (Springer-Verlag, Berlin, 1966).


\bibitem{Arfken}
\textsc{G.~B. Arfken and H.~J. Weber}. \emph{Mathematical Methods for Physicists}. 6$^{\textrm{th}}$ ed. (Academic Press, London, 2005).


\bibitem{Tannor}
\textsc{D.~J. Tannor}. \emph{Introduction to Quantum Mechanics: A Time-Dependent Perspective} (University Science Books, Sausalito, Calif, 2007).


\bibitem{PPT}
\textsc{A.~M. Perelomov, V.~S. Popov and M.~V. Terent'ev}. Ionization of atoms in an alternating magnetic field{.  \emph{Sov. Phys. J.E.T.P.} \textbf{23} no.~5 (1966), pp. 924--934}. Translation of \textit{J. Exptl. Theoret. Phys. (U.S.S.R)} \textbf{50} (1966), pp. 1939--1409.


\bibitem{NIST-handbook}
\href{http://dlmf.nist.gov/}{NIST Digital Library of Mathematical Functions}, release 1.0.6 (2013). 


\bibitem{NIST-hardcopy}
\textsc{F.~W.~J. Olver, D.~W. Lozier, R.~F. Boisvert and C.~W. Clark}, editors. \emph{{N}{I}{S}{T} Handbook of Mathematical Functions} (Cambridge University Press, New York, 2010). Print companion to \cite{NIST-handbook}.


\bibitem{Jackson}
\textsc{J.~D. Jackson}. \emph{Classical Electrodynamics}. $3^\textrm{rd}$ ed. (Wiley, New Jersey, 1999).


\bibitem{gradshteyn}
\textsc{I.~S. Gradshteyn and I.~M. Ryzhik}. \emph{Table of Integrals, Series and Products}. $7^\textrm{th}$ ed. (Academic, Oxford, 2007).


\bibitem{MagnusAndOberhettinger}
\textsc{W.~Magnus and F.~Oberhettinger}. \emph{Formulas and Theorems for the Special Functions of Mathematical Physics}, vol.~52 of \emph{Grundlehren Der Mathematischen Wissenschaften} (Chelsea Publishing Company, New York, 1949).


\bibitem{Askey}
\textsc{G.~Andrews, R.~Askey and R.~Roy}. \emph{Special Functions}, vol.~71 of \emph{Encyclopedia of Mathematics and Its Applications} (Cambridge University Press, Cambridge, 2001).


\bibitem{Erdelyi-I}
\textsc{A.~Erd{\'e}lyi, W.~Magnus, F.~Oberhettinger and F.~G. Tricomi}, editors. \emph{Higher transcendental functions}, vol.~I (McGraw-Hill, London, 1953). Part of the Bateman Manuscript Project.


\bibitem{copeland}
\textsc{T.~Copeland}. The inverse {M}ellin transform, {B}ell polynomials, a generalized {D}obinski relation, and the confluent hypergeometric functions. Retrieved on 21 September 2011 from \textit{Shadows of Simplicity} at \href{http://tcjpn.wordpress.com/2011/11/16/a-generalized-dobinski-relation-and-the-confluent-hypergeometric-fcts/}{http://tcjpn.wordpress.com/2011/11/16/a-generalized-dobinski-relation-and-the-confluent-hypergeometric-fcts/}.


\bibitem{mathoverflow}
\textsc{\href{http://mathoverflow.net/users/12178}{T. Copeland.}} Pochhammer symbol of a differential, and hypergeometric polynomials. Retrieved on 21 September 2011 from \textit{MathOverflow} at \href{http://mathoverflow.net/questions/107191}{http://mathoverflow.net/questions/107191}.


\end{thebibliography}
